\documentclass[twocolumn]{aastex63}

\shorttitle{The FLASHES Survey I}
\shortauthors{O'Sullivan et al.}

\usepackage{enumerate}

\graphicspath{{./}{figures/}}
\newcommand\LyA{\texorpdfstring{Ly$\alpha~$}{Lyman-Alpha}}

\accepted{2020/03/24}
\published{2020/04/29 in The Astrophysical Journal}

\begin{document}
\noindent{\textbf{arXiv Preprint:} This is the version of the article as submitted to The Astrophysical Journal, before editing by the journal. IOP Publishing Ltd is not responsible for any errors or omissions in this version of the manuscript or any version derived from it. The Version of Record is available online at \url{https://doi.org/10.3847/1538-4357/ab838c}}

\title{The FLASHES Survey I: Integral Field Spectroscopy of the CGM around 48 $z\simeq2.3-3.1$ QSOs}

\correspondingauthor{Donal B. O'Sullivan}
\email{dosulliv@caltech.edu}

\author{Donal B. O'Sullivan}
\affiliation{California Institute of Technology}

\author{Christopher Martin}
\affiliation{California Institute of Technology}

\author{Mateusz Matuszewski}
\affiliation{California Institute of Technology}

\author{Keri Hoadley}
\affiliation{California Institute of Technology}

\author{Erika Hamden}
\affiliation{University of Arizona}

\author{James D. Neill}
\affiliation{California Institute of Technology}

\author{Zeren Lin}
\affiliation{California Institute of Technology}

\author{Prachi Parihar}
\affiliation{California Institute of Technology}

\begin{abstract}

We present the pilot study of the Fluorescent Lyman-Alpha Structures in High-z Environments (FLASHES) Survey; the largest integral-field spectroscopy survey to date of the circumgalactic medium at $z=2.3-3.1$. We observed 48 quasar fields with the Palomar Cosmic Web Imager \citep{Matuszewski+10}  { to an average ($\mathrm{2\sigma}$) limiting surface brightness of $\mathrm{6\times10^{-18}~erg~s^{-1}~cm^{-2}~arcsec^{-2}}$ (in a $1''$ aperture and $\sim20\mathrm{\AA}$ bandwidth.)} Extended HI Lyman-$\alpha$ emission is discovered around  {37/48 of the observed quasars, ranging in projected radius from $14-55$ proper kiloparsecs (pkpc), with one nebula exceeding $100$ pkpc in effective diameter. The dimming-adjusted circularly averaged surface brightness profile peaks at $1\times10^{-15}~\mathrm{erg~s^{-1}~cm^{-2}~arcsec^{-2}}$ at $\mathrm{R_\perp\sim20~pkpc}$ and integrated luminosities range from $0.4-9.4\times10^{43}~\mathrm{erg~s^{-1}}$. The emission appears to have an eccentric morphology and an average covering factor of $\sim30-40\%$ at small radii. On average, the nebular spectra are red-shifted with respect to both the systemic redshift and \LyA peak of the quasar spectrum.} The integrated spectra of the nebulae mostly have single or double-peaked profiles with global dispersions ranging from $143-708$ $\mathrm{km~s^{-1}}$, though the individual Gaussian components of lines with complex shapes mostly have dispersions $\leq 400$ $\mathrm{km~s^{-1}}$, and the flux-weighted velocity centroids of the lines vary by thousands of $ \mathrm{km~s^{-1}}$ with respect to the QSO redshifts. Finally, the root-mean-square velocities of the nebulae are found to be consistent  {with those expected from gravitational motions in dark matter halos of mass $\mathrm{Log_{10}(M_h[M_\odot])\simeq12.2^{+0.7}_{-1.2} }$.} We compare these results to existing surveys at higher and lower redshift.\\

\end{abstract}

\section{Introduction} \label{sec:intro}

To understand the evolution of galaxies and their properties, it is critical to understand their environments. Our current picture of galaxy formation takes place in a universe dominated by cold dark matter \citep{BlumenthalFaber84}. In this picture, dark matter structures collapse in a hierarchical manner, dragging with them the baryonic material that eventually forms and fuels galaxies. A key element of this framework is the interplay between galaxies and their environments; galaxies form and evolve through a series of interactions with both the circumgalactic and intergalactic medium (CGM and IGM; e.g., \citealt{bond96,fukugita98}). A long history of accretion, outflows and merger events underlies the properties of galaxies that we observe today (e.g, \citealt{keres05,dekel+09,fumagalli11,correa15}). \\

With the development of highly sensitive integral field spectrographs, there is now the opportunity to contribute substantial direct observational evidence to the discussion around high-redshift galaxy environments, which has so far taken place largely in the realms of theory and simulation. The sensitivity, spatial resolution and spectral flexibility of these new instruments enable exploratory surveys which map the density, morphology, composition and kinematics of the CGM.  { Several integral-field spectroscopy (IFS) studies focused on individual targets have produced remarkable insights into galactic environments at high redshift ($z\gtrsim2$). \cite{Umehata_2019} reported the discovery of giant \LyA{} filaments, spanning more than a megaparsec, embedded in a $z=3.1$ protocluster. Several kinematic studies of extended nebulae around QSOs have revealed evidence for intergalactic gas spiraling into dark matter halos \citep{Martin+19, Martin16, ArrigoniaBattaia18_InfallingGas}. A number of studies over the past 5-6 years have revealed giant \LyA{} nebulae around individual high-redshift galaxies and QSOs ( e.g. \cite{Cai2018, Martin14a, Martin14b}) as well as connecting pairs of QSOs \citep{ArrigoniBattaia_QSOPairs}. Multi-phase observations of similar nebulae have also begun to emerge \citep{Cantalupo19, MarquesChaves2019}. However, to fully characterize the morphology, composition and dynamics of the CGM, large samples with multi-wavelength observations are needed.} \\

\begin{figure}[t]
\centering
\includegraphics[width=0.45\textwidth]{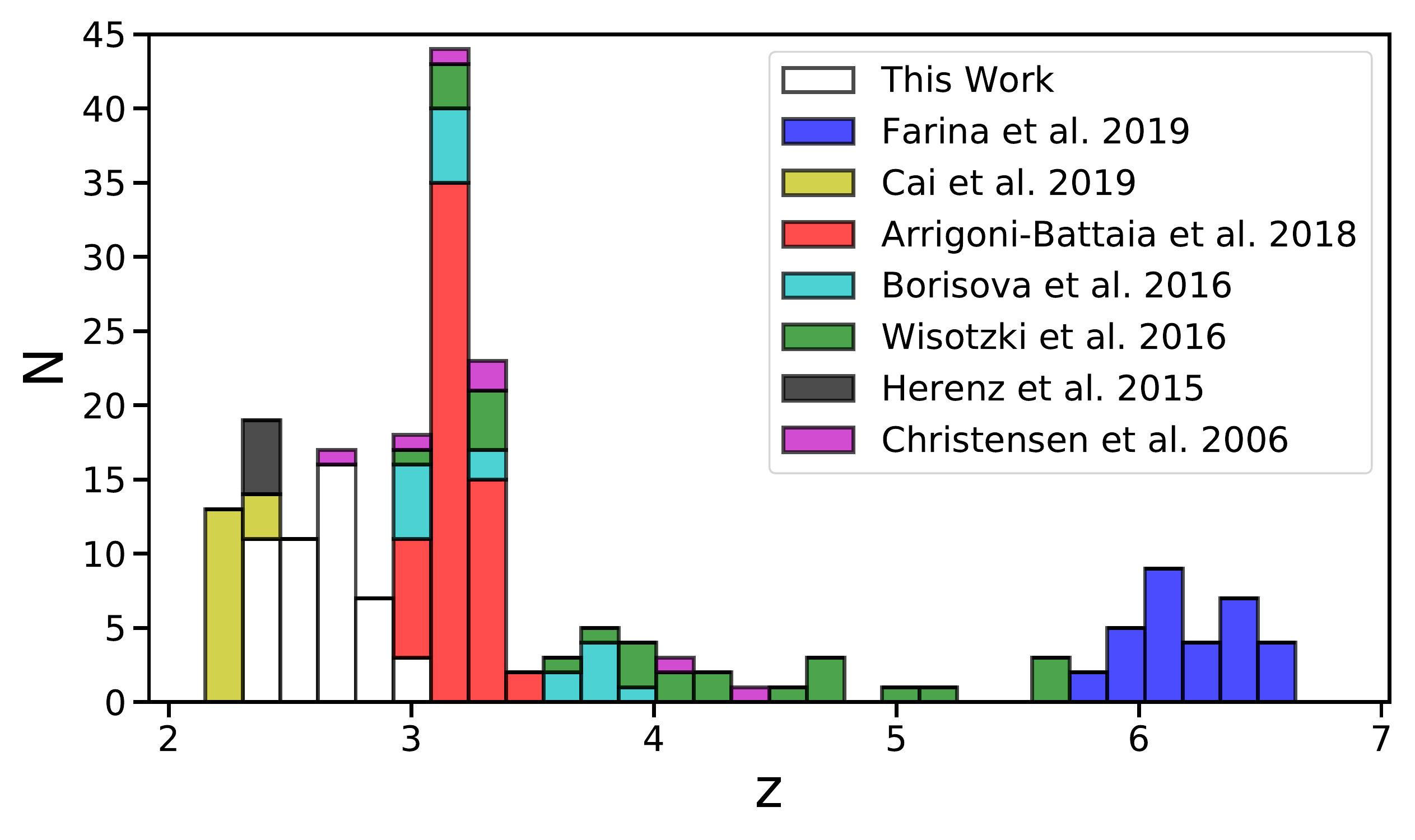}
\caption{ \small{ IFS surveys of extended emission around high redshift galaxies. Surveys are shown as stacked histograms representing the number of targets in each. } }
\label{fig:zVsSB}
\end{figure}

 {\cite{Christensen06} and \cite{Herenz15} provide some of the first examples of IFU surveys of high-redshift QSO environments. These studies focused on extended \LyA{} with sample sizes of seven and five, detecting extended emission in 4/7 and 1/5 targets, respectively.} More recently, teams using the Multi-Unit Spectroscopic Explorer (MUSE) on the Very Large Telescope \citep{Caillier14} have produced surveys of \LyA emission around quasars and galaxies at $z\gtrsim3$ with  sample sizes on the order of tens of targets.  \cite{Borisova16} (hereafter B16) studied 17 bright radio-quiet quasars (and 2 radio-loud) at $z\sim3.5$, finding ubiquitous ``giant" \LyA nebulae on scales larger than $100$ pkpc, with clear asymmetries and a circularly-averaged radial profile following power laws. \cite{ArrigoniBattaia19} (hereafter A19) studied 61 QSOs with a median redshift of $3.17$, finding \LyA nebulae extending on the order of tens of kpc around their quasars. The nebulae they discover have some spread in their degree of spatial symmetry, and they find their radial profiles are best fit by an exponential profile with a scale length $r_H\sim15$ pkpc. They compare this to a narrow-band study at $z\sim2$ \citep{ArrigoniBattaia16}, but with the actual centroid of \LyA emission varying by thousands of $\mathrm{km~s^{-1}}$ from systemic QSO redshifts, it is not clear how reliable narrow-band imaging is without prior knowledge of the emission wavelength.  \cite{Wisotzki16} performed an ultra-deep exposure of the \textit{Hubble} Deep Field South with MUSE, reaching a ($1\sigma$) limiting surface brightness of $1\times10^{-19}$erg s$^{-1}$cm$^{-2}$ arcsec$^{-2}$. They report detections of extended \LyA halos around 21 of the 26 total $z=3-6$ galaxies in their sample, on spatial scales of $R_{\perp} \sim \mathcal{O}(10$ pkpc$)$. The remaining 5 non-detections represent the faintest galaxies in the sample, and thus the are thought to be a matter of insufficient Signal-to-Noise (S/N), making the overall result consistent with the ubiquitous \LyA halos reported in B16. More recently, \cite{Cai2019} observed $16$ QSOs with redshifts $z=2.1-2.3$ using the Keck Cosmic Web Imager \citep{Morrissey18} (KCWI) and report extended emission around all of them, although 2/16 of the nebulae are reported to have projected sizes smaller than $50$ pkpc. The authors find that the nebulae are more asymmetric and lower in surface brightness than the $z>3$ MUSE studies.\\

\begin{figure}[t]
\centering
\includegraphics[width=0.45\textwidth]{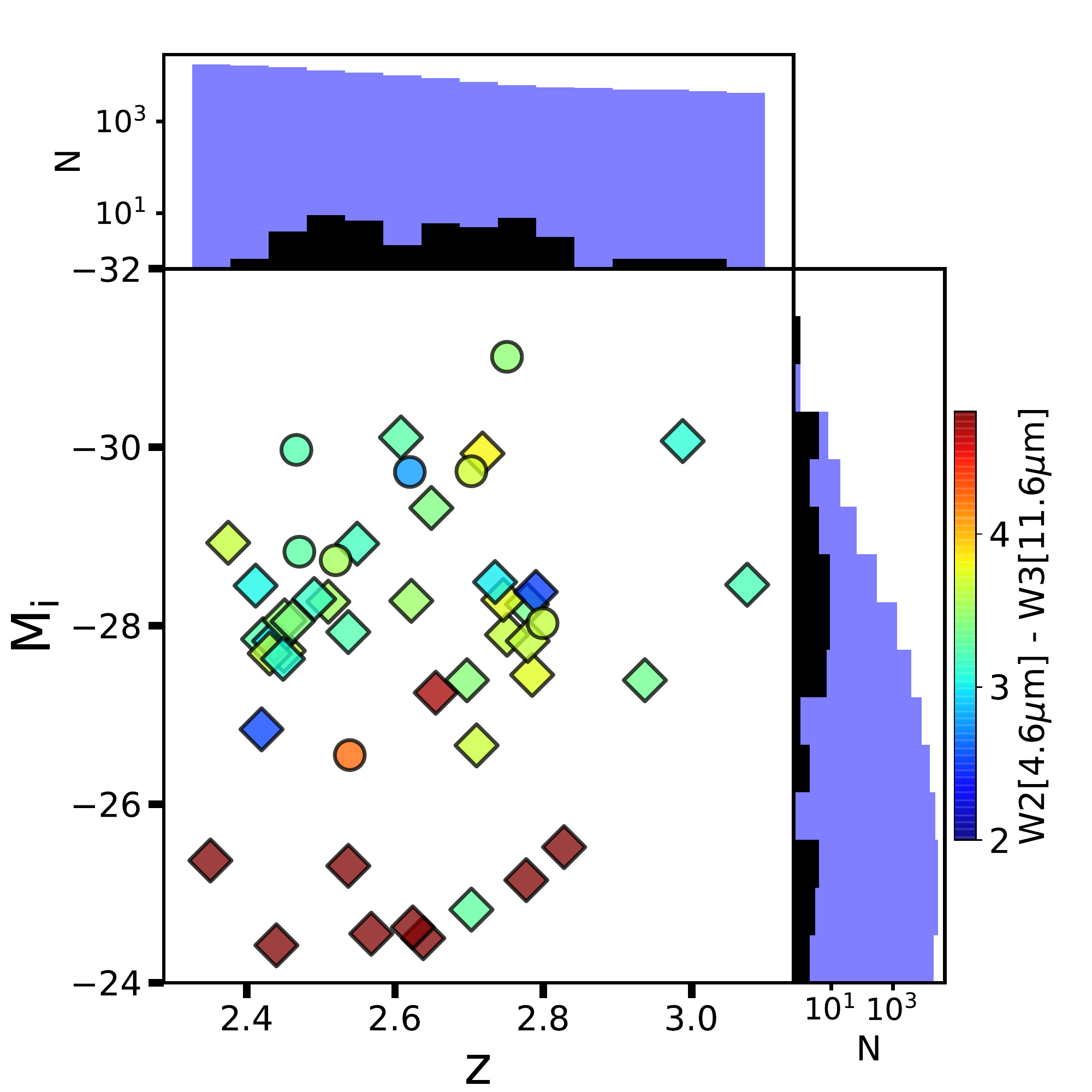}
\caption{ \small{ The FLASHES pilot sample in redshift ($z$) vs. absolute i-band magnitude (M$_i$). Circles indicate targets for which the value of $M_i$ is estimated from the given apparent magnitude, while diamonds indicate those for which a value of M$_i$ was provided in the SDSS DR12Q. The colorbar indicates the WISE infrared color W2-W3 ($4.6-11.6\mathrm{\mu m}$) color.  {The blue histograms represent the distributions of $z$ and $M_i$ in the SDSS DR12Q in the same redshift range.}} }
\label{fig:WISE}
\end{figure}

We have utilized the Palomar Cosmic Web Imager (PCWI) \citep{Matuszewski+10} to conduct a pilot study of the gaseous environments of quasars spanning a redshift range of $z=2.3-3.1$, filling a gap in existing observations (see Figure \ref{fig:zVsSB}). This survey, which we call FLASHES (Fluorescent Lyman-Alpha Structures in High-z Environments), consists of a broad pilot survey component, presented in this paper, and follow-up deep survey, to be presented in a future paper. The pilot survey aims to map \LyA emission from the CGM around the full sample of $48$ quasars at redshifts $2.3 \leq z \leq 3.1$ to a $2\sigma$ surface-brightness limit of $\sim5\times10^{-18}$ erg s$^{-1}$cm$^{-2}$arcsec$^{-2}$ in a $\sim1\mathrm{~arcsec^2}$ aperture for a pseudo-Narrow-band (pNB) image with a typical bandwidth of $20\mathrm{~\AA}$ \textbf{(a limit of $\sim6\times10^{-18}$ erg s$^{-1}$cm$^{-2}$arcsec$^{-2}$ was achieved in the final pNB images.)} Based on existing observational work (e.g., A19, B16), this is expected to be sufficient to map CGM \LyA emission within a $50-100$ proper kiloparsecs (pkpc) of the quasars, and enable us to constrain the morphology and kinematics of the CGM in this redshift range. In addition, we search for the presence of the gaseous filaments that are theorized to feed gas from the cosmic web into dark matter halos. Recent observations have offered tantalizing direct evidence supporting cold-flow accretion from multiple filaments forming `cold inflow disks' \citep{Martin15,Martin16,Martin+19}. A larger sample of observations will allow us to test the validity of such models and their utility in constraining the gas dynamics associated with cold-flow accretion. A subset of these targets will be followed up with deep KCWI exposures for the latter component of the survey, targeting \LyA emission at surface brightness levels an order of magnitude fainter than in the pilot survey, as well as targeting emission from metals such as $\lambda \mathrm{CIV} 1549$ and $\lambda \mathrm{HeII} 1640$ which probe the multi-phase structure of the CGM. \\

In this paper, we focus exclusively on the FLASHES pilot survey. In Section~\ref{methodology} we describe the survey methodology, target selection, and choice of observables. In Section \ref{observations} we present a summary of the observations and data. In Section \ref{reduction} we describe the data reduction with the standard PCWI pipeline and a newly developed Python3 package, `CWITools'. In Section \ref{analysis} we describe the data analysis required to extract and characterize the nebular emission. In Section \ref{results} we present the core observational results: emission maps, kinematic maps, spectra, symmetries and radial profiles. Finally, in Section \ref{discussion} we discuss the implication of our results, sensitivity limits, and comparisons to existing work, before summarizing our findings in Section \ref{conclusions}. For calculations of the luminosity distance and physical plate scales (pkpc per pixel) throughout the paper, we use a $\Lambda$CDM cosmology with $H_0=70\mathrm{kms^{-1}Mpc^{-1}}$, $\Omega_b=0.3$ and $\Omega_\Lambda=0.7$.

\startlongtable
\begin{deluxetable*}{llcccccccc}
\tablecaption{Summary of FLASHES Pilot Observations \label{tab:pilot_targets}}
\tablewidth{0pt}
\tablehead{\colhead{ID} & \colhead{Target Name} &  \colhead{Coordinates} & \colhead{$\mathrm{z_{QSO}}$} &
\colhead{$\mathrm{M_i}$} & \colhead{Seeing} & \colhead{Clouds} & \colhead{Exp}& \colhead{$\mathrm{SB_{2\sigma}}$\tablenotemark{g}} \\
 &  & \colhead{hh:mm:ss.ss $\pm$dd:mm:ss.ss} & ($\pm2\sigma$) &
\colhead{ABmag} & \colhead{arcsec} &  & \colhead{min}  & ($10^{-18}$cgs)
}

\startdata
1	&	HS1700+6416\tablenotemark{a}	&	17:01:01.00 +64:12:09.10	&	2.7375	$\pm$	0.0010	&	-31.01	&	1.3	&	CR	&	222	&	0.48	\\
2	&	SDSS1112+1521	&	11:12:52.45 +15:21:23.50	&	2.7898	$\pm$	0.0003	&	-28.38	&	1.1	&	CR	&	60	&	1.07	\\
3	&	SDSS0834+1238	&	08:34:08.63 +12:38:36.54	&	2.7465	$\pm$	0.0013	&	-28.29	&	1	&	CR-P	&	60	&	1.06	\\
4	&	SDSS1011+2941\tablenotemark{a}	&	10:11:56.00 +29:41:42.00	&	2.6400			&	-30.11	&	1.3	&	CR	&	60	&	1.27	\\
5	&	SDSS0735+3744	&	07:35:35.44 +37:44:50.42	&	2.7514	$\pm$	0.0003	&	-27.90	&	1.3	&	CR	&	60	&	0.94	\\
6	&	SDSS0103+1316\tablenotemark{a}	&	01:03:11.27 +13:16:17.70	&	2.6985	$\pm$	0.0010	&	-29.93	&	1.9	&	CR	&	56	&	1.08	\\
7	&	SDSS0958+4703	&	09:58:45.42 +47:03:24.43	&	2.4907	$\pm$	0.0003	&	-28.30	&	2.1	&	CR	&	60	&	1.61	\\
8	&	SDSS0132+3326\tablenotemark{d}	&	01:32:44.60 +33:26:55.42	&	2.4205	$\pm$	0.0003	&	-26.84	&	1.4	&	CR-F	&	60	&	0.81	\\
9	&	SDSS0837+1459	&	08:37:12.89 +14:59:17.38	&	2.5100	$\pm$	0.0004	&	-28.27	&	2	&	CR	&	48	&	1.36	\\
10	&	SDSS2241+1225	&	22:41:45.11 +12:25:57.24	&	2.6222	$\pm$	0.0006	&	-28.28	&	1.7	&	CR	&	60	&	1.23	\\
11	&	SDSS1626+4858\tablenotemark{e}	&	16:25:59.89 +48:58:17.49	&	2.7347	$\pm$	0.0007	&	-28.49	&	1.1	&	CR-P	&	54	&	1.03	\\
12	&	SDSS2328+0443\tablenotemark{c}	&	23:28:28.48 +04:43:46.84	&	2.5681	$\pm$	0.0001	&	-24.55	&	1.7	&	CR-P	&	60	&	1.17	\\
13	&	SDSS1002+2008	&	10:02:55.43 +20:08:02.56	&	2.6555	$\pm$	0.0006	&	-27.25	&	1.2	&	CR	&	60	&	1.62	\\
14	&	SDSS1218+2414	&	12:18:10.98 +24:14:10.90	&	2.3752	$\pm$	0.0008	&	-28.93	&	1.5	&	CR-P	&	40	&	1.84	\\
15	&	SDSS0108+1635\tablenotemark{a}	&	01:08:06.40 +16:35:50.00	&	2.6399	$\pm$	0.0003	&	-29.32	&	1.6	&	P	&	70	&	1.29	\\
16	&	SDSS0753+4030	&	07:53:26.11 +30:40:38.63	&	2.9304	$\pm$	0.0004	&	-28.72	&	1.5	&	TC	&	60	&	1.06	\\
17	&	SDSS0057+0346	&	00:57:37.78 +03:46:45.03	&	2.4365	$\pm$	0.0005	&	-27.83	&	2.1	&	CR	&	60	&	1.43	\\
18	&	SDSS0012+3344	&	00:12:15.26 +33:44:00.33	&	2.4502	$\pm$	0.0003	&	-27.72	&	2	&	CR-P	&	60	&	1.96	\\
19	&	SDSS0013+1630\tablenotemark{d}	&	00:13:55.86 +16:30:51.78	&	2.5907	$\pm$	0.0002	&	-27.93	&	1.6	&	CR	&	60	&	0.89	\\
20	&	SDSS0006+1614	&	00:06:39.47 +16:14:59.30	&	2.4216	$\pm$	0.0005	&	-27.85	&	1.8	&	TC	&	60	&	1.29	\\
21	&	SDSS0730+4340	&	07:30:02.80 +43:40:03.04	&	2.9367	$\pm$	0.0005	&	-27.39	&	1.8	&	CR	&	60	&	0.64	\\
22	&	SDSS0822+1626	&	08:22:00.22 +16:26:52.87	&	2.4541	$\pm$	0.0005	&	-28.06	&	1.4	&	CR-P	&	60	&	1.39	\\
23	&	SDSS1428+2336	&	14:28:10.96 +23:36:40.21	&	2.7792	$\pm$	0.0004	&	-27.83	&	1.2	&	CR-F	&	60	&	0.98	\\
24	&	SDSS0851+3148\tablenotemark{c}	&	08:51:24.79 +31:48:55.72	&	2.6384	$\pm$	0.0005	&	-24.50	&	1.3	&	CR-P	&	60	&	1.36	\\
25	&	SDSS0214+1912\tablenotemark{a}	&	02:14:29.71 +19:12:37.40	&	2.4710			&	-28.83	&	1.8	&	CR-P	&	70	&	1.19	\\
26	&	SDSS0015+2927	&	00:15:53.14 +29:27:21.45	&	3.0755	$\pm$	0.0003	&	-28.46	&	1.3	&	CR	&	60	&	0.72	\\
27	&	SDSS2339+1901\tablenotemark{b}	&	23:39:44.60 +19:01:52.00	&	2.6200			&	-29.72	&	1.8	&	CR-P	&	60	&	1.54	\\
28	&	SDSS0300+0222\tablenotemark{b}	&	03:00:46.02 +02:22:45.24	&	2.5240			&	-28.73	&	2	&	CR	&	68	&	1.26	\\
29	&	SDSS0639+3819	&	06:39:01.60 +38:19:15.24	&	2.5393	$\pm$	0.0003	&	-26.55	&	1.3	&	CR	&	60	&	1.87	\\
30	&	SDSS2338+1504\tablenotemark{e} 	&	23:38:23.16 +15:04:45.22	&	2.4121	$\pm$	0.0009	&	-28.45	&	1.8	&	CR-P	&	56	&	1.26	\\
31	&	SDSS0321+4132	&	03:21:08.45 +41:32:20.86	&	2.4457	$\pm$	0.0007	&	-29.97	&	1.8	&	TC	&	70	&	1.10	\\
32	&	SDSS0211+3117	&	02:11:39.25 +31:17:24.67	&	2.7854	$\pm$	0.0005	&	-27.45	&	1.7	&	CR-P	&	60	&	1.12	\\
33	&	SDSS0118+1950	&	01:18:39.93 +19:50:27.86	&	2.7780	$\pm$	0.0002	&	-28.24	&	1.1	&	CR	&	60	&	1.10	\\
34	&	SDSS0144+0838	&	01:44:14.08 +08:38:20.40	&	2.4307	$\pm$	0.0008	&	-27.69	&	1.8	&	TC	&	60	&	1.10	\\
35	&	SDSS1532+3059	&	15:32:58.24 +30:59:06.59	&	2.5492	$\pm$	0.0004	&	-28.92	&	1.3	&	CR-P	&	40	&	1.43	\\
36	&	SDSS2151+0921	&	21:51:55.30 +09:21:14.07	&	2.4493	$\pm$	0.0005	&	-27.63	&	1.3	&	CR-P	&	56	&	1.22	\\
37	&	SDSS0303+3838	&	03:03:09.16 +38:38:57.20	&	2.7989	$\pm$	0.0004	&	-28.03	&	1.1	&	CR	&	60	&	1.52	\\
38	&	SDSS0126+1559	&	01:26:36.12 +15:59:29.94	&	2.6969	$\pm$	0.0004	&	-27.39	&	1.5	&	P	&	64	&	1.45	\\
39	&	SDSS2234+2637\tablenotemark{c}	&	22:34:53.07 +26:37:25.00	&	2.7774	$\pm$	0.0009	&	-25.15	&	1.5	&	CR-P	&	60	&	1.09	\\
40	&	SDSS2259+2326	&	22:59:04.02 +23:26:43.91	&	2.4622	$\pm$	0.0012	&	-28.05	&	1.3	&	CR	&	60	&	1.41	\\
41	&	SDSS0041+1925	&	00:41:09.83 +19:25:19.85	&	2.7096	$\pm$	0.0007	&	-26.66	&	1.5	&	TC	&	60	&	1.10	\\
42	&	SDSS1552+1757	&	15:52:00.50 +17:57:22.70	&	2.7034	$\pm$	0.0003	&	-24.82	&	1.2	&	CR	&	80	&	1.22	\\
43	&	SDSS0205+1902\tablenotemark{b}	&	02:05:27.51 +19:02:29.10	&	2.7030			&	-29.73	&	1.6	&	TC	&	70	&	1.25	\\
44	&	SDSS1258+2123\tablenotemark{c}	&	12:58:11.25 +21:23:59.70	&	2.6245	$\pm$	0.0003	&	-24.62	&	1.3	&	CR	&	70	&	1.13	\\
45	&	SDSS2340+2418\tablenotemark{c,d}	&	23:40:39.74 +24:18:59.15	&	2.3513	$\pm$	0.0007	&	-25.37	&	1.4	&	CR-F	&	60	&	0.50	\\
46	&	SDSS0107+1104\tablenotemark{c,d}	&	01:07:14.66 +11:04:46.10	&	2.5369	$\pm$	0.0010	&	-25.31	&	1.4	&	CR	&	60	&	1.00	\\
47	&	SDSS2350+3135\tablenotemark{c}	&	23:50:36.46 +31:35:05.02	&	2.8285	$\pm$	0.0020	&	-25.52	&	1.5	&	CR	&	60	&	1.43	\\
48	&	SDSS0137+2405\tablenotemark{c} 	&	01:37:58.65 +24:05:41.01	&	2.4398	$\pm$	0.0012	&	-24.42	&	1.8	&	CR-P	&	60	&	1.97	\\
\enddata

\tablecomments{ $\mathrm{z_{QSO}}$, the QSO's systemic redshift, is from DR12Q where available and SDSS or 2MASS elsewhere. $\mathrm{M_i}$, the QSO's absolute i-band magnitude, is taken from DR12Q where given and derived using the luminosity distance elsewhere. For cloud cover: CLR - Clear, PC - Patchy Clouds, TC-Thin Cirrus, F-Fog.}

\tablenotetext{a}{Literature target}
\tablenotetext{b}{Target selected from SIMBAD to fill observing schedule}
\tablenotetext{c}{Dust-obscured targets, indicated by $\mathrm{W2-W3}$ WISE color.}
\tablenotetext{d}{Observed without Nod-and-Shuffle technique.}
\tablenotetext{e}{Radio-loud QSO}
\tablenotetext{g}{ {Limiting surface brightness in a $1\mathrm{~arcsec^2}$ aperture in a single $\mathrm{0.55~\AA}$ cube layer.}}

\end{deluxetable*}

\section{Survey Methodology}\label{methodology}

\subsection{Choice of Observables}

At a redshift of $z=2$, \cite{Bertone13} estimate that 80\% of the energy emitted by the diffuse  material of the CGM/IGM is carried by emission lines, with the remaining 20\% in continuum emission. The Hydrogen Lyman series - and primarily \LyA - is the main contributor to this, carrying 20\% of the line emission energy budget. Metal lines serve as better tracers for a wider range of over-densities or temperatures. They are typically an order of magnitude fainter than \LyA and depend strongly on gas metallicity and phase (\citealt{BertoneSchaye12}). The ubiquity and brightness of \LyA make it a clear choice for the pilot survey's goal of detecting and mapping the cool-warm phase of the CGM. With \LyA, we can constrain the morphology, density and baryonic mass of detected nebulae. Targets of interest can then be followed-up in the deep study component of the survey, targeting metal lines such as HeII and CIV, in order to get a more complete picture of the multi-phase CGM.

\subsection{Target Selection}

The FLASHES sample is primarily selected from SDSS DR12Q - the QSO Catalog from the 12th Data Release of the Sloan Digital Sky Survey \citep{Alam+2015}. Targets were chosen within the redshift range of $z\simeq2.3$ and $z\simeq3.1$ based on the observability of \LyA given the wavelength range accessible to the medium resolution grating of PCWI. An effort was made to select targets evenly across this redshift range though operational constraints such as the number of required instrument settings on a singe night or the times at which various targets were observable from Palomar at low airmass, limited this effort. An effort was also made to select a range of absolute i-Band (rest-frame optical) magnitudes,  {as opposed to focusing on the brightest quasars}, in order to explore any dependence of the nebular emission on QSO brightness.  {However, the FLASHES sample is still somewhat biased towards brighter QSOs when compared to the full distribution in the SDSS DR12Q.} The distributions of the pilot sample in redshift and absolute i-band magnitude, alongside the distribution of these values in the SDSS DR12Q, is shown in Figure~\ref{fig:WISE}. A WISE color cut of of $\mathrm{W2[4.8\mu m]-W3[11.6\mu m]>4.8}$, was used to identify heavily dust-obscured targets within the SDSS DR12Q which were expected to exhibit extended \LyA emission, as discussed in \cite{Bridge13}. The FLASHES pilot sample includes 9 of these dust-obscured targets, indicated in Figure ~\ref{fig:WISE} by the colorbar. We note that these 9 targets are not classical `Type II' QSOs, as their spectra do contain broad line emission despite exhibiting heavily suppressed continuum emission. Over the course of the pilot survey, we included four additional targets from \cite{TrainorSteidel12} which were known to exhibit extended emission but lacked any IFS observations. Table~\ref{tab:litTargs} shows the names of these targets in both papers, for reference.\\

\begin{deluxetable}{rrl}
\caption{Targets included from \cite{TrainorSteidel12} \label{tab:litTargs}}
\tablehead{ \colhead{ID}  &  \colhead{Name} & \colhead{Name (Source)}}
\startdata
 1 &  HS1700+6416 &  HS1700+6416\\
 4 &  SDSS1011+2941 &  Q1009+29 (CSO 38)\\
 6 &  SDSS0103+1316 &  Q0100+13 (PHL 957)\\
 15 &  SDSS0108+1635 &  HS0105+1619 
\enddata
\end{deluxetable}

In this work, a distinction is made between the total detection rate and the `blind' detection rate, which excludes these five targets. Three targets were selected by searching the SIMBAD Astronomical Database based on coordinates and redshift to fill gaps in our observing schedule where no suitable targets were available from the SDSS DR12Q. Finally, two soft constraints were applied in our selection. First, targets with few obscuring foreground stars and galaxies were preferred, as blended and nearby sources can make the data analysis step of isolating the nebular emission prohibitively complicated. Second, where radio data was available, radio-quiet sources were preferred. One of the goals of FLASHES is to study gas dynamics and cold inflows from the cosmic web, and the presence of jets associated with radio-loud quasars would complicate this analysis. Of the 48 pilot targets, only two are detected in radio and classify as radio loud  {using the criterion $R=\mathrm{f_{\nu}^{1.4GHz}/f_{\nu}^{4400\AA} \gtrsim 10}$  \citep{kellerman89}}. Table ~\ref{tab:pilot_targets} provides a breakdown of all of the pilot survey targets, coordinates and sources. \\

The FLASHES target selection is multi-pronged, and there are biases in the methodology towards radio-quiet quasars with fields relatively clear of nearby/foreground sources. Any biases in the SDSS DR12Q will also be inherited. As such, the authors caution that while this is the first large sample of its kind in this redshift range, the results of this work should not be quickly or trivially extrapolated to the wider galaxy population.

\section{Observations and Ancillary Data} \label{observations}

We observed $48$ QSO fields between 2015 and 2018 on the 5-meter Hale telescope at Palomar using PCWI. PCWI is an image-slicer IFS mounted at the Cassegrain focus of the 5-meter Hale telescope at Palomar Observatory. The instrument field of view is $60''\times40''$ (approximately $480\times320~\mathrm{pkpc^2}$ at $z\sim2-3$). The longer axis is composed of 24 slices with a width of $\sim2.5''$ and an in-slice pixel size of $\sim0.55''$.  {Typical seeing at Palomar is $\sim1.5''$ full-width at half-maximum, so individual exposures are slit-limited along the y-axis and seeing-limited along the x-axis. Exposures can be dithered to increase the sampling rate along the y-axis.} Gratings and filters are interchangeable on PCWI. Our pilot observations used the medium resolution Richardson grating, which has a resolution of $R\simeq2500$ and operates over a bandpass of $400-600~\mathrm{nm}$. With a spectral plate scale of $0.55$~\AA/px, the minimum resolution element $\Delta\lambda\sim2$\AA~ is sampled above the Nyquist rate. For all observations, we use a filter with a bandpass of $350-580~\mathrm{nm}$. For 44/48 of the targets, Nod-and-Shuffle (N\&S) mode of PCWI was used (see \cite{Matuszewski+10} for details). In short, N\&S allows for highly accurate sky subtraction, almost entirely free of systematic residuals, at the cost of bandwidth and statistical noise. The standard pilot observation consists of three 40 minute N\&S observations (20 minutes on sky, 20 minutes on source), stacked for a total of 1 hour on source and 1 hour on sky. Seeing conditions at Palomar are generally such that the full-width at half-maximum (FWHM) of a point source is $1-2''$. To increase the spatial sampling, the second and third N\&S observations are dithered by $\pm1''$ perpendicular to the direction of the slices. N\&S mode was not used for four targets in this sample (see Table~\ref{tab:pilot_targets}). This was done on one observing run in the interest of spending more telescope time on source rather than on sky, but the increase in systematic sky residuals was not deemed worth it for future observations. For these, an A-B-B-A pattern was used to alternate between 20-minute science frames and 10-minute sky frames. Lastly, one target (HS1700+6416) has a significantly longer total exposure time, as it was one of the earliest targets to be observed. However, as it still represents an initial exploration, it is included in the Pilot sample. \\

The goal for each target was $60$ minutes on source and $60$ minutes on sky. For four targets we obtained 56, 54, 48 and 40 minutes in total due to time lost to poor weather.  {Some fields are not centered exactly QSO, due in part to guiding constraints (the guider and instrument fields of view have a fixed offset and orientation) and in part to position foreground sources in such a way that they could be masked/subtracted. Because of this, a small number of the fields shown in Figure~\ref{fig:targplots} have blank areas on one side.} Multi-wavelength ancillary data were obtained for each target when available. Near- and far-UV data were obtained from GALEX (\cite{Bianchi+2011}). Photometric u, g, r, i and z-band magnitudes were obtained from the Sloan Digital Sky Survey's Photometric Catalog's 12th data release (SDSS DR12 - \cite{Alam+2015}). 2MASS J, H and K-band magnitudes as well as WISE 3.35$\mu$m, 4.6$\mu$m, 11.6$\mu$m, and 22.1$\mu$m magnitudes were obtained from the AllWISE Data Release \citep{Cutri+2013}. Finally, 1.4GHz radio fluxes were obtained from the FIRST Survey \citep{Helfand+2015}. All magnitudes and fluxes were converted to AB magnitudes for consistency. These data are presented in Appendix \ref{app:seds}.

\section{Data Reduction} \label{reduction}

\subsection{Standard Pipeline Reduction}
Initial data reduction is performed using the standard PCWI Data Reduction Pipeline\footnote{PCWI DRP: https://github.com/scizen9/pderp}, which converts raw, 2D science frames into flux-calibrated, three-dimensional cubes with real-world coordinate systems in RA, DEC and wavelength. A detailed description of PCWI calibration products, with useful reference images, is available in \cite{Matuszewski+10}. \\

All frames are initially cosmic-ray subtracted, and bias subtracted. As PCWI is a Cassegrain-mounted instrument, there are sometimes slight offsets in the data due to gravitational flexure. These are corrected using a 2D cross-correlation method before the construction of 3D data products. The pipeline then maps from the 2D space of raw images to the 3D image coordinates ($x,y,z$) and on-sky/wavelength coordinates ($\alpha,\delta,\lambda$) using a `continuum bars' image and an `arc-flat' image, which have known spatial and spectral features, respectively. The uneven illumination of the image slicer is then corrected for in two steps - first correcting the profile within each slice, and then correcting the slice-to-slice variation. Finally, a spectrophotometric standard star observation is used to convert detector counts to physical flux units. The final product of this pipeline is a three-dimensional, flux calibrated data cube for each individual exposure. For the four targets observed without N\&S mode, sky subtraction was performed by extracting 2D sky spectra from the adjacent sky frames and scaling them on a slice-by-slice basis. 

\subsection{Cube Correction and Coadding}

The large volume of data in this survey and complex nature of the 3D IFS data required the development of a toolkit for common reduction and analysis functions. \texttt{CWITools}\footnote{CWITools: https://github.com/dbosul/CWITools} is a \texttt{Python3} toolkit written specifically for the analysis of PCWI and KCWI data. It is available publicly on GitHub and will be presented in more detail in a future paper. \\

Before co-adding, individual exposure cubes are first corrected by adjusting their world-coordinate system and trimming them. The RA/DEC coordinate system is corrected for each frame using the known location of a visible source in the field (typically the target QSO, though occasionally an adjacent star). The actual position of the source is measured in image coordinates, and then the coordinate system is updated such that that location accurately points to the known RA/DEC. This does not correct for any errors in rotation, though these are expected to be negligible. In a similar way, the wavelength axis is corrected using the positions of known sky emission lines. Finally, the cube is trimmed to only the wavelength range that which is shared by all slices  {(as each slice has a slightly different bandpass)}, and edge pixels are trimmed off the spatial axes. The corrected and cropped input cubes are then coadded.\\

\texttt{CWITools} uses a custom-built method for this  {which calculates the footprint of each input pixel on the coadd frame, and distributes flux onto the coadd grid accordingly.The on-sky footprint of each input frame is calculated, and a new world-coordinate-system representing the coadd frame is constructed so that it encompasses all input data and has an aspect ratio of 1:1. The wavelength axes of the input cubes are first aligned using linear interpolation to perform any sub-pixel shifts (variance is propagated by convolving with the square of the convolution matrix used to shift the data.) With the cubes aligned in wavelength, the problem of coadding then becomes two-dimensional. To calculate the footprint of each input pixel on the coadd grid, the vertices of each input pixel is represented as a vector of four $(x^v_i,y^v_i)$ coordinates, where the $i$ subscript denotes the input coordinate system and the superscript $v$ denotes that they represent the pixel vertices (not the center). These vertices are then transformed into a vector of on-sky coordinates (i.e. a vector of $\alpha^v$, $\delta^v$ coordinates) and from there into a vector of coadd frame coordinates, ($x^v_o, y^v_o$). A polygon representing the footprint of the input pixel is then created in coadd coordinates, and the overlapping area with each pixel in the coadd grid is calculated. The flux from the input pixel is then redistributed accordingly, following:}

\begin{equation}
    F_{coadd}(x, y) = \sum_{x_i} \sum_{y_i} F_{input}(x_i, y_i) f(x, y, x_i, y_i)
\end{equation}

 {where $f(x, y, x_i, y_i)$ is the fraction of the footprint of the input pixel ($x_i, y_i$) that falls on the output pixel ($x,y$). Because the pixels are represented as flexible polygons, this method allows for the input of frames with arbitrary spatial resolution and position-angle. It also allows a variance estimate to be propagated following:}

\begin{equation}
    V_{coadd}(x, y) = \sum_{x_i} \sum_{y_i} V_{input}(x_i, y_i) f^2(x, y, x_i, y_i)
\end{equation}

 {However, dividing up the pixel this way is implicitly performing linear interpolation, which introduces covariance between the pixels. We discuss how this is handled in Section~\ref{analysis:covar}. The final pixel size in the coadded image has a 1:1 aspect ratio with the same plate scale as the x-axis of the input data ($\simeq0.55~\mathrm{arcsec~px^{-1}}$)}.\\

\section{Data Analysis}\label{analysis}

\begin{deluxetable*}{lcccc|lcccc}
\tablehead{ \colhead{ID}  & \colhead{$\mathrm{\lambda_c}$} & \colhead{$\mathrm{\delta \lambda}$} & \colhead{$\mathrm{\delta}$} & $\mathrm{SB_{2\sigma}}\tablenotemark{a}$ &\colhead{ID}  & \colhead{$\mathrm{\lambda_c}$} & \colhead{$\mathrm{\delta \lambda }$} & \colhead{$\mathrm{\delta v}$} & $\mathrm{SB_{2\sigma}}\tablenotemark{a}$ \\ \hline
 & \colhead{$\mathrm{\AA}$} & \colhead{$\mathrm{\AA}$} & \colhead{$\mathrm{km~s^{-1}}$} & $\mathrm{10^{-18} ~cgs}$ & & \colhead{$\mathrm{\AA}$} & \colhead{$\mathrm{\AA}$} & \colhead{$\mathrm{km~s^{-1}}$} & $\mathrm{10^{-18} ~cgs}$}
\startdata
1	&	4555	&	21	&	1383	&	2.92	&	25	&	4240	&	32	&	2264	&	7.22	\\
2	&	4605	&	13	&	846	&	4.44	&	26	&	4980	&	10	&	602	&	2.76	\\
3	&	4569	&	25	&	1641	&	6.02	&	27	&	4407	&	27	&	1837	&	9.11	\\
4	&	4437	&	20	&	1352	&	5.49	&	28	&	4295	&	13	&	908	&	4.83	\\
5	&	4559	&	19	&	1250	&	4.39	&	29	&	4304	&	25	&	1742	&	9.28	\\
6	&	4522	&	16	&	1061	&	4.84	&	30	&	4166	&	17	&	1224	&	5.95	\\
7	&	4240	&	20	&	1415	&	10.43	&	31	&	4208	&	20	&	1425	&	5.56	\\
8	&	4160	&	20	&	1442	&	4.85	&	32	&	4615	&	15	&	975	&	4.14	\\
9	&	4283	&	29	&	2031	&	7.98	&	33	&	4590	&	19	&	1241	&	4.93	\\
10	&	4430	&	23	&	1557	&	6.43	&	34	&	4211	&	18	&	1282	&	5.00	\\
11	&	4522	&	29	&	1923	&	5.78	&	35	&	4351	&	13	&	896	&	4.40	\\
12	&	4313	&	15	&	1043	&	5.40	&	36	&	4188	&	15	&	1074	&	5.20	\\
13	&	4456	&	12	&	807	&	5.44	&	37	&	4620	&	12	&	779	&	4.39	\\
14	&	4128	&	27	&	1962	&	9.52	&	38	&	4488	&	15	&	1002	&	6.13	\\
15	&	4441	&	15	&	1013	&	5.92	&	39	&	4587	&	23	&	1504	&	5.89	\\
16	&	4786	&	34	&	2131	&	6.48	&	40	&	4211	&	23	&	1638	&	6.63	\\
17	&	4205	&	27	&	1926	&	7.99	&	41	&	4497	&	23	&	1534	&	6.66	\\
18	&	4193	&	10	&	715	&	6.15	&	42	&	4497	&	23	&	1534	&	6.36	\\
19	&	4365	&	18	&	1237	&	4.81	&	43	&	4537	&	25	&	1653	&	6.04	\\
20	&	4189	&	17	&	1217	&	5.55	&	44	&	4396	&	19	&	1296	&	6.15	\\
21	&	4806	&	21	&	1310	&	2.94	&	45	&	4066	&	20	&	1475	&	4.85	\\
22	&	4237	&	26	&	1840	&	8.43	&	46	&	4308	&	19	&	1323	&	4.67	\\
23	&	4611	&	15	&	975	&	4.22	&	47	&	4650	&	24	&	1548	&	8.09	\\
24	&	4442	&	27	&	1823	&	7.79	&	48	&	4194	&	16	&	1144	&	7.82	\\
\enddata
\caption{Final pNB image parameters for the FLASHES Pilot Observations.}
\tablenotetext{a}{Limiting surface brightness in $1\mathrm{~arcsec^{-2}}$ aperture in units of $10^{-18}\mathrm{~erg~s^{-1}~cm^{-2}~arcsec^{-2}}$.}
\label{tab:pnbParams}
\end{deluxetable*}

In this section we describe the steps taken to extract extended \LyA emission in the CGM and produce scientific products from the data. We initially search for extended emission using a two-dimensional channel map method, which trades spectral resolution for an increased signal-to-noise ratio (SNR). Once emission is identified, we then analyze it in three dimensions to obtain kinematics and spectra.

\subsection{Generation of pseudo-Narrow-Band Images}
 {In order to identify extended emission, an initial exploration of the cubes is performed using pNB images, which are narrow-band images formed by collapsing wavelength layers of the data cube. For the purpose of studying extended emission, continuum emission must be subtracted. For each pNB image, a white-light (WL) image is formed by summing $\sim50$\AA~ on either side of the current pNB bandpass. Pixels within a circular region of radius $\sim1.5''$ around the QSO are then used to calculate a set of scaling factors between the images. The scaling factors are sigma-clipped at $\pm3\sigma$ and the resulting mean is taken as the global scaling factor for the WL image. The WL image is then scaled and subtracted from the pNB image in a circular region out to $\sim5''$. Nearby continuum sources are identified using the SDSS catalog from the built-in catalog function of the SAOImage DS9 tool \citep{saoDS9}. These sources are masked and excluded from all subsequent analysis in this work. The masks are shown as black regions in the pNB panels of Figure~\ref{fig:targplots}.} \\

 {Variance images are also produced for the pNB images, using the propagated error on the coadded cubes as input. To prioritize the extraction of faint emission on large spatial scales, the data is smoothed with a simple $5\times5$ (pixels) box kernel. This smoothing, increases the covariance in the data. In the next Section, we describe an empirical variance calibration method which scales the propagated variance estimates to account for covariance. }

\subsection{Covariance in the pNB Images}\label{analysis:covar}

\begin{figure}[t]
\centering
\includegraphics[width=0.45\textwidth]{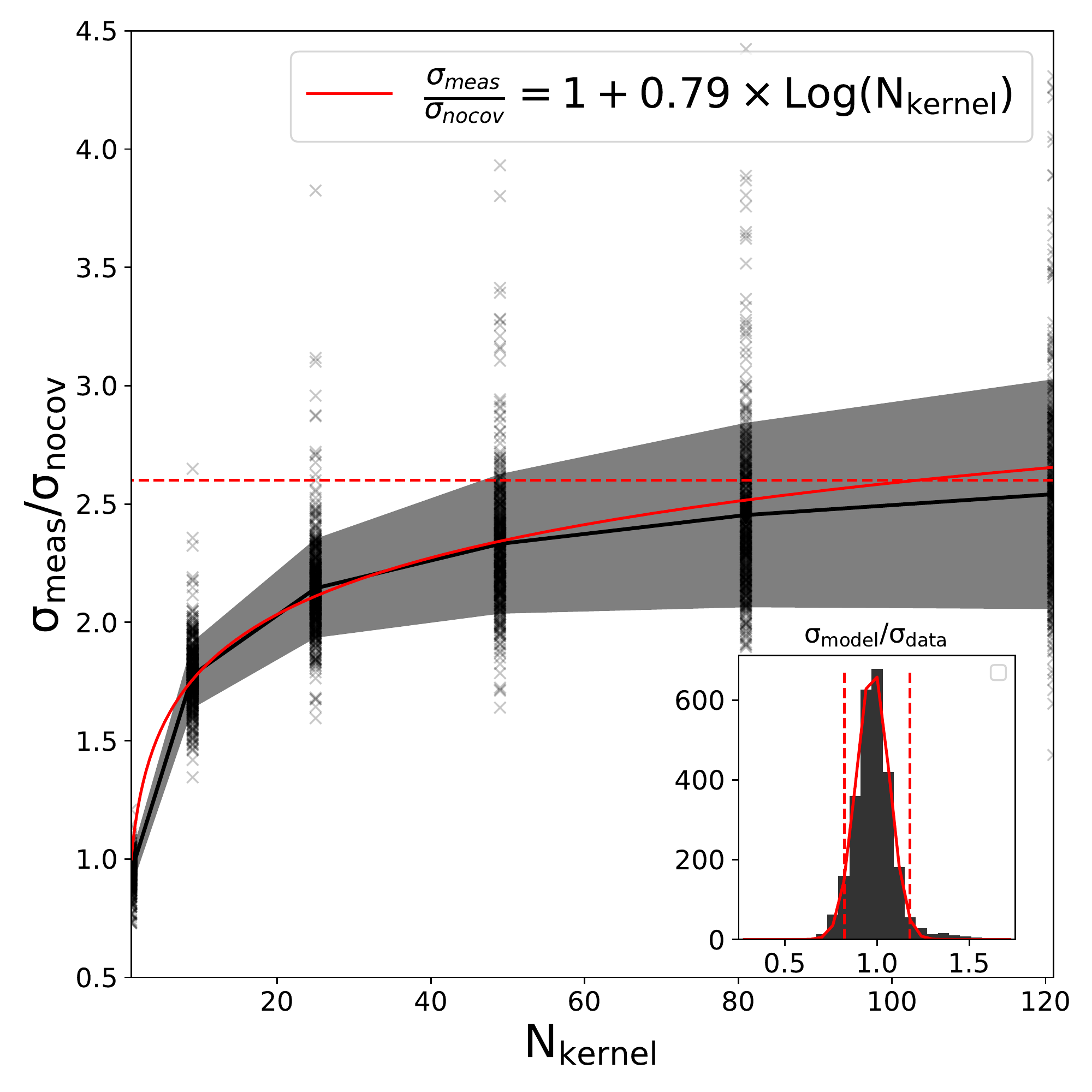}
\caption{ \small{ Calibration of variance measurement in FLASHES pilot data. Black crosses indicate individual calibration measurements. The solid black curve indicates the averaged profile, while the grey-shaded region represents the $\pm1\sigma$ uncertainty. The solid red curve indicates the functional fit to $\sigma_{meas}/\sigma_{nocov}=(1+\alpha_v Log(N_k))$, with $\alpha_v=0.79$, and the horizontal dashed red line indicates the approximate asymptote for the relationship at large $N_k$ ($\beta_v\simeq 2.6$.)} }
\label{fig:varcal}
\end{figure}

 {IFS data contains covariance between adjacent pixels from resampling onto a regular wavelength or spatial axis, distributing flux onto a new pixel grid when coadding, and any subsequent smoothing or binning steps. This complicates efforts to use standard error analysis when smoothing or summing flux from data cubes. It is extremely complex, and typically beyond the scope of standard data reduction pipelines, to analytically determine the exact form of this covariance.  Because of this, the variance produced by such pipelines underestimate the true noise of the data. As a first step, following the approach of similar studies in the field \citep{Borisova16, ArrigoniBattaia19}, we re-scale the propagated variance on each coadded cube to match the global noise properties of the cube. This is done comparing the variance with the distribution of voxel values in the cube background (i.e. after masking sources and emission-lines.) The variance rescaling factor at this stage is approximately $\sim1.5$ for all cubes.\\}

 {However, we also have to take into account the covariance added by smoothing. For smoothing with a uniform box kernel $K(m,l)$ of side $N_k$ and where $K=1$ for all $m,l$, the propagated variance assuming independent variables is:}
\begin{equation}
    V_{nocov}'(x,y) = \frac{\sum_{m=-N_k+1}^{m=N_k-1} \sum_{l=-N_k+1}^{l=N_k-1} V(x+m, y+l)}{N_k^2}
\end{equation}
 {The numerator here is a binning operation, while the denominator is a fixed normalization factor which does not add to the covariance. As such, to account for the covariance introduced by this smoothing operation, we adopt the methodology used by the data reduction pipelines for the Calar Alto Legacy Integral Field Area survey \citep{CALIFADRP} and the SDSS-IV MaNGA IFU Galaxy Survey \citep{mangaDRP}. These pipelines estimate the covariance in spatially summed/binned spectra using an empirical calibration of variance as a function of the kernel size. This is done by binning data by $\mathrm{N_k\times N_k}$ pixels, measuring the noise in the binned signal ($\mathrm{\sigma_{meas}}$) and comparing it to the error derived under the assumption of no covariance ($\mathrm{\sigma_{nocov}}$). The relationship between these variables is then fit with a functional form following $\mathrm{\sigma_{meas}=(1+\alpha_{v} Log(N_{k}^2))\sigma_{nocov}}$, and used to calibrate future variance estimates. For large $\mathrm{N_k}$, where most of the data under the kernel is uncorrelated, this functional form beaks down and instead follows a simple scaling form of $\mathrm{\sigma_{meas}\simeq \beta_{v} \sigma_{nocov}}$.} \\

 { To perform this calibration, We generate a set of pNB images at continuum wavelengths, such that they contain no extended emission and contain mostly empty background after WL subtraction. From our 48 coadded data cubes, we can generate $\sim440$ such pNB images. We then measure the noise after smoothing these images with box kernels of size $N_k=1, 3, 5, 7$ and $9$, and compare it to the noise estimate from the simply propagated variance. When measuring the noise of the smoothed image, we use only values for which the convolution with the smoothing kernel did not rely on zero padding (this would underestimate the noise.) We then fit the above functional form to find $\alpha_{v}=0.79$ and $\beta_v\simeq2.6$. Figure~\ref{fig:varcal} shows the result of this calibration. The inset shows the error on the calibration itself; the model estimates the variance to within $\pm18\%$ ($\pm2\sigma$). Since we smooth our pNB images with a $5\times5$ box kernel, we rescale the variance by a factor of $f_{var}=(2.11)^2\simeq4.45$. We also scale the variance following this form when calculating the integrated SNR of an extended region.}\\

\subsection{Optimizing the pNB Image Parameters}\label{sec:2D}

\begin{figure}[t]
\centering
\includegraphics[width=0.45\textwidth]{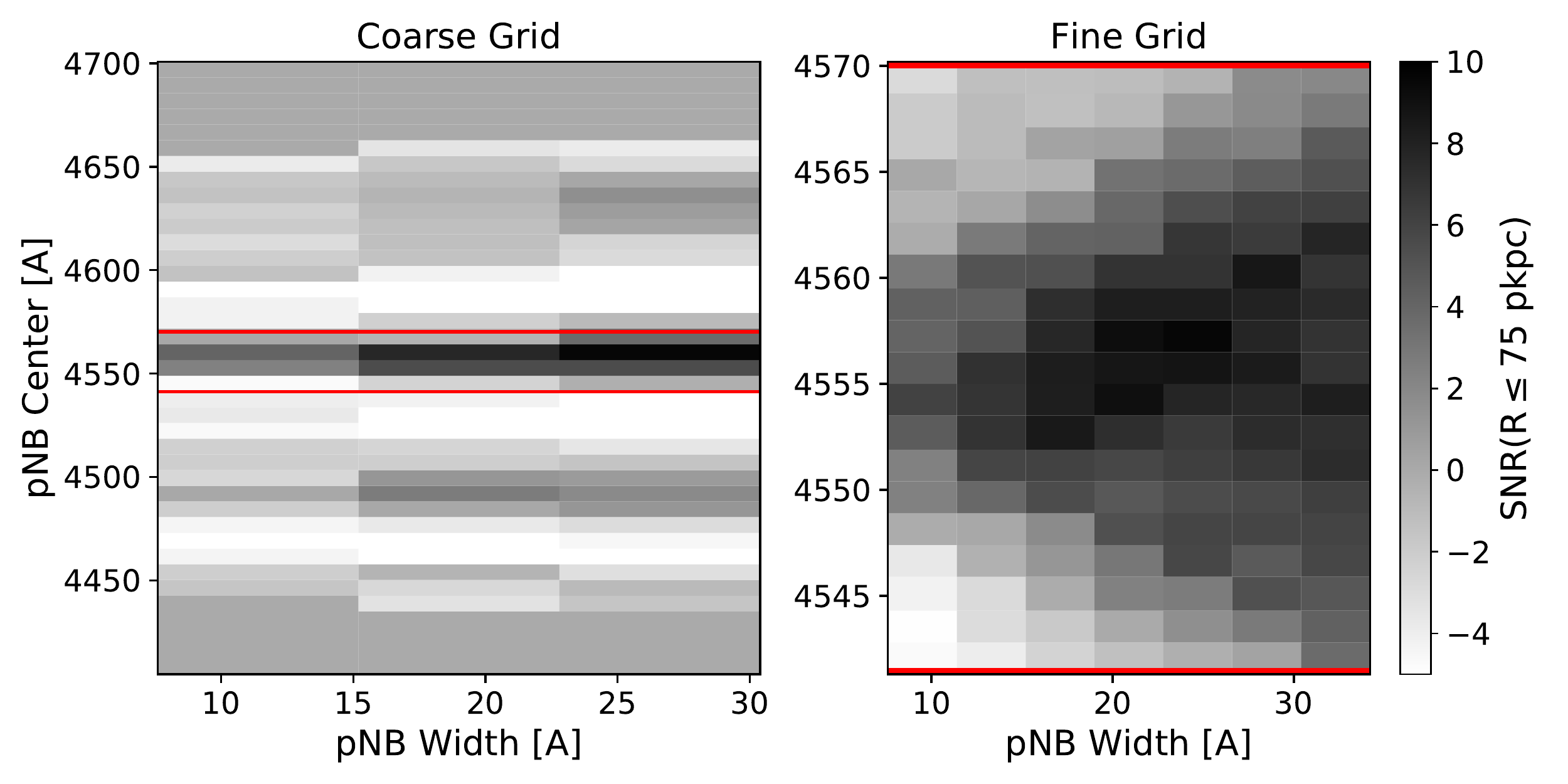}
\caption{ \small{ Example of the optimization of pNB image parameters - wavelength center and bandpass- for target HS1700+6416 (ID 1). The colormap in both panels shows the integrated SNR in a circular region of radius $75$pkpc around the QSO. The left panel shows the initial coarse grid, which searches a wide velocity range of $\pm10000\mathrm{~km~s^{-1}}$. The red lines indicate the $\pm1000\mathrm{~km~s~^{-1}}$ zoom-in for the high resolution grid, shown on the right. } }
\label{fig:pnbsearch}
\end{figure}

 { To identify extended emission in the cubes, a 2D approach using pNB images is adopted. This approach is chosen over a 3D voxel-by-voxel search so that the signal can be integrated along the wavelength axis. The basic approach is to generate sets of pNB images with varying combinations of wavelength center and width and measure the integrated SNR in the vicinity of the QSO (within a projected radius of $75$ pkpc). The pNB parameters which optimize the signal in the vicinity of the QSO is then chosen. This is done in two steps:  first using a coarse grid with a large range of wavelength/center combinations to find an approximate central wavelength, and then using a fine grid with a smaller range. The pNB centers in the coarse grid range over $\pm10000\mathrm{~km~s^{-1}}$ in velocity from the peak of Ly$\alpha$ emission in the QSO spectrum, in steps of $\mathrm{\delta v_{center}=1000~km~s^{-1}}$. This large range is motivated by the findings of previous works such as \cite{Borisova16} that the centroid of Ly$\alpha$ emission can vary by thousands of kilometers per second from the QSO redshift. The pNB velocity width for the coarse grid is varied over a range of $500-2500\mathrm{~km~s^{-1}}$ in steps of $\mathrm{\delta v_{width}=500~km~s^{-1}}$. This process runs semi-automatically, as regular visual inspection of the pNB images is needed to ensure the integrated SNR is not influenced by systematic errors (under-subtraction of the QSO, bright sky line residuals, etc).\\}

 {Once an approximate wavelength center is identified from the first step, a higher resolution grid is generated to refine the best center/width. This time, the pNB velocity centers range over a smaller range of $\pm 1000\mathrm{~km~s^{-1}}$ with a step-size of $100\mathrm{~km~s^{-1}}$, and the bandwidth ranges from $500-2500\mathrm{~km~s^{-1}}$ in steps of $250\mathrm{~km~s^{-1}}$. This grid is used to identify the optimal center and width for the pNB image for each target. Figure~\ref{fig:pnbsearch} shows an example of the coarse and fine grids generated during this process for one target (ID 3 / HS1700+6416). For targets in which there is no clear choice of center/width, the default setting is to be centered on the peak of QSO Ly$\alpha$ emission with a velocity width of $1500\mathrm{~km~s^{-1}}$. Table~\ref{tab:pnbParams} shows the final parameters for each pNB image alongside the $2\sigma$ limiting surface brightness achieved in a $1~\mathrm{arcsec^2}$ aperture.}\\

\subsection{Extracting Emission from pNB Images}

 {When the optimal center and width of the pNB are identified, the final data products are produced, including the WL image, non-subtracted pNB image, subtracted pNB image, source mask, variance map, and SNR map. These are used to identify regions of extended emission. The data is initially segmented by a threshold of $\mathrm{SNR\geq 2\sigma}$. The integrated SNR of each region is then calculated (taking covariance into account after summing under the region) and an integrated SNR threshold of $\mathrm{SNR_{int}} \geq 4.5\sigma$ is applied. The search is limited to a $250\time250\mathrm{~pkpc^2}$ box around the QSO. If no regions are found of a sufficient SNR, the target is counted as a non-detection. If there are detected regions, the total integrated SNR of all regions is measured and used to determine the order of the targets (from highest to lowest.)}

\subsection{Characterizing 2D Morphology} \label{analysis:morphology}
in order to highlight different characteristics, we measure the size of the nebulae in three ways. First, we use the maximum extent of the nebula from its flux-weighted centroid, $\mathrm{R_{max}}$. Secondly, we define an \emph{effective} radius to be the radius of an equivalent circular area, i.e., $\mathrm{R_{eff}=\sqrt{Area/\pi}}$. We emphasize that $\mathrm{R_{eff}}$ is not a true radius, but a characteristic scale. Finally, we measure the flux-weighted root-mean-square radius, $\mathrm{R_{rms} = \sqrt{\langle R^2 \rangle_f}}$, using the flux values under the 2D nebular mask. While $\mathrm{R_{max}}$ and $\mathrm{R_{eff}}$ give a sense of the maximum and average extent of the nebula, respectively, $\mathrm{R_{rms}}$ size gives a sense of the characteristic scale at which most of the emission is concentrated. \\

Beyond measurement of size, the 2D morphology is characterized by three parameters; eccentricity (i.e. asymmetry), displacement, and covering factor. To quantify the symmetry of the nebulae and allow for direct comparison with existing literature, we adopt the same measurement of spatial symmetry as presented in A19. This parameter, $\mathrm{\alpha}$, is derived from the second-order spatial moments and reflects the ratio of the semi-minor axis ($b$) to the semi-major axis ($a$) of the emission (i.e., $\alpha=b/a$). We then convert it to an elliptical eccentricity parameter ($\mathrm{e}$), which we find to be more intuitive, following:
\begin{equation}
   \mathrm{e = \sqrt{1 - b^2/a^2} = \sqrt{1 - \alpha^2}}
\end{equation}
The displacement, which we denote $\mathrm{d_{QSO}}$, is the projected physical distance (in proper kiloparsecs) between the flux-weighted centroid of the nebular emission (under the mask) and the quasar. 

\subsection{Radial Profiles}\label{analysis:rprofs}

Radial surface brightness profiles are measured from a minimum projected radius of $18$ pkpc to a maximum radius of $150~\mathrm{pkpc}$ in logarithmic bins of 0.1 dex. All of the detected emission in this sample falls within this range. The 2D object mask is not applied when calculating the circularly averaged surface-brightness profile, but the locations of known and subtracted continuum sources are masked.  {For non-detections, the wavelength of any CGM Ly$\alpha$ emission is not known, so it may not be contained in the bandpass of the pNB image. For this reason, the averaged radial profile including non-detections may slightly underestimate the true radial profile.} The covering factor is calculated using the same radial bins, and defined as the fraction of pixels in each annular region above an SNR of $2\sigma$. 

\subsection{Luminosities}

The integrated luminosity of each nebula is calculated following
\begin{equation}
   \mathrm{ L_{tot} = 4 \pi D_{L}^2(z) (\delta\theta)^2    \Sigma_{x}\Sigma_y SB(x,y)M(x,y) }
\end{equation}
where $\mathrm{SB(x,y)}$ is the 2D surface-brightness map in units of $\mathrm{erg~s^{-1}~cm^{-2}~arcsec^{-2}}$, $\mathrm{M(x,y)}$ is the binary 2D mask defined earlier, $\delta\theta$ is the angular size of a pixel, and $D_L(z)$ is the luminosity distance at the redshift of the target. We note that integrated luminosities are sensitive to the surface-brightness threshold used to define $\mathrm{M(x,y)}$, any comparison to luminosities reported in other works should consider the differences in cosmic dimming-adjusted surface brightness limits.

\subsection{Point-source Subtraction in 3D}\label{sec:3D}

\texttt{CWITools} performs 3D point-spread function (PSF) subtraction in a similar fashion to \cite{Borisova16}, which is a basic extrapolation of the pNB method, described earlier, to 3D. A white-light image is formed by summing all of the wavelength layers of the cube, which is then used to identify the positions of any point sources. For each point source above a certain signal-to-noise threshold, the following routine is repeated: For each wavelength layer in the cube, a broad-band (i.e., white-light) image centered on the current wavelength layer is formed by summing over a  large spectral range ($\sim100$\AA). This image is then scaled and subtracted from the wavelength layer using the method described in Section~\ref{sec:2D}. The underlying assumption of this technique is that the shape of the PSF will be dominated by white-light, not nebular emission. In the case of obscured quasars with faint continuum or quasars with particularly bright extended emission, the wavelength range containing nebular emission may be masked to prevent it being used for the white-light image. A small inner radius roughly equal to the seeing ($\sim1''$) is used to calculate the scaling factors, and the scaled WL image is subtracted out to a larger radius, typically a few times the seeing ($\sim5''$). Once this PSF subtraction is completed for all detectable point sources, any remaining continuum or scattered light is subtracted (if necessary) using a low-order ($k=1$ or $2$) polynomial fit to the spectrum in each spaxel. If strong nebular emission is identified, it can be masked during this fitting process to avoid over-fitting. Finally, the PSF cores of bright sources that have been subtracted are masked to prevent noisy residuals influencing any measurements later on. As with the 2D pNB images, the positions of known continuum sources are identified and masked using sources from the 12th SDSS Data Release \citep{Alam+2015}.\\

\begin{figure*}
    \centering
    \includegraphics[width=0.8\textwidth]{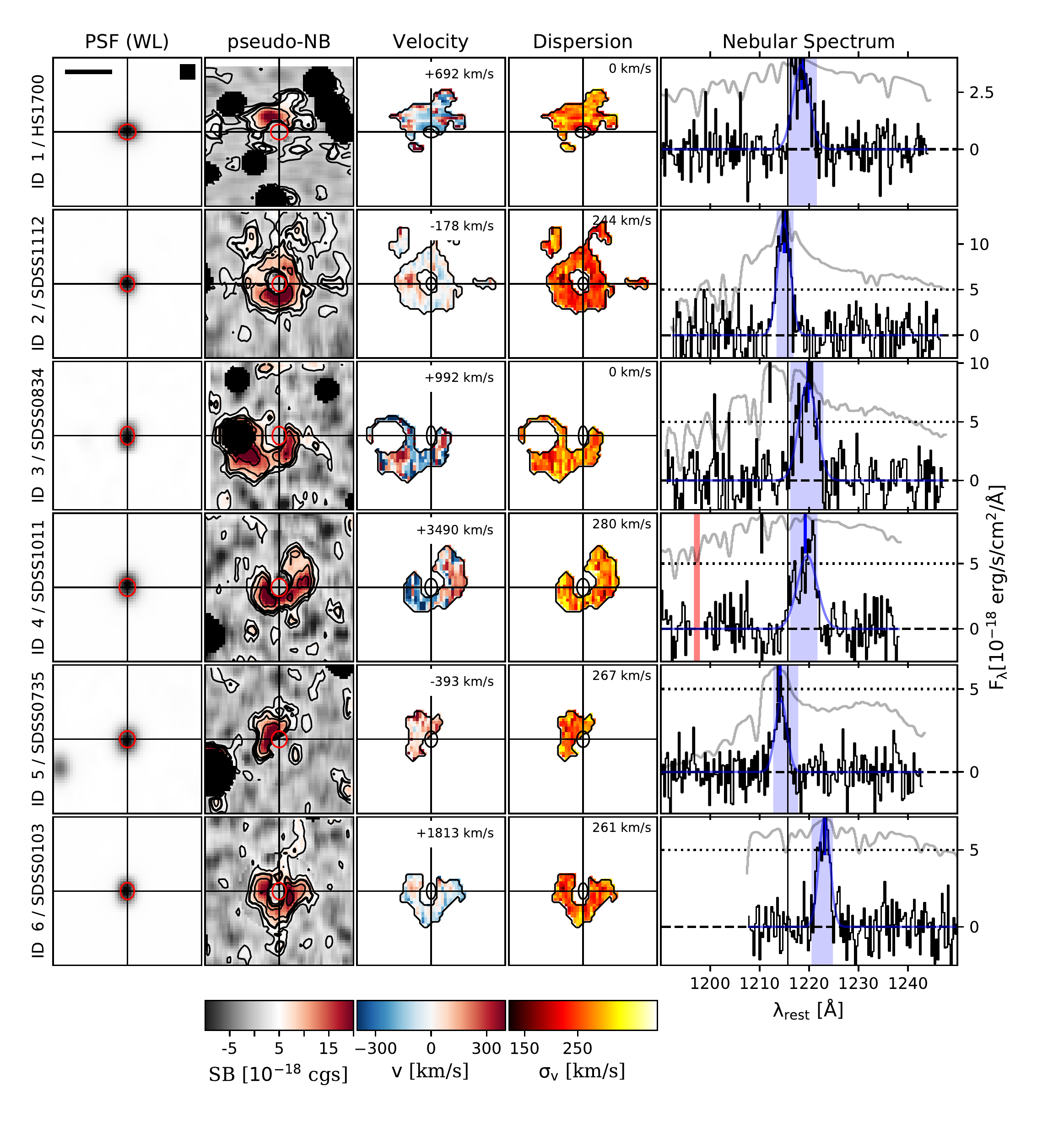}
    \caption{FLASHES Pilot survey observations (ID 1-6). Each tile is $250\times250~\mathrm{pkpc^2}$ in size, centered on the QSO. The left-most four columns show a white-light image, \LyA surface brightness, velocity, and dispersion. Surface brightness is in cgs units, $\mathrm{erg~ s^{-1}~cm^{-2}~arcsec^{-2}}$. The black bar in the top white-light image shows $\mathrm{100~pkpc}$ and the black square shows the box kernel used to smooth the WL and pNB data. Foreground sources in each field have been masked, with the masked regions shown in black. The rightmost column shows integrated nebular spectra (black) and scaled QSO spectra (grey). The spectra are summed over the object masks and shown in units of $\mathrm{10^{-17}~erg~s^{-1}~cm^{-2}~}$\AA$\mathrm{^{-1}}$. Spectra are shown rest-frame wavelength, according to the systemic QSO redshift. Blue lines indicate the flux-weighted centers of nebular emission, while black lines indicate the peak of QSO emission. A very bright mercury sky emission line ($\mathrm{Hg~\lambda4358.3}$) is masked in some spectra and shown here as a vertical red band wherever it appears. Empty regions (shown in white) in the pNB images are outside the field of view. An ellipse representing the FWHM of the QSO's PSF is shown in each tile. Red ellipses are used for smoothed PSF (in the WL and pNB images) while black ellipses are used for the unsmoothed PSF (moment maps.) }
    \label{fig:targplots}
\end{figure*}

\begin{figure*}
    \figurenum{5}
    \centering
    \plotone{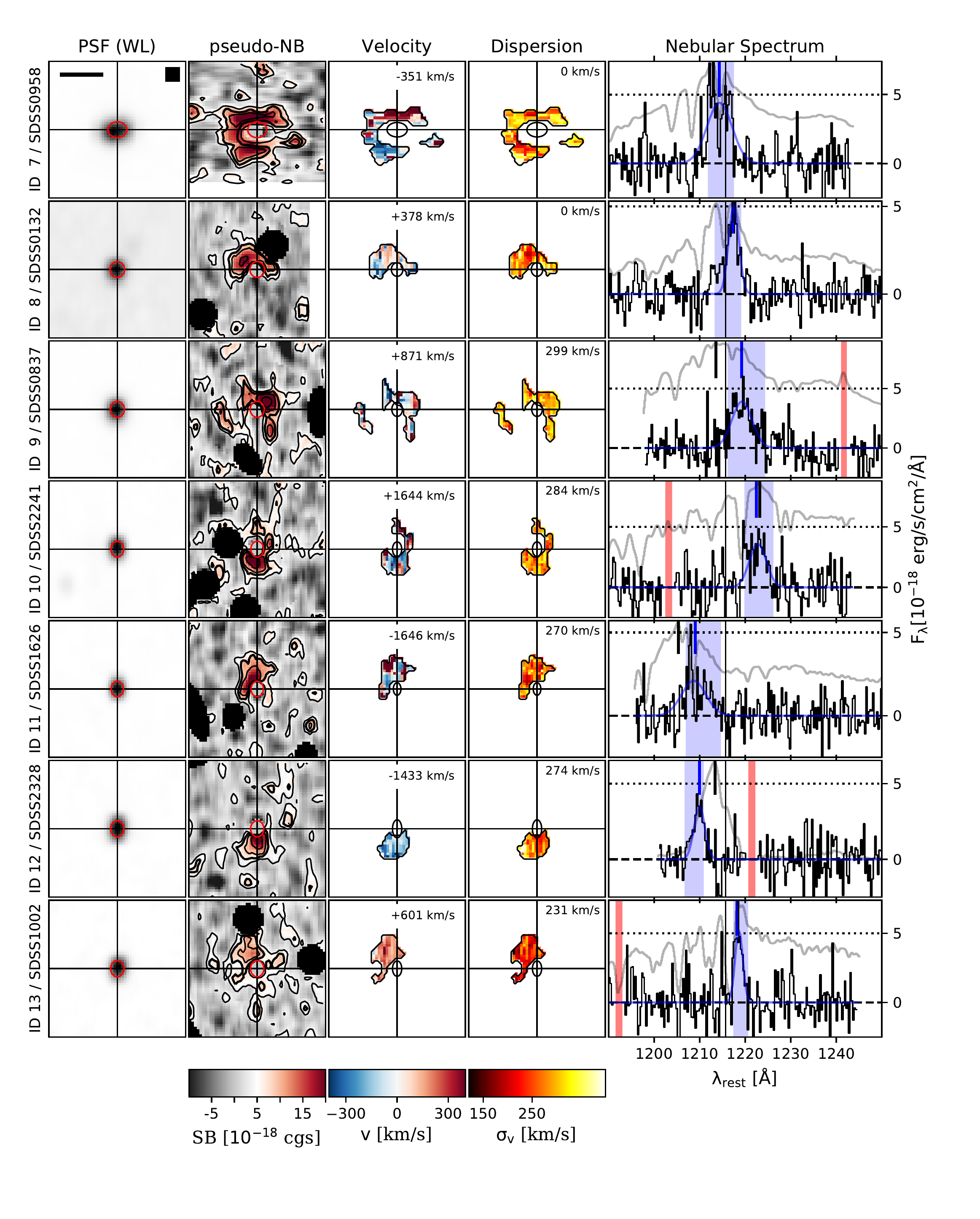}
    \caption{ (continued) FLASHES Pilot survey observations (ID 7-13).}
\end{figure*}

\begin{figure*}
    \figurenum{5}
    \centering
    \plotone{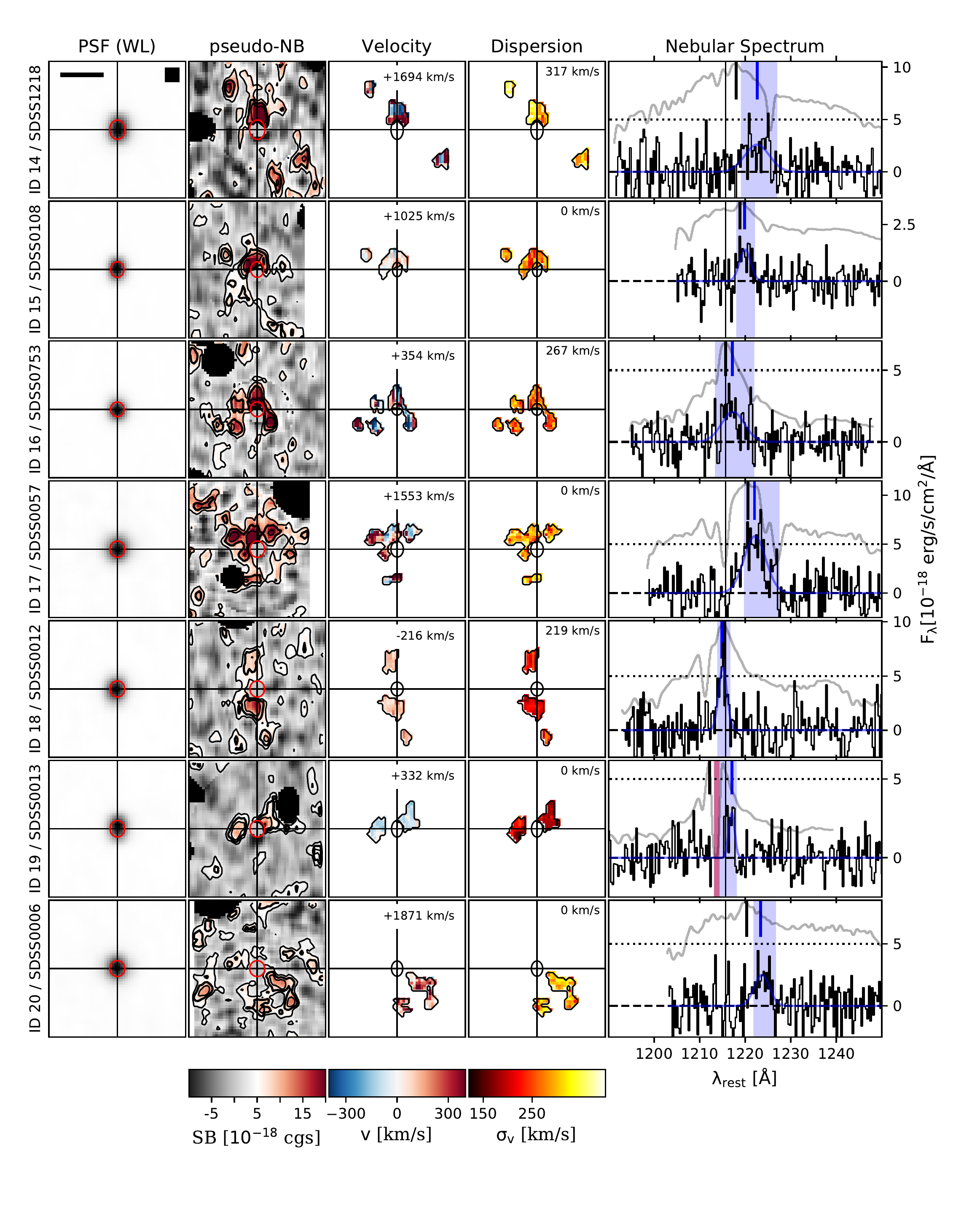}
    \caption{(continued) FLASHES Pilot survey observations (ID 14-20).}
\end{figure*}

\begin{figure*}
    \figurenum{5}
    \centering
    \plotone{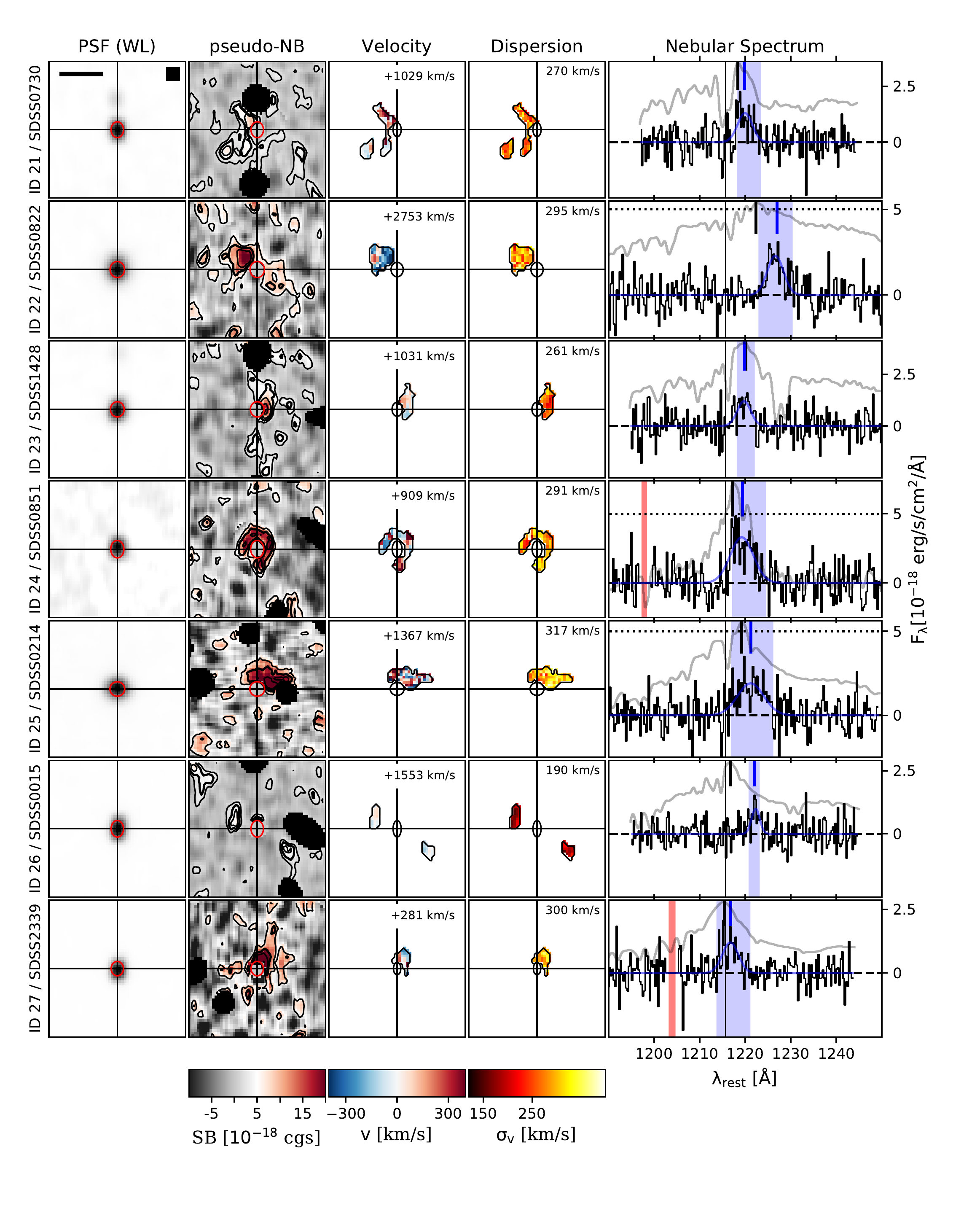}
    \caption{(continued) FLASHES Pilot survey observations (ID 21-27).}
\end{figure*}

\begin{figure*}
    \figurenum{5}
    \centering
    \plotone{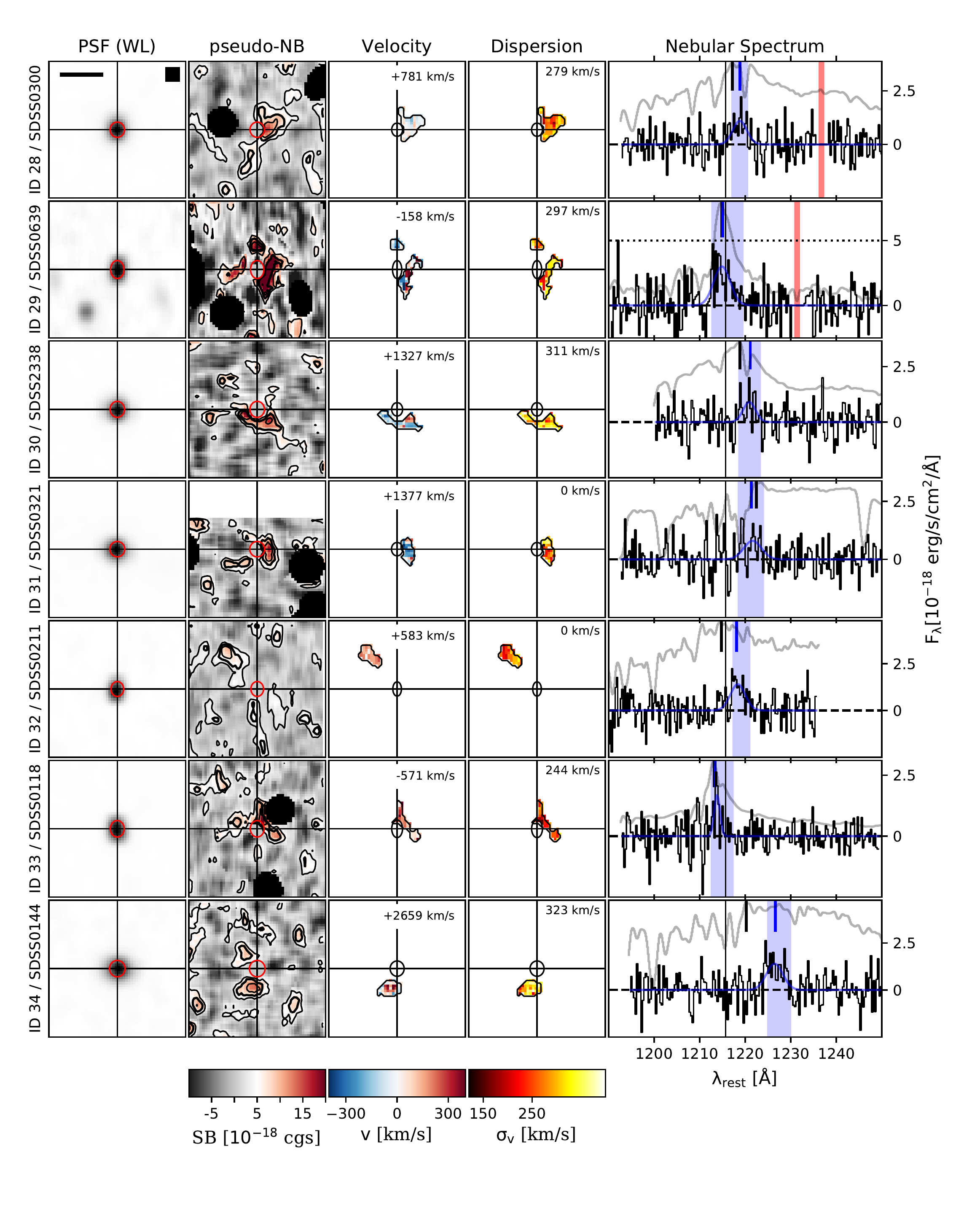}
    \caption{(continued) FLASHES Pilot survey observations (ID 28-34).}
\end{figure*}

\begin{figure*}
    \figurenum{5}
    \centering
    \plotone{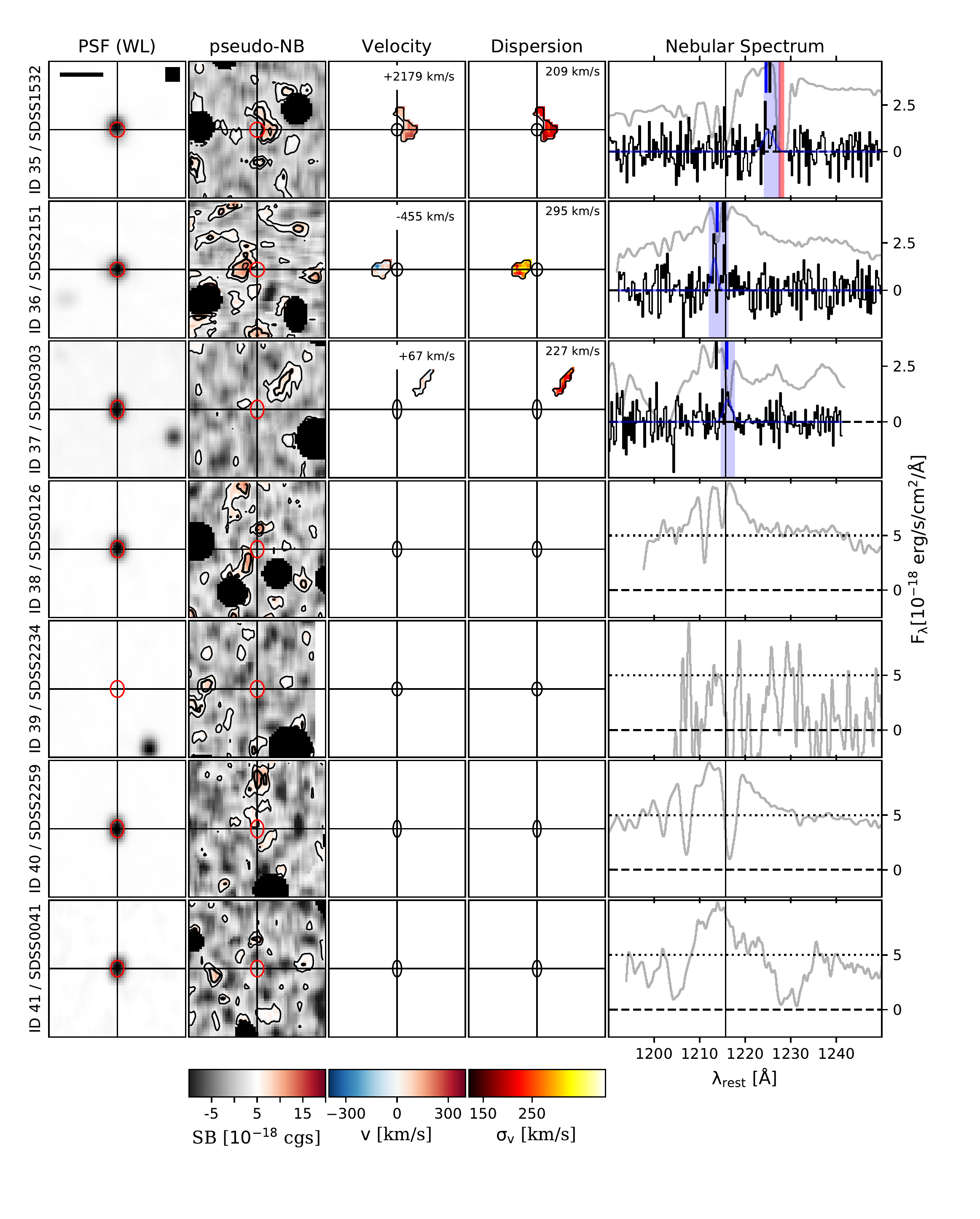}
    \caption{(continued) FLASHES Pilot survey observations (ID 35-41).}
\end{figure*}

\begin{figure*}
    \figurenum{5}
    \centering
    \plotone{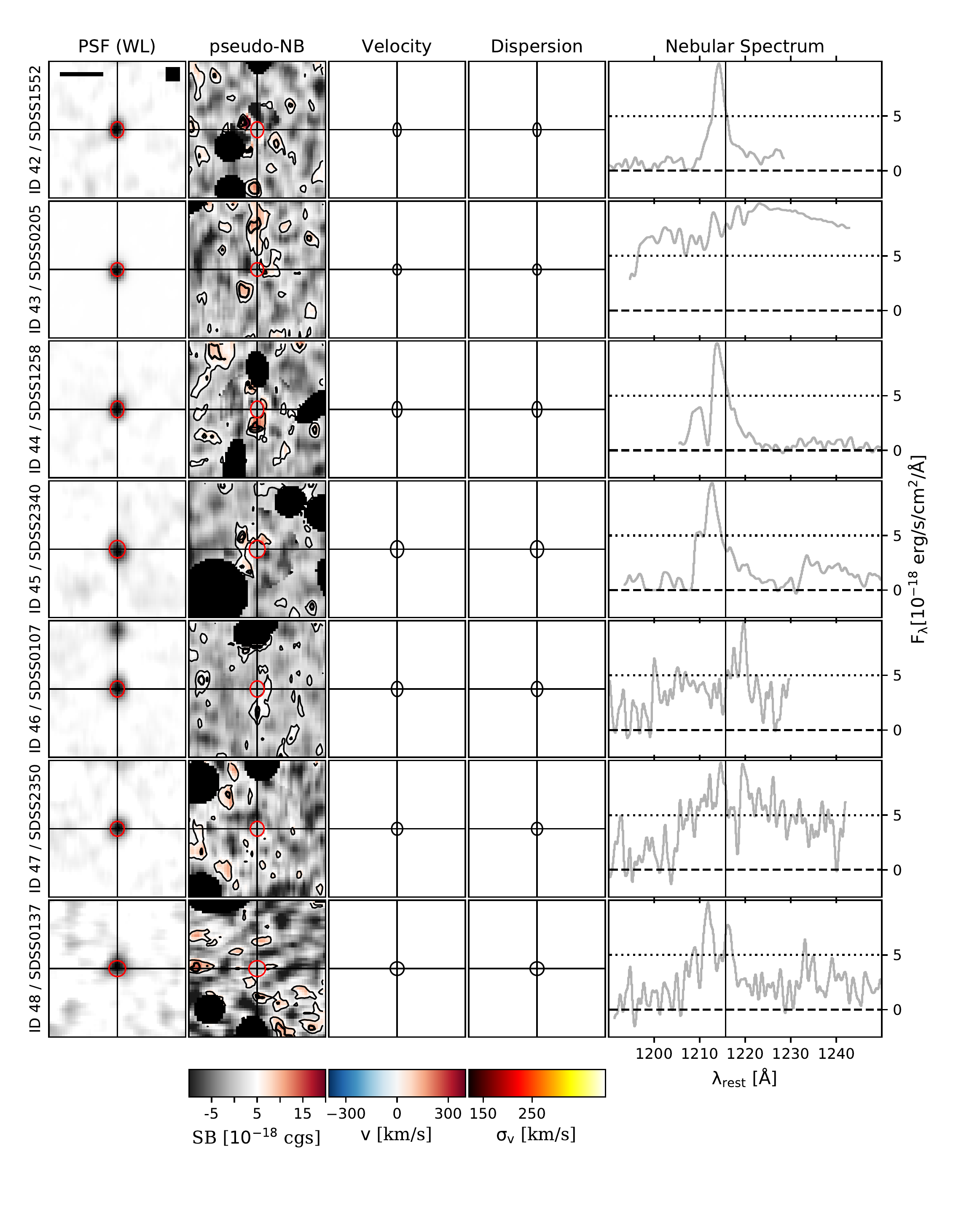}
    \caption{(continued) FLASHES Pilot survey observations (ID 42-48).}
\end{figure*}

\subsection{Integrated Nebular Spectra and Line-Fitting} \label{sec:spectralmodeling}

To create an approximate 3D mask encompassing the emission, the spatial object mask, $\mathrm{M(x,y)}$ is extended along the wavelength axis over the same range as was used to form the final pNB image. Nebular spectra are obtained by summing over the spatial axes under the 3D mask. The spectra are fit with both a simple Gaussian model, a model consisting of multiple ($1-4$) Gaussian components, and a simple linear model. To determine which model best represents the data, we calculate the Bayesian Information Criterion (BIC) for each, following

\begin{equation}
    \mathrm{BIC} = n \ln(RSS/n) + k\ln{n}
\end{equation}

where $n$ is the number of independent variables (i.e., length of the spectrum), $k$ is the number of free parameters in the model, and $RSS$ is the residual sum of squares of the model. Lower BIC values indicate a better representation of the data. Weights representing the \emph{relative} likelihood of a set of models can be derived from the BIC values as:

\begin{equation}
    w_i = \frac{\exp{(-\frac{1}{2}\Delta_i(BIC))}}{\Sigma_j \exp{(-\frac{1}{2}\Delta_j(BIC))}}
\end{equation}

where $\Delta_i(\mathrm{BIC})$ is the difference between the $i$th BIC value and the minimum BIC value of the set \citep{Wagenmaker2004}. The value $0\leq w_i \leq 1$ corresponds to the relative likelihood that the $i$th model is the best representation of the data (among those considered). The linear model is included as a simpler alternative in order to validate the single-component Gaussian models; if only Gaussian models were considered, the BIC would still indicate a single-component Gaussian as the best fit for pure noise. This multiple-component fit provides an important piece of contextual information when interpreting the global dispersions of the nebulae - as complex line shapes with multiple components can appear quite broad when viewed as a single Gaussian, or otherwise treated as a single kinematic component (e.g., by calculating the second moment). \\

To create a stacked \LyA spectrum from the individual detections, spectra are linearly interpolated onto a rest-frame wavelength grid ranging from $1200$ \AA~ to $1230$ \AA~ with a sampling rate of $0.14$ \AA/px (approximately the PCWI sampling around \LyA at $z=3$). As there is more than one measure of redshift (e.g. the flux-weighted center of emission vs. the systemic QSO redshift), we create four versions of the stacked spectrum, each using a different central wavelength (i.e., redshift) to convert to rest-frame units: (i) the systemic QSO redshift given in DR12Q, (ii) the flux-weighted center of \LyA emission, (iii) the peak of \LyA emission in the QSO spectrum, and (iv) the HeII $\lambda1640$ redshift from DR12Q.  {The stacked spectra are simple averages of the individual detections, though non-detections are necessarily excluded as no integrated nebular spectrum can be measured. }

\subsection{2D Moment Maps}

Two-dimensional first and second flux-weighted z-moment maps are calculated as:\\
\begin{equation} \label{eq:mu1}
\mu_{\lambda,1} = \frac{ \sum_{k}{ \lambda(k) I_{k} } }{\sum_{k}{I_{k} } }  \\
\end{equation}
\begin{equation} \label{eq:mu2}
\mu_{\lambda,2} =  \sqrt{ \frac{ \sum_{k}{ (\lambda_k-\mu_1)^2 I_{k} } }{\sum_{k}{I_{k} } }}  \\
\end{equation}

where $I_{k}$ is the $k^{th}$ wavelength layer of the intensity cube and $\lambda_k$ is the wavelength at that layer. $I$ and $\mu$ are both two-dimensional arrays with the spatial indices ($i,j$) omitted for simplicity (i.e., $I_k = I_{ijk}$).  {No smoothing is applied to the 3D data prior to calculating the moments.}\\

As a statistical moment is not well defined for a distribution with negative weights, some non-negative threshold must be applied to the spectra before calculating the first or second moment. For bright signals, a high SNR threshold can be applied which rejects virtually all noise while also retaining enough signal for an accurate measurement. However, for fainter signals, it can be challenging to find a threshold which satisfies both of these requirements. A simple positive threshold (i.e., $\mathrm{F_\lambda >0}$) can be applied, positive fluctuations in the background noise will then bias the calculation. For the calculation of the first moment, $\mu_1$, an iterative approach can be used to overcome this. The effect of evenly distributed noise (in a well background-subtracted signal) will be to bias the result towards the center of whichever wavelength window is used. If the wavelength window is centered on the true first moment, then this biasing effect will be negligible. As such, if we perform this calculation iteratively, updating the center of the window each time to the new value of $\mu_1$, the window center will eventually converge on the true value. If the size of the wavelength window used for the calculation is also reduced as the solution converges, this further mitigates any biasing effect from unevenly distributed noise. We use this method to determine the first moment (i.e. velocity center) of the spectra in each spaxel, with a starting window size of $25$\AA~(to fully explore the range used for the pNB images), reduced in steps of $\Delta \lambda=1$\AA~until a minimum window size of $10$\AA~ is reached. Appendix~\ref{app:iterative_moments} provides an illustration of this method.  \\

For the second moment, a convergent method cannot be used to the same effect, as the influence of normally distributed noise on the second moment is to unilaterally increase its value. Instead, we apply a basic non-negative threshold and treat the derived values as upper limits. The spatially resolved maps still provide our only insights into the 2D distribution of the second moment, and as such are valuable despite this limitation. We can rely on line-fitting of the integrated nebular spectrum (see Section~\ref{sec:spectralmodeling}) for more robust measurements of the global dispersions of the nebulae. \\

Once the moments are calculated, \LyA velocity and dispersion maps can be derived as:
\begin{equation}
 \mathrm{v(i,j)=\Big(\frac{\mu_{\lambda,1}(i,j)-\lambda_0}{\lambda_0}\Big)c}
\end{equation} 
\begin{equation}
 \mathrm{ \sigma_v(i,j)=\Big(\frac{\mu_{\lambda,2}(i,j)}{\lambda_0}\Big)c }
\end{equation} 
where $\mathrm{\lambda_0}$ is the flux-weighted average wavelength of the integrated nebular spectrum, $\mathrm{\mu_{\lambda,1}(i,j)}$ is the first moment in wavelength (Eq.~\ref{eq:mu1}) at that position, and $c$ is the speed of light in vacuum. For each nebula, we also calculate the flux-weighted, one dimensional root-mean-square velocity along the line of sight, $\mathrm{v_{rms}=\sqrt{\langle v^2 \rangle_f}}$. To be clear, this is the root-mean-square of velocities in individual spaxels \emph{relative} to the flux-weighted average velocity of the nebula. Finally, we measure the offset between the the flux-weighted average velocity of the nebula and three key wavelengths; the wavelength of \LyA at the systemic redshift of the QSO ($\mathrm{\lambda_{\alpha,QSO}}$), the wavelength of the peak of \LyA emission in the QSO spectrum ($\mathrm{\lambda_{\alpha,peak}}$) and the wavelength of \LyA at the HeII $\lambda1640$ redshift of the QSO ($\mathrm{\lambda_{\alpha, HeII}}$).\\
\begin{equation}
 \mathrm{ \Delta v_{QSO} =\Big(\frac{\lambda_{\alpha,QSO}-\lambda_0}{\lambda_0}\Big)c }
\end{equation}
\begin{equation}
 \mathrm{ \Delta v_{peak} =\Big(\frac{\lambda_{\alpha,peak}-\lambda_0}{\lambda_0}\Big)c}
\end{equation}
\begin{equation}
 \mathrm{ \Delta v_{HeII} =\Big(\frac{\lambda_{\alpha,HeII}-\lambda_0}{\lambda_0}\Big)c}
\end{equation}

\startlongtable
\begin{deluxetable*}{rrrrrrrrrrrrr}
\tablecaption{Measured CGM properties from the FLASHES Pilot Survey. \label{tab:neb_props}}

\tablewidth{0pt}
\tablehead{ \colhead{ID} &   \colhead{$\mathrm{L_{43}}$\tablenotemark{a}} & \colhead{$\mathrm{R_{eff}}$} &  \colhead{$\mathrm{R_{rms}}$}& \colhead{$\mathrm{R_{max}}$} &   \colhead{$\mathrm{d_{QSO}}$} &  \colhead{$e$}& \colhead{$\mathrm{n_{reg}}$}&  \colhead{$\mathrm{z_{Ly\alpha}}$}&     \colhead{$\mathrm{\Delta v_{QSO}}$}&        \colhead{$\mathrm{\Delta v_{peak}}$}&     \colhead{$\mathrm{\sigma_v}$} &  \colhead{$\mathrm{N_{G}}$} \\
&  \colhead{$\mathrm{erg~s^{-1}}$} &   \colhead{pkpc} &  \colhead{pkpc} &  \colhead{pkpc} &    \colhead{pkpc}& \colhead{(0-1)}&  & $(\pm2\sigma)$ & \colhead{$\mathrm{km~s^{-1}}$} &  \colhead{$\mathrm{km~s^{-1}}$}&  \colhead{$\mathrm{km~s^{-1}}$}&    \colhead{(0-4)}
}

\startdata
    1&       4.3&      41.4&      28.3&      63.0&      21.8&      0.80&         2&     2.746$\pm$0.0008&      +692&      -119&       395&         2\\
    2&       9.4&      54.8&      47.2&     110.2&       9.0&      0.67&         3&     2.788$\pm$0.0010&      -178&      -171&       325&         2\\
    3&       8.0&      45.1&      40.7&      83.8&      41.0&      0.82&         1&     2.759$\pm$0.0010&      +992&     +1624&       459&         2\\
    4&       6.3&      43.0&      37.9&      69.4&      10.5&      0.88&         1&     2.651$\pm$0.0012&      +871&     +2179&       473&         3\\
    5&       3.4&      30.2&      22.4&      45.5&      19.5&      0.87&         1&     2.747$\pm$0.0014&      -393&       +93&       263&         1\\
    6&       5.4&      38.5&      32.9&      55.9&       8.2&      0.79&         1&     2.721$\pm$0.0012&     +1813&       -65&       265&         2\\
    7&       6.1&      42.5&      45.4&      93.1&       7.0&      0.56&         2&     2.487$\pm$0.0012&      -351&      +459&       707&         2\\
    8&       2.5&      30.8&      25.8&      53.3&      21.4&      0.84&         1&     2.425$\pm$0.0010&      +378&      +852&       288&         3\\
    9&       3.9&      33.9&      39.7&      77.2&       7.7&      0.72&         2&     2.520$\pm$0.0014&      +871&      +708&       560&         1\\
   10&       3.2&      28.8&      28.1&      68.0&      11.9&      0.82&         1&     2.642$\pm$0.0016&     +1644&      -110&       474&         2\\
   11&       3.0&      28.2&      22.8&      45.7&      25.5&      0.86&         1&     2.714$\pm$0.0018&     -1646&      +928&       664&         2\\
   12&       2.0&      25.4&      17.0&      31.1&      27.0&      0.60&         1&     2.551$\pm$0.0016&     -1433&      -738&       277&         1\\
   13&       2.1&      28.7&      23.8&      56.3&      34.1&      0.73&         1&     2.663$\pm$0.0016&      +601&       -60&       202&         1\\
   14&       2.6&      27.4&      58.6&     119.8&      20.9&      0.96&         3&     2.394$\pm$0.0022&     +1694&     +1132&       654&         2\\
   15&       2.4&      25.6&      22.9&      54.0&      19.5&      0.78&         2&     2.652$\pm$0.0016&     +1025&       +59&       337&         1\\
   16&       5.5&      31.0&      36.8&      72.5&      10.2&      0.69&         4&     2.935$\pm$0.0016&      +354&      +360&       627&         2\\
   17&       3.6&      31.4&      36.3&      76.1&      17.1&      0.72&         3&     2.454$\pm$0.0012&     +1553&      +389&       571&         4\\
   18&       1.8&      28.4&      48.9&      90.3&      13.8&      0.97&         3&     2.448$\pm$0.0018&      -216&      -192&       214&         1\\
   19&       1.3&      26.2&      32.7&      52.1&       9.8&      0.99&         2&     2.595$\pm$0.0018&      +332&      +719&       137&         1\\
   20&       1.5&      27.7&      30.9&      49.1&      60.7&      0.70&         2&     2.443$\pm$0.0022&     +1871&      +944&       430&         1\\
   21&       1.2&      25.3&      34.1&      54.2&      28.1&      0.96&         2&     2.950$\pm$0.0018&     +1029&      +406&       374&         2\\
   22&       1.8&      23.4&      16.3&      31.6&      33.9&      0.51&         1&     2.486$\pm$0.0016&     +2753&      +888&       424&         2\\
   23&       1.0&      18.7&      15.5&      36.5&      22.1&      0.93&         1&     2.792$\pm$0.0018&     +1031&       -50&       390&         2\\
   24&       2.9&      26.0&      26.6&      45.4&       5.9&      0.64&         1&     2.650$\pm$0.0012&      +909&      +564&       608&         2\\
   25&       2.7&      25.1&      20.9&      42.8&      30.5&      0.87&         1&     2.487$\pm$0.0018&     +1367&      +418&       707&         2\\
   26&       0.7&      19.2&      57.6&      79.0&       4.3&      $\sim$1.00&         2&     3.097$\pm$0.0022&     +1553&     +1280&       240&         1\\
   27&       2.1&      16.6&      12.3&      27.4&      21.0&      0.67&         1&     2.623$\pm$0.0020&      +281&      +317&       408&         1\\
   28&       1.1&      21.8&      17.5&      36.0&      27.6&      0.71&         1&     2.533$\pm$0.0020&      +781&      +273&       419&         1\\
   29&       2.4&      21.8&      30.1&      59.1&      15.7&      0.94&         2&     2.537$\pm$0.0020&      -158&       +53&       488&         2\\
   30&       1.1&      19.9&      23.3&      41.9&      20.0&      0.98&         1&     2.427$\pm$0.0030&     +1327&      +674&       360&         1\\
   31&       0.7&      16.9&      13.0&      25.3&      23.7&      0.91&         1&     2.462$\pm$0.0022&     +1377&      -554&       499&         2\\
   32&       0.7&      17.6&      14.0&      25.9&      79.4&      0.89&         1&     2.793$\pm$0.0022&      +583&      +605&       390&         3\\
   33&       1.2&      18.2&      21.8&      40.5&      22.6&      0.97&         1&     2.771$\pm$0.0022&      -571&       +46&       196&         1\\
   34&       0.7&      18.1&      12.9&      24.2&      34.8&      0.76&         1&     2.461$\pm$0.0016&     +2659&     +1309&       427&         2\\
   35&       0.7&      18.6&      19.1&      37.0&      21.8&      0.92&         1&     2.575$\pm$0.0022&     +2179&      -153&       276&         1\\
   36&       0.6&      16.2&      11.7&      19.0&      24.3&      0.75&         1&     2.444$\pm$0.0032&      -455&      -348&       143&         2\\
   37&       0.4&      13.6&      15.8&      28.3&      71.9&      0.98&         1&     2.800$\pm$0.0020&       +67&      +613&       194&         1\\
\enddata

\tablecomments{From left to right: target ID, target name, luminosity ($\mathrm{L_{43}}$), sizes ($\mathrm{R_{eff}}$, $\mathrm{R_{rms}}$, $\mathrm{R_{max}}$) , displacement ($\mathrm{d_{QSO}}$), eccentricity ($\mathrm{e}$), systemic redshift ($\mathrm{z_{QSO}}$), redshift of CGM \LyA emission ($\mathrm{z_{Ly\alpha}}$), velocity offset from systemic redshift ($\mathrm{\Delta v_{QSO}}$), velocity offset from peak of \LyA emission in the QSO spectrum ($\mathrm{\Delta v_{peak}}$), dispersion as fit by a single Gaussian ($\mathrm{\sigma_v}$), and best-fit number of Gaussian components ($\mathrm{N_{G}}$). }

\tablenotetext{a}{$\mathrm{L_{43}=L/10^{43}~erg~s^{-1}}$}

\end{deluxetable*}

\section{Results} \label{results}

\begin{figure}[t]
\centering
\includegraphics[width=0.45\textwidth]{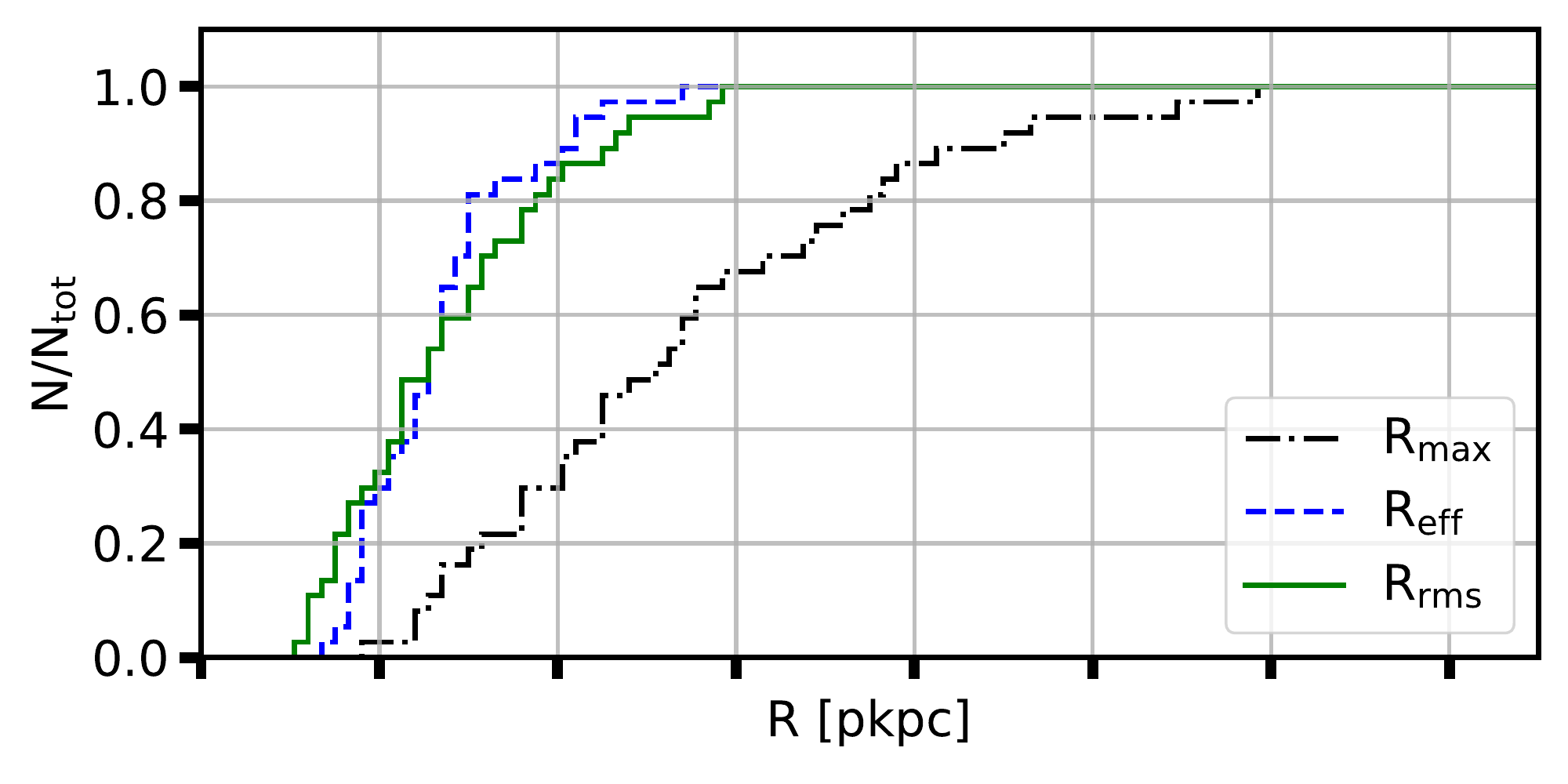}
\caption{ \small{ {Cumulative distributions of the sizes of the detected nebulae in the FLASHES Pilot sample, as measured using: effective radius ($\mathrm{R_{eff}}$), maximum radial extent ($\mathrm{R_{max}}$), and flux-weighted RMS radius ($\mathrm{R_{rms}}$).}} }
\label{fig:cumulativeR}
\end{figure}

In this section we present the 2D morphologies, eccentricities, radial profiles, kinematic properties and integrated spectra of the nebulae detected in the FLASHES Pilot sample. For the survey as a whole we present an averaged radial profile, covering factors, and distributions of kinematics. In order to provide a more complete physical picture of each QSO environment, the basic observational data (surface brightness, velocity, dispersion and integrated spectra) are displayed side-by-side in the extended Figure~\ref{fig:targplots} for each target. 
 
\subsection{Size and Luminosity}

The leftmost column of Figure~\ref{fig:targplots} shows the pNB images generated following Section~\ref{sec:2D}.  {An average limiting surface brightness of $\sim6\times10^{-18}$ erg s$^{-1}$cm$^{-2}$arcsec$^{-2}$ (in a $1~\mathrm{arcsec^2}$ aperture) was achieved. The individual limiting surface brightnesses are presented alongside the observational details in Table~\ref{tab:pilot_targets}.} We detect nebulae (i.e. regions of emission with $\mathrm{SNR_{int}>4.5}$) around  {37 of the 48} objects in our sample. Of these, only one has an effective diameter $\mathrm{D_{eff}=2R_{eff}\geq100}$ pkpc. Excluding the four targets obtained from literature, which were previously known to contain extended emission, we find a detection rate of  {33/44}. The nebulae are found to have projected radii on the order of tens of proper kiloparsecs, with $\mathrm{R_{eff}}\simeq13-55~\mathrm{pkpc}$ and $\mathrm{R_{rms}}\simeq12-59~\mathrm{pkpc}$.  {The maximum radial extent of the nebulae are found to be span a much larger range than the effective radii ($\mathrm{R_{max}}\simeq19-120~\mathrm{pkpc}$)} indicating some degree of asymmetry. We plot the cumulative distribution functions for each measurement in Figure ~\ref{fig:cumulativeR}.  Table~\ref{tab:sizes} below summarizes the distributions of these three parameters. The integrated luminosities range from  {$\mathrm{L_{min}}=0.4\times10^{43}~\mathrm{erg~s^{-1}}$ to $\mathrm{L_{max}}=9.4\times10^{43}~\mathrm{erg~s^{-1}}$, with mean $\mathrm{L_{avg}}=2.7\times10^{43}~\mathrm{erg~s^{-1}}$ and standard deviation $\mathrm{\sigma_{L}}=\pm2.13\times10^{43}~\mathrm{erg~s^{-1}}$}.

\begin{deluxetable}{lccccc}
\caption{Distributions of Measured Sizes \label{tab:sizes}}

\tablehead{   &  \colhead{Min(R)} & \colhead{Max(R)} & \colhead{Mean(R)} & \colhead{Median(R)}  & \colhead{$\mathrm{\sigma(R)}$}\\ 
&  pkpc & pkpc & pkpc & pkpc  & pkpc }
\startdata
 $\mathrm{R_{max}}$&  19&    120&  42&   42&   32\\
 $\mathrm{R_{eff}}$&  14&    55&   21&   23&   14\\
 $\mathrm{R_{rms}}$&  12&    59&   22&   22&   16
\enddata
\end{deluxetable}

\subsection{2D Morphology}

From a quick glance at the pNB images in Figure~\ref{fig:targplots}, it is clear that there is quite a spread in the spatial symmetry of the nebulae. As discussed in Section~\ref{analysis:morphology}, we quantify this using the eccentricity parameter, $0 < e \leq 1$.  { The value for each target is presented in Table~\ref{tab:neb_props}. The detected nebulae are found to exhibit eccentric morphologies, ranging from a minimum of $e=0.51$ to a maximum of $e\sim1$, with a mean (and median) of $0.82$) and a standard deviation $\sigma_e=0.13$. A number of targets with $e\simeq0.9-1.0$ appear to be the result of two or more co-linear patches of emission (IDs 14, 18, 19, and 26). To provide some context, we present the number of distinct spatial components in each object mask alongside the eccentricity in Table~\ref{tab:neb_props}. It is important to remember that what is being measured here is the collective eccentricity of the detected regions, and that - with deeper sensitivity - fainter emission filling the space between and around these regions might be detected, which would lower the eccentricity. We explore the relationship between the surface brightness threshold and measured eccentricity in more detail in Section~\ref{discussion}.}\\

 {The distance between flux-weighted center of mass of the detections and the QSO has a mean value of $d_{QSO,avg}=18\mathrm{~pkpc}$ and also a spread of $\sigma(d_{QSO})=18~\mathrm{pkpc}$. Of the 37 detections, 34 have centers of mass within $50\mathrm{~pkpc}$ of the QSO, while three (IDs 20, 32, 37) have large displacements. While ID 20 appears to have some connection to the QSO, IDs 32 and 37 appear similar in nature to the displaced emission seen in A19's target 25.} \\

\subsubsection{Radial Profiles}

\begin{figure}[t]
\centering
\includegraphics[width=0.45\textwidth]{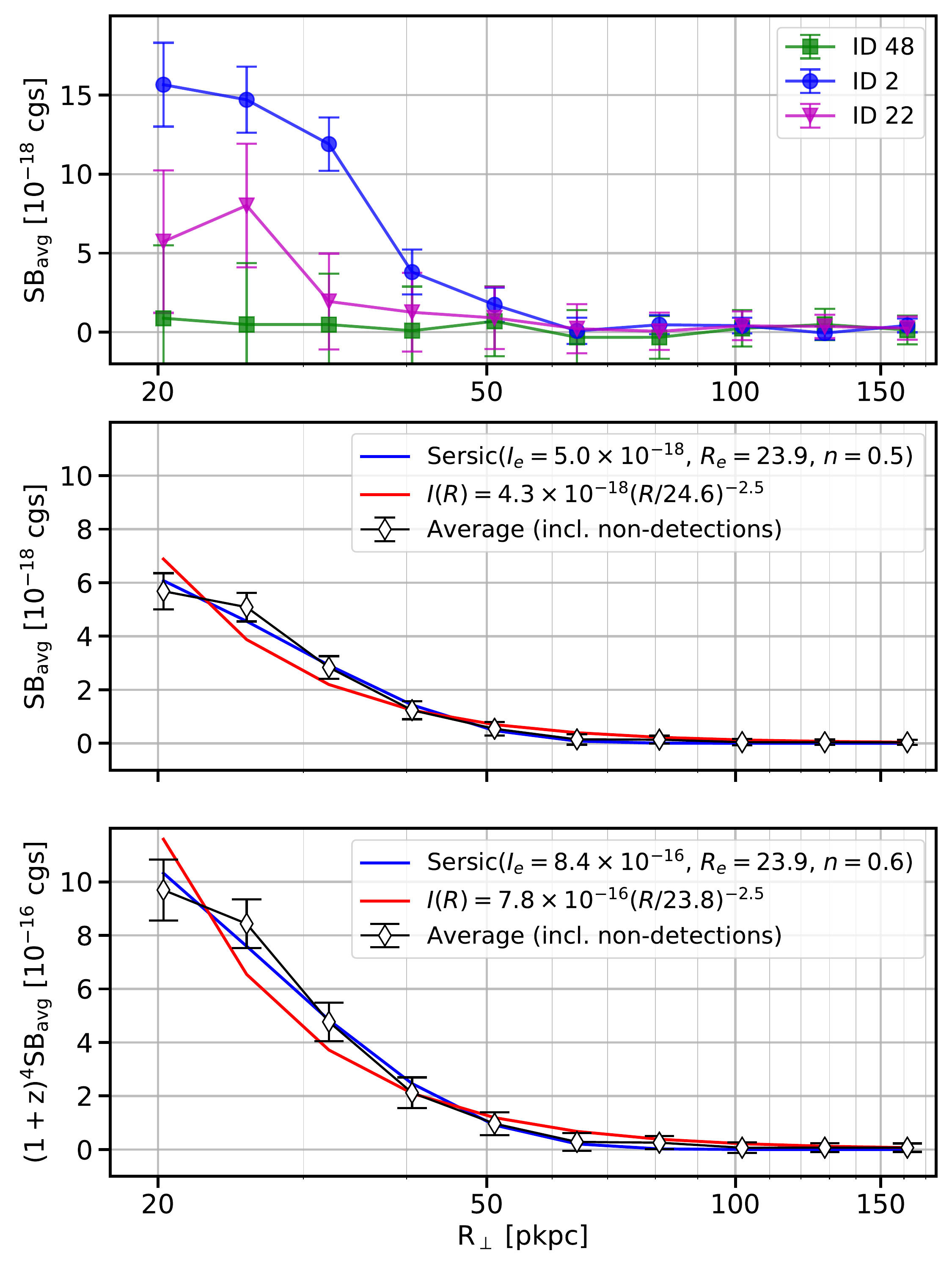}
\caption{ \small{ Circularly averaged radial profiles of the detected CGM, centered on the QSOs.  {The top panel shows three examples; a bright detection (ID2), an intermediate detection (ID 22) and a non-detection (ID 48). The middle panel shows the averaged profiles in observed surface brightness, with a S{\'e}rsic fit and a power-law fit. The bottom panel shows the average of the profiles after scaling each by $(1+z)^4$ to correct for cosmological surface brightness dimming, with the same fits. The x-axis is shown in log-scale.}} }
\label{fig:radial}
\end{figure}

\begin{figure}
\centering
\includegraphics[width=0.45\textwidth]{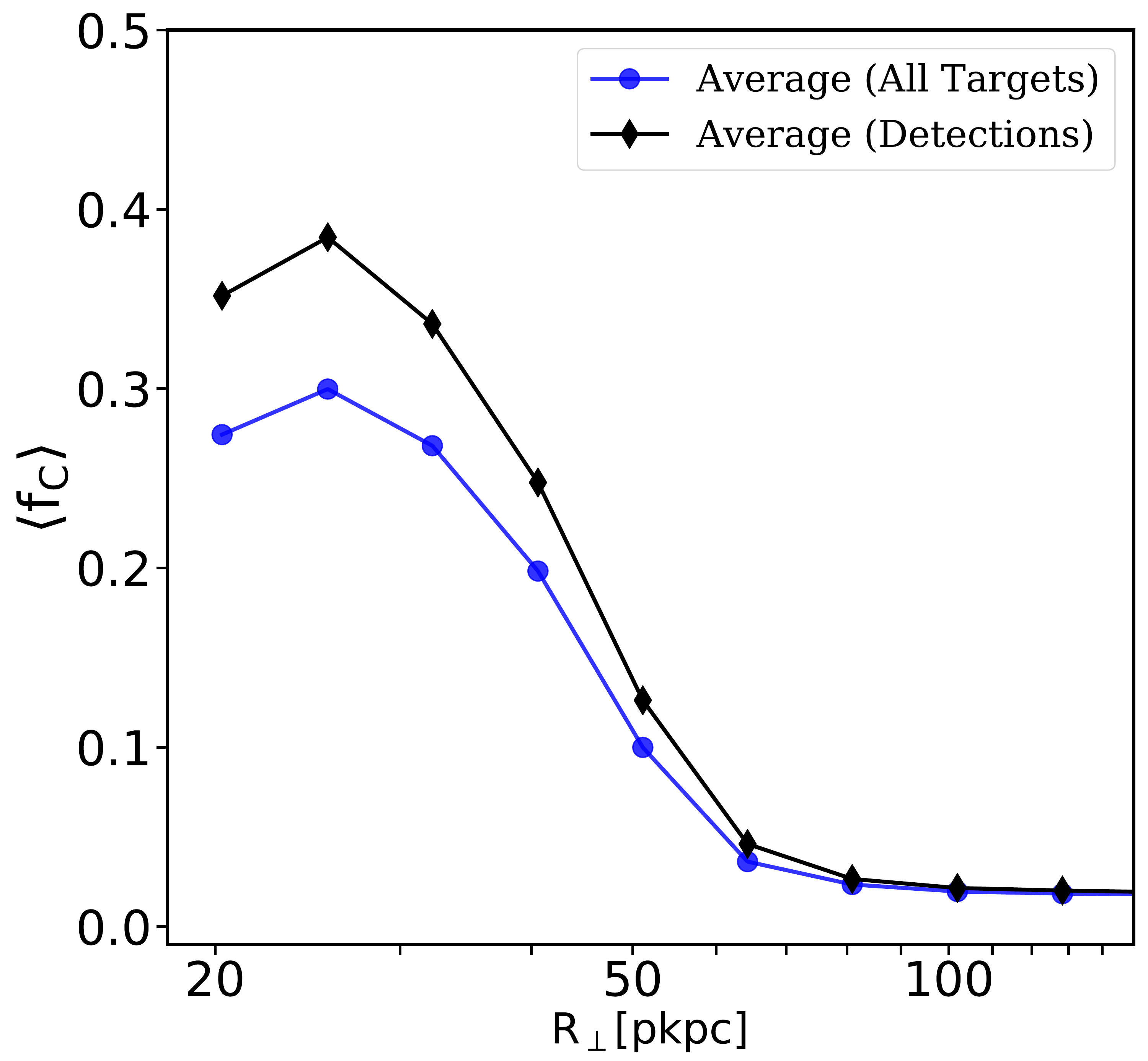}
\caption{ \small{  {Covering fraction of $SNR\geq 2\sigma$ as a function of radius. The average for all fields in the sample is shown with black diamond markers, while the average of all detections is shown with blue circle markers.}} }
\label{fig:fcprof}
\end{figure}

Figure \ref{fig:radial} shows the average radial surface-brightness profiles of the FLASHES Pilot survey  {observations. The average radial surface brightness profile peaks at around $\mathrm{SB_{max}^{obs}}\simeq6\times10^{-18}~\mathrm{erg~s^{-1}~cm^{-2}~arcsec^{-2}}$ ($\mathrm{SB^{(1+z)^4}_{max}}\simeq1\times10^{-15}~\mathrm{erg~s^{-1}~cm^{-2}~arcsec^{-2}}$). On average, the bulk of emission appears to fall within $50$ pkpc of the quasar. We fit two models to each profile; a power law model with the form $I(R)=I_0 (R/R_e)^\alpha$ and a S{\'e}rsic profile. For both observed and adjusted profiles, the emission appears to be best described by a profile with S{\'e}rsic index $n\simeq0.5-0.6$, and half-light radius $R_{e}\simeq24$ pkpc. The observed profile has intensity (surface brightness) at the half-light radius $I_e=5\times10^{-18}~\mathrm{erg~s^{-1}~cm^{-2}~arcsec^{-2}}$ , while the adjusted profile has $I_e^{(1+z)^4}\simeq0.8\times10^{-15}~\mathrm{erg~s^{-1}~cm^{-2}~arcsec^{-2}}$}. We note that an exponential profile is a S{\'e}rsic wih $n=1$, and that a parameter space of $n=0.1-6.0$ was explored during the fitting process using a stochastic optimizer (differential evolution from SciPy - \citep{SciPy, DifferentialEvolution}) which is less susceptible to local minima than standard gradient descent algorithms. \\

Figure~\ref{fig:fcprof} shows the covering factor as a function of projected radius  {for the same two sample-wide averages}. There is a stark contrast between the peak value of $\sim30\%$ for the sample-wide average and the near unity covering factor reported by A19. Even among the detections, the  {average covering factor does not exceed $40\%$}. We discuss these findings further in Section \ref{discussion}.\\

\subsection{2D Kinematic (Moment) Maps}

The second column from the left in Figure~\ref{fig:targplots} shows 2D \LyA velocity (first wavelength moment) maps, generated as discussed in Section \ref{analysis}. The majority of velocities fall within  $\pm300$ $\mathrm{km~s^{-1}}$ of the flux-weighted mean velocity of each nebula.  {The vast majority of the targets do not exhibit any clear kinematic structure. However, two targets (ID 4 and ID 7) stand out from the rest of the sample in this regard. The Eastern side of the extended emission around target 4 appears to be mostly blue-shifted, while the Western side appears to be mostly red-shifted. For target 7, the South/South-East side of the nebula appears to be broadly red-shifted while the North/North-West side is mostly blue-shifted. Determining the significance of such structures is non-trivial given the spectral resolution and spatial covariance in the data. We thus present a full discussion on tests for kinematic coherence in the data in Section~\ref{analysis:kinematics}.}\\

The third column from the left in Figure~\ref{fig:targplots} shows two-dimensional maps of the second wavelength/velocity moment (i.e. velocity dispersion). What appears immediately obvious is that the average dispersions of the nebulae vary significantly, over a range of $\sim200-400$ $\mathrm{km~s^{-1}}$. Within the individual nebulae it is difficult to recognize any clear patterns. It is worth repeating here (as discussed in Section~\ref{analysis}) that these dispersions are upper limits and are influenced by the size of the wavelength window used to calculate them. To obtain more accurate dispersion maps, deep observations are required as they will allow line fitting techniques to be used on a spaxel-by-spaxel basis. \\

\begin{figure}[t]
\centering
\includegraphics[width=0.5\textwidth]{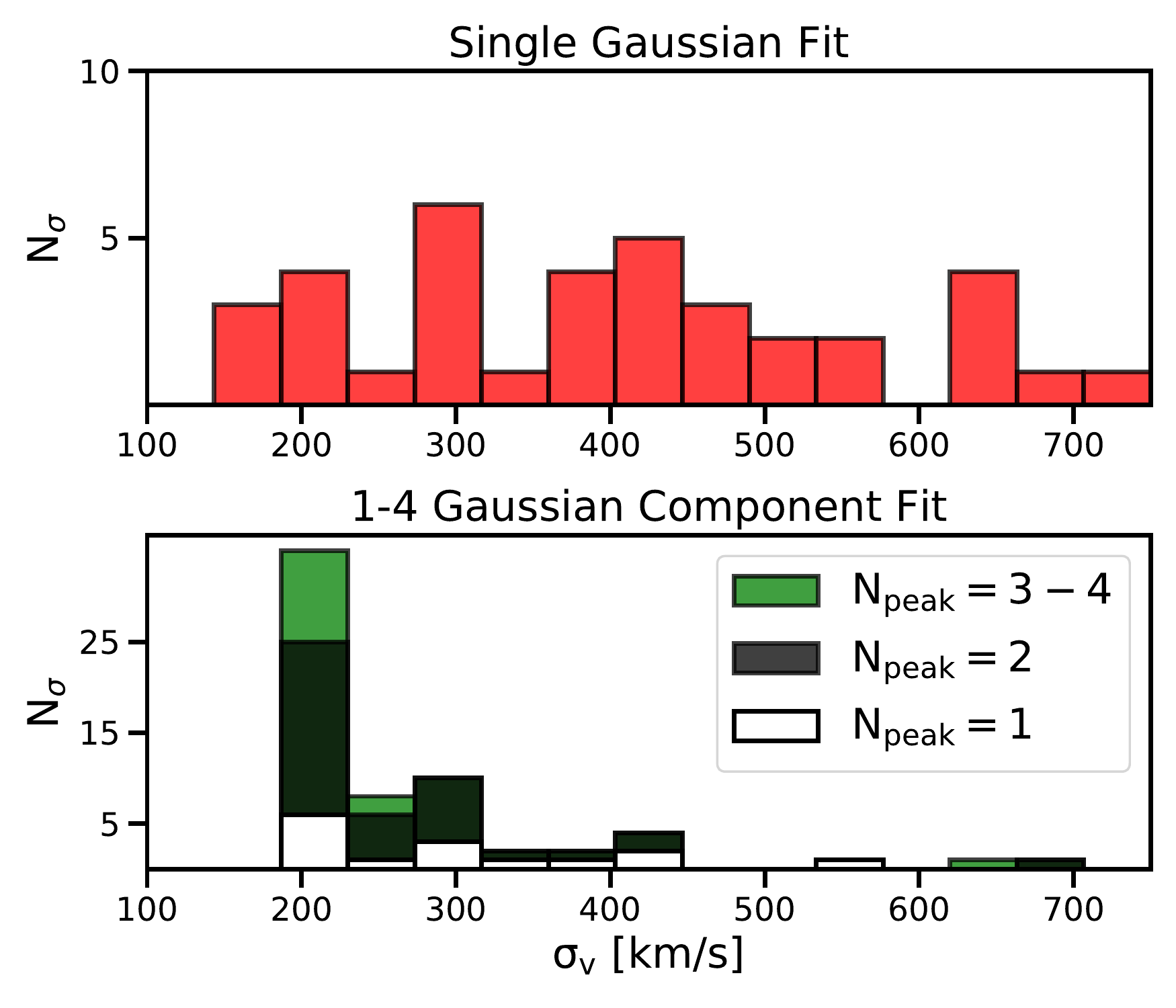}
\caption{ \small{ Top panel: global dispersions of the detected nebulae as measured from a single-component Gaussian fit. Bottom panel: dispersions of individual features when fitting spectra with a composite model of 1-4 Gaussian components.} }
\label{fig:disphist}
\end{figure}

\begin{figure}[t]
\centering
\includegraphics[width=0.5\textwidth]{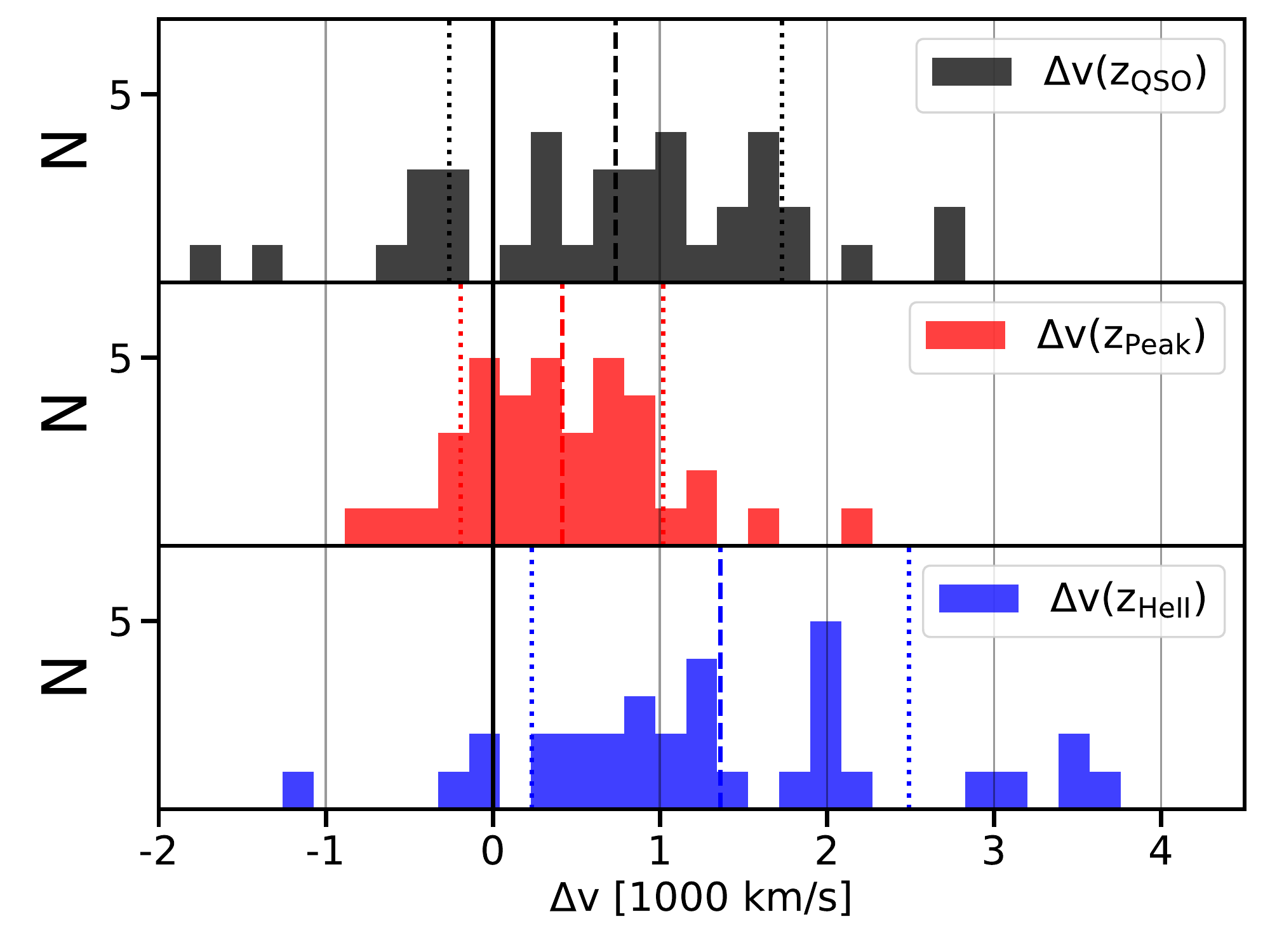}
\caption{ Distributions of CGM \LyA velocity offsets with respect to different redshifts. The top panel shows velocity with respect to the best-fit SDSS/DR12Q QSO redshift. The middle panel shows velocity with respect to the peak of \LyA emission in the QSO spectrum. The bottom panel shows velocity offset with respect to the HeII $\lambda1640$ redshift from SDSS. }
\label{fig:veldist}
\end{figure}

Figure~\ref{fig:veldist} shows the distributions of three velocity offsets, $\mathrm{\Delta v_{QSO}}$, $\mathrm{\Delta v_{peak}}$ and $\mathrm{\Delta v_{HeII}}$. The distribution of velocity offsets with respect to the systemic redshift ($\mathrm{\Delta v_{QSO}}$) is spread over a wide range, from $\mathrm{\Delta v_{QSO}^{min}}=-1647~ \mathrm{km~s^{-1}}$ to $\mathrm{\Delta v_{QSO}^{max}}=+2754~ \mathrm{km~s^{-1}}$ with a median of $\mathrm{\Delta v_{QSO}^{med}}=+871~\mathrm{km~s^{-1}}$ and a standard deviation of $\mathrm{\sigma(\Delta v_{QSO})}=994~ \mathrm{km~s^{-1}}$. The distribution of offsets with respect to the peak of \LyA emission in the QSO spectrum is more concentrated, ranging from $\mathrm{\Delta v_{peak}^{min}}=-738~ \mathrm{km~s^{-1}}$ to $\mathrm{\Delta v_{peak}^{max}}=+2179~ \mathrm{km~s^{-1}}$ with a median value of $\mathrm{\Delta v_{peak}^{med}}=+390~ \mathrm{km~s^{-1}}$ and a standard deviation of $\mathrm{\sigma(\Delta v_{peak})}=606~ \mathrm{km~s^{-1}}$. Finally, the spread in velocity with respect to $\mathrm{z_{HeII}}$ is the widest of all, ranging from $\mathrm{\Delta v_{HeII}^{min}}=-1090~ \mathrm{km~s^{-1}}$ to $\mathrm{\Delta v_{HeII}^{max}}=+3709~ \mathrm{km~s^{-1}}$ with a standard deviation of $\mathrm{\sigma(\Delta v_{HeII})}=1130~ \mathrm{km~s^{-1}}$. The median of distribution is also significantly redshifted ($\mathrm{\Delta v_{HeII}^{med}}=+1195~ \mathrm{km~s^{-1}}$). \\

The rightmost column in Figure~\ref{fig:targplots} shows the integrated nebular spectra, extracted from the data cubes by first applying the 2D emission mask and summing over the spatial axes. Fits to the data indicate that  {33/37} of the profiles can be decently described by a  {one- or two-component} Gaussian fit, with  {four} targets exhibiting more complex line structure. We note that, as these spectra are spatially integrated, the line shape may be a result of the superposition of spatially separated components as well as being influenced by \LyA radiative transfer within a single, unresolved emitter. Given that the global dispersion will be heavily influenced by the presence of multiple kinematic components, we present two sets of measurements in Figure~\ref{fig:disphist} in order to distinguish between the extrinsic (i.e., superposition of spatially separated components) and the intrinsic (i.e., line broadening) dispersion. The former is measured as the width of single-component Gaussian fits (top panel).  { These dispersions range from $\mathrm{\sigma_{v}^{min}}=143~ \mathrm{km~s^{-1}}$ to $\mathrm{\sigma_{v}^{max}}=708~ \mathrm{km~s^{-1}}$, with a mean of $\mathrm{\sigma_{v}^{avg}}=399~ \mathrm{km~s^{-1}}$ and a $1\sigma$ spread in this distribution of $154~ \mathrm{km~s^{-1}}$}. The latter is indicated by the dispersions of the individual Gaussian components wherever a multi-Gaussian (i.e. 1-4 Gaussian components) is the best-fit model. With few exceptions, these dispersions are found to be $<400$ km/s. The single-component dispersion and the best-fit number of peaks are presented in Table~\ref{tab:neb_props}.

\subsection{Stacked \LyA Profiles}

\begin{figure}[t]
\centering
\includegraphics[width=0.45\textwidth]{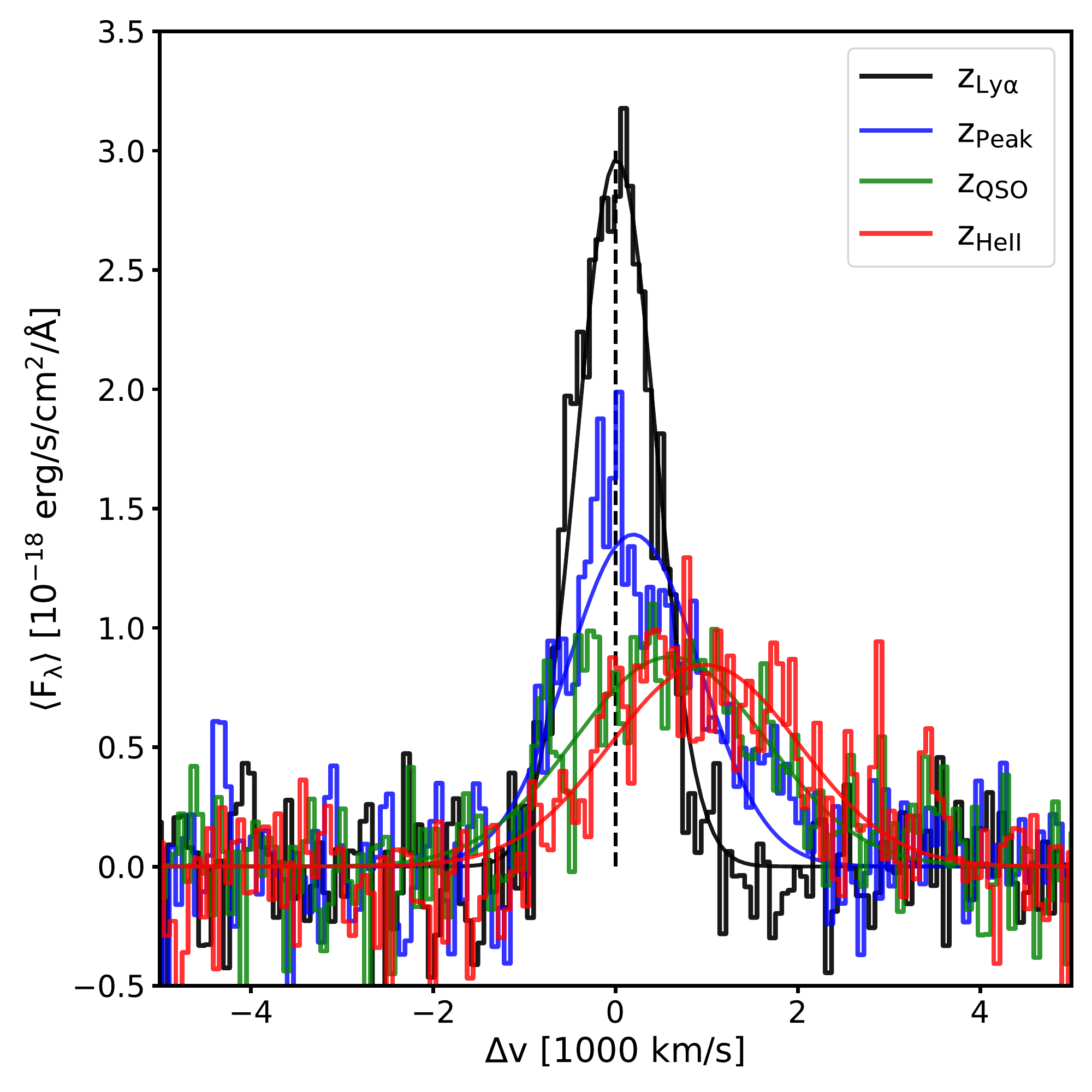}
\caption{ \small{Stacked \LyA profiles of the CGM detections in the FLASHES pilot survey. Different colors indicate different redshifts used to convert from observed to rest-frame wavelengths: the redshift of the CGM \LyA emission itself, the redshift of the peak of \LyA emission in the QSO (blue), the QSO's systemic redshift from DR12Q (green), and the HeII $\lambda1640$ redshift from SDSS (red). } }
\label{fig:stackedspectrum}
\end{figure}

\begin{deluxetable}{cccc}
\caption{Properties of Stacked \LyA Profiles \label{tab:stackedvels}}
\tablehead{  \colhead{Redshift} & \colhead{$\mathrm{F_\lambda}$\tablenotemark{a}} & \colhead{$\mathrm{v_{avg}}$} & \colhead{$\mathrm{\sigma_{v}}$} \\
& \colhead{$\mathrm{erg~s^{-1}cm^{-2}}$\AA$^{-1}$} & \colhead{$\mathrm{km~s^{-1}}$} & \colhead{$\mathrm{km~s^{-1}}$}}
\startdata
$\mathrm{z_{Ly\alpha}}$ & 3.0  &  8    &    430\\
$\mathrm{z_{peak}}$     & 1.4  &  +311  &   721\\
$\mathrm{z_{QSO}}$      & 0.9  &  +754  &   1049\\
$\mathrm{z_{HeII}}$     & 0.8  &  +1367  &   1035\\
\enddata
\tablenotetext{a}{Amplitude of Gaussian fit.}
\end{deluxetable}

Figure~\ref{fig:stackedspectrum} shows stacked \LyA profiles of the detected CGM emission in the pilot sample, converted to rest-frame units using (i) the redshift of the CGM \LyA emission in each field ($\mathrm{z_{Ly\alpha}}$), (ii) the redshift corresponding to the peak of \LyA emission in the QSO spectra ($\mathrm{z_{peak}}$), (iii) the SDSS/DR12Q best-fit systemic redshift of the QSO ($\mathrm{z_{QSO}}$), and (iv) the redshift of HeII emission in the QSO spectrum ($\mathrm{z_{HeII}}$).  {The averaged line profiles exhibit typical Gaussian shapes, with widths reflecting the velocity distributions of the emission relative to each redshift.} Table ~\ref{tab:stackedvels} presents the amplitude, mean and standard deviation of each stacked profile. With the exception of the $\mathrm{z_{Ly\alpha}}$-aligned profile, all of the stacked spectra have a clear redward bias.

\section{Discussion} \label{discussion}

\subsection{From non-detections to Giant \texorpdfstring{\LyA}{Lyman-Alpha} Nebulae}

\begin{figure}[t]
\centering
\includegraphics[width=0.45\textwidth]{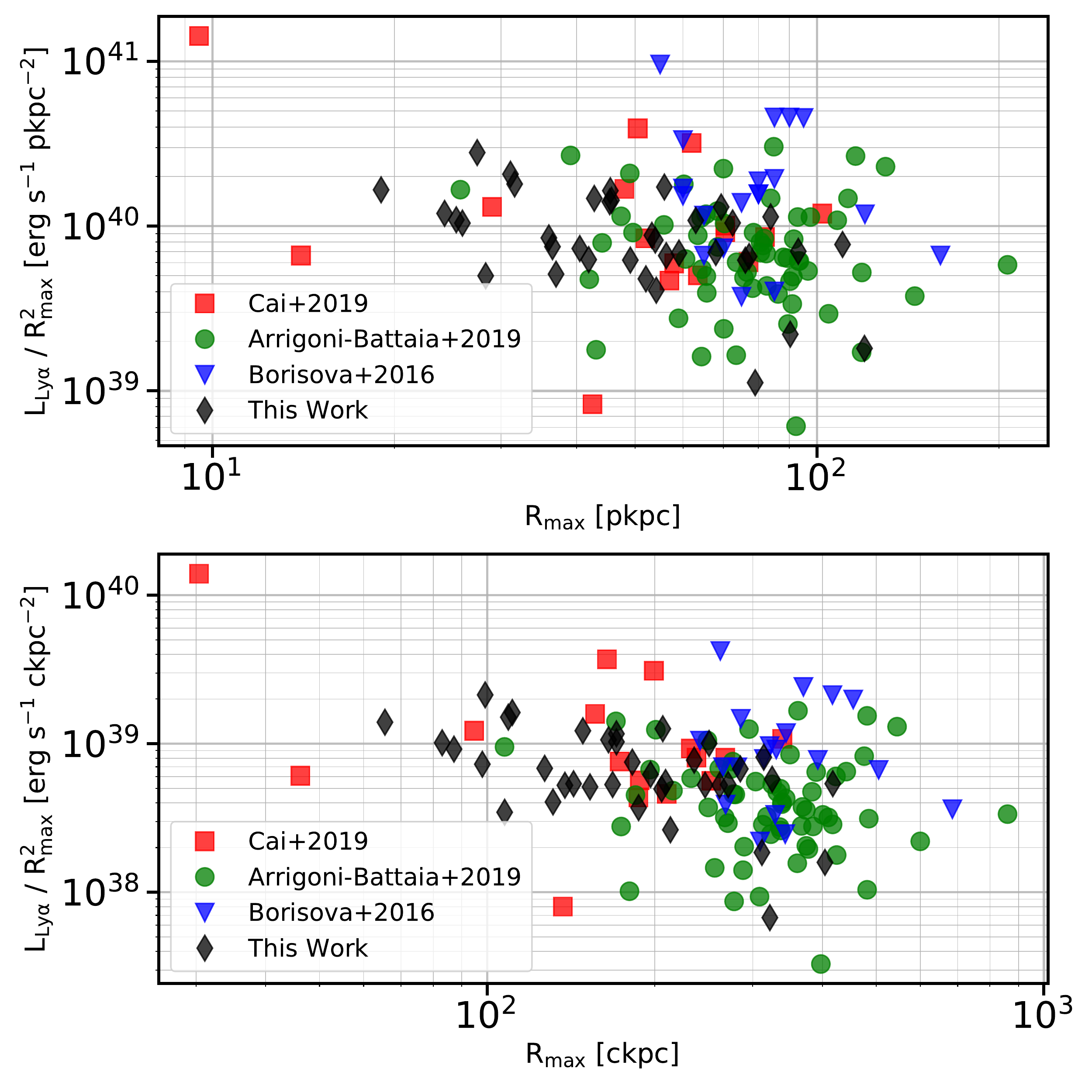}
\caption{ \small{  \textbf{Comparison of  $\mathrm{L_{Ly\alpha}/R_{max}^2}$ as a function of $\mathrm{R_{max}}$ for different surveys. The top panel shows the comparison for sizes in proper kiloparsecs, while the bottom panel shows the same comparison for comoving kiloparsecs. The quantity $\mathrm{L_{Ly\alpha}/R_{max}^2}$ should depend only on the intrinsic radial surface brightness profile of the emission, so comparing nebula of equal size under this metric provides an equitable comparison of the average surface brightness of \emph{detected} regions.}}}
\label{fig:luminosities}
\end{figure}

\cite{Borisova16} report ubiquitous giant nebulae ($R_{max}\geq50\mathrm{~pkpc}$) in their sample of 19 quasars at $z\sim3.5$, with a limiting sensitivity of $10^{-18}\mathrm{~erg~s^{-1}~cm^{-2}~arcsec^{-2}}$ in a $1~\mathrm{arcsec^2}$ aperture in a $1.25\mathrm{\AA}$ layer. \cite{ArrigoniBattaia19} report ubiquitous nebulae on scales of tens to hundreds of $\mathrm{pkpc}$ around their sample of 61 $z\sim3.1$ quasars with similar sensitivity to B16. \cite{Cai2019} report nebulae with projected diameters greater than 50 $\mathrm{pkpc}$ for 14/16 $z\simeq2.1-2.3$ QSOs, again at comparable sensitivity but at significantly lower redshift. Our work now reveals nebulae around  {$37/48$} $z\simeq2.3-3.1$ quasars on  { spatial scales of tens of $\mathrm{pkpc}$}. Because our detection method used wavelength integrated data, there is no perfect one-to-one sensitivity comparison with the above surveys. The average dimming-adjusted radial profile measured here appears to be  {almost an order of magnitude} fainter than that reported in A19, with a peak brightness of $\mathrm{SB_{max}^{adj}}\simeq10^{-15}$ compared to $10^{-14}~\mathrm{erg~s^{-1}~cm^{-2}~arcsec^{-2}}$.  {C19 reports a median surface brightness profile which is also significantly fainter than both A19 and B16, albeit slightly brighter than our average.}\\

\begin{figure}[t]
\centering
\includegraphics[width=0.45\textwidth]{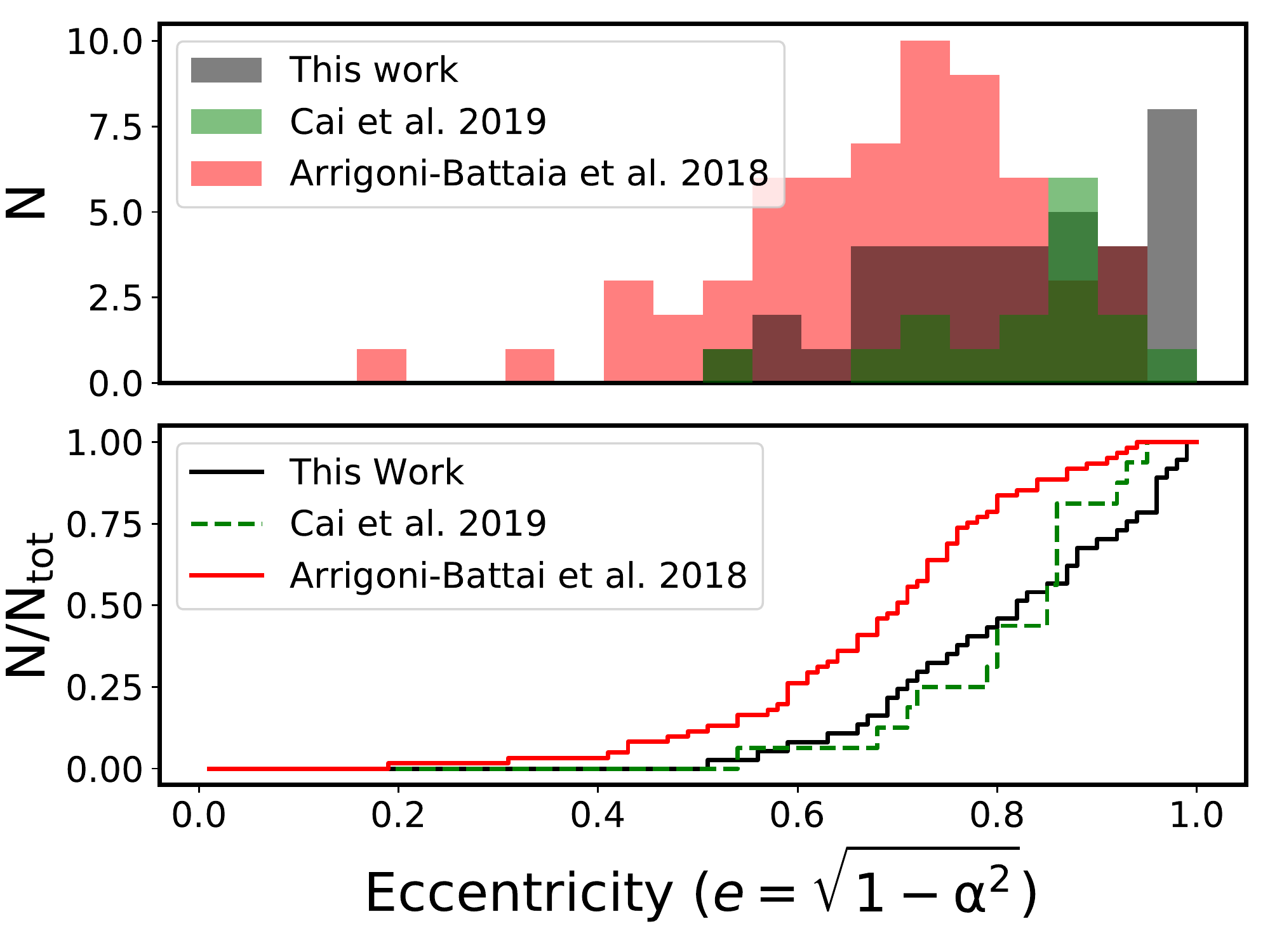}
\caption{ \small{ Cumulative distribution of nebular eccentricities for the FLASHES survey, A19 and C19. The solid black line shows the distribution as measured from the FLASHES $1\sigma$ contour object masks. The dashed black line shows those measured from $2\sigma$ contour object masks. The red line shows the distribution of values presented in A19. The green dashed line shows the distribution from C19. } }
\label{fig:cumulativeSym}
\end{figure}

 {From this picture, it appears possible that there is some redshift evolution from $z\sim3$ to $z\sim2$ towards lower average Ly$\alpha$ surface brightness in the vicinity of QSOs. However, when comparing averaged radial surface brightness profiles, there is a degeneracy between the covering fraction of emitting gas and the average surface brightness of that gas; a faint nebula covering a large area factor may have the same circularly averaged radial surface-brightness profile as a small but bright nebula. Assuming that the luminosity, $\mathrm{L}$ grows approximately as $\mathrm{L(R)\propto R^2}$, where $\mathrm{R}$ is the radius, the quantity $\mathrm{L(R)/R^2}$ should depend only on the intrinsic radial surface brightness profile of the emitting gas. Comparing this quantity for nebulae of similar size then provides a comparison of the average intrinsic surface brightness within the nebular region, which can be used to distinguish between the two above scenarios. In Figure~\ref{fig:luminosities} we compare the detected emission from A19, B16, C19 and this paper in the parameter space of $\mathrm{L(R)/R^2}$ vs. $\mathrm{R}$, where we have used $\mathrm{R_{max}}$ as a proxy for size because it is readily available in all studies. \textbf{We perform this comparison both for sizes measured in pkpc and comoving kiloparsecs ($\mathrm{ckpc}$). No obvious overall difference emerges between the studies. From this comparison, we find that there is no systematic difference in the average intrinsic surface brightness of the detected regions at different redshifts; i.e. nebulae of similar size have similar average brightness.} This implies that the driving factor between the fainter circularly averaged profiles in this work (and C19) is the lower covering fraction of detected gas, rather than globally fainter emission. Although they overlap with the other surveys, the average surface brightness measured by B16 does appear systematically higher than the other surveys, possibly because their sample focused on brighter QSOs (although we measure no significant relationship between QSO magnitude and Ly$\alpha$ luminosity here.) }

\begin{figure}[t]
\centering
\includegraphics[width=0.45\textwidth]{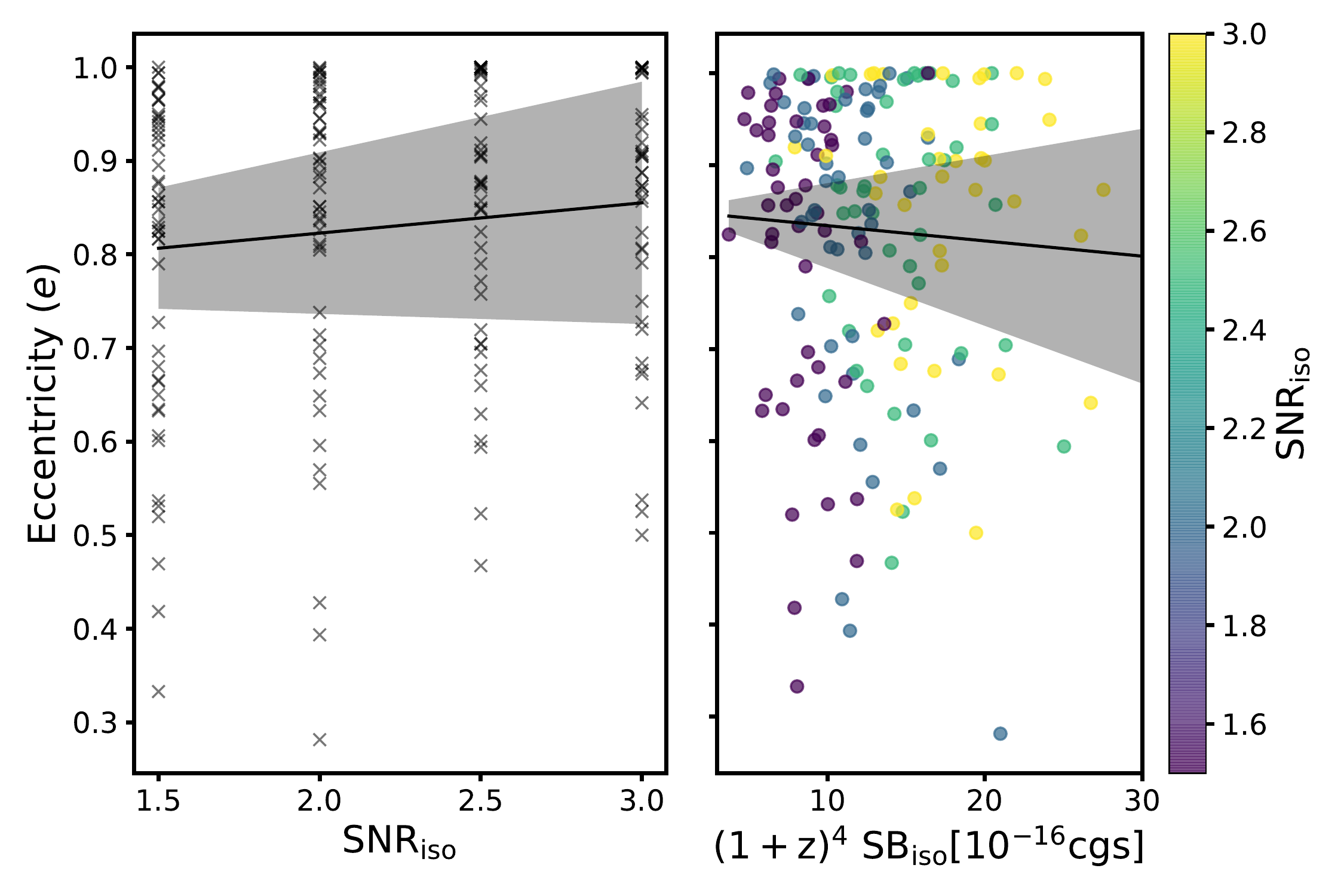}
\caption{ \small{ Change in eccentricity as a function of the increase in limiting surface brightness. Contours show a Gaussian kernel density estimate and the black line with shaded region shows the best-fit linear model with $\mathrm{\pm2\sigma}$ slope uncertainty. The linear regression shows a strong correlation in which eccentricity increases as the surface brightness threshold increases. } }
\label{fig:ecc_v_sb}
\end{figure}

\subsection{Asymmetry of the \texorpdfstring{\LyA}{Lyman-Alpha} Emission}\label{discussion:ecc}

It is clear from visual inspection of the pNB images in Figure~\ref{fig:targplots} alone that there is a pronounced degree of asymmetry in many of the detected nebulae. The distribution of values of the eccentricity parameter ($e$) supports this impression, with a mean value of  {$0.82$}. Figure~\ref{fig:cumulativeSym} compares the cumulative distributions of $\mathrm{e}$ for the FLASHES pilot sample with those presented in C19 and A19 (none were presented in B16). We use the two-sample Kolmogorov-Smirnov (K-S) test to compare the distributions of $e$, and find that we can reject the null hypothesis (that the two samples are from the same underlying distribution) when comparing to A19 ($p<0.002$ - see Table~\ref{tab:symtest} for exact values). However, when comparing to C19 using the two sample K-S test, we cannot reject the null hypothesis ($p\simeq0.44$). Table ~\ref{tab:symtest} summarizes the results of the K-S tests.  {The means of the distributions for A19, C19 and this work, respectively, are $e_{A19}=0.69$, $e_{C19}=0.82$ and $e_{F}=0.82$, with $1\sigma$ spreads in each distribution of $\sigma(e_{A19})=0.15$, $\sigma(e_{C19})=0.1$ and $\sigma(e_{F})=0.13$.}  \\

\begin{figure*}
\centering
\includegraphics[width=\textwidth]{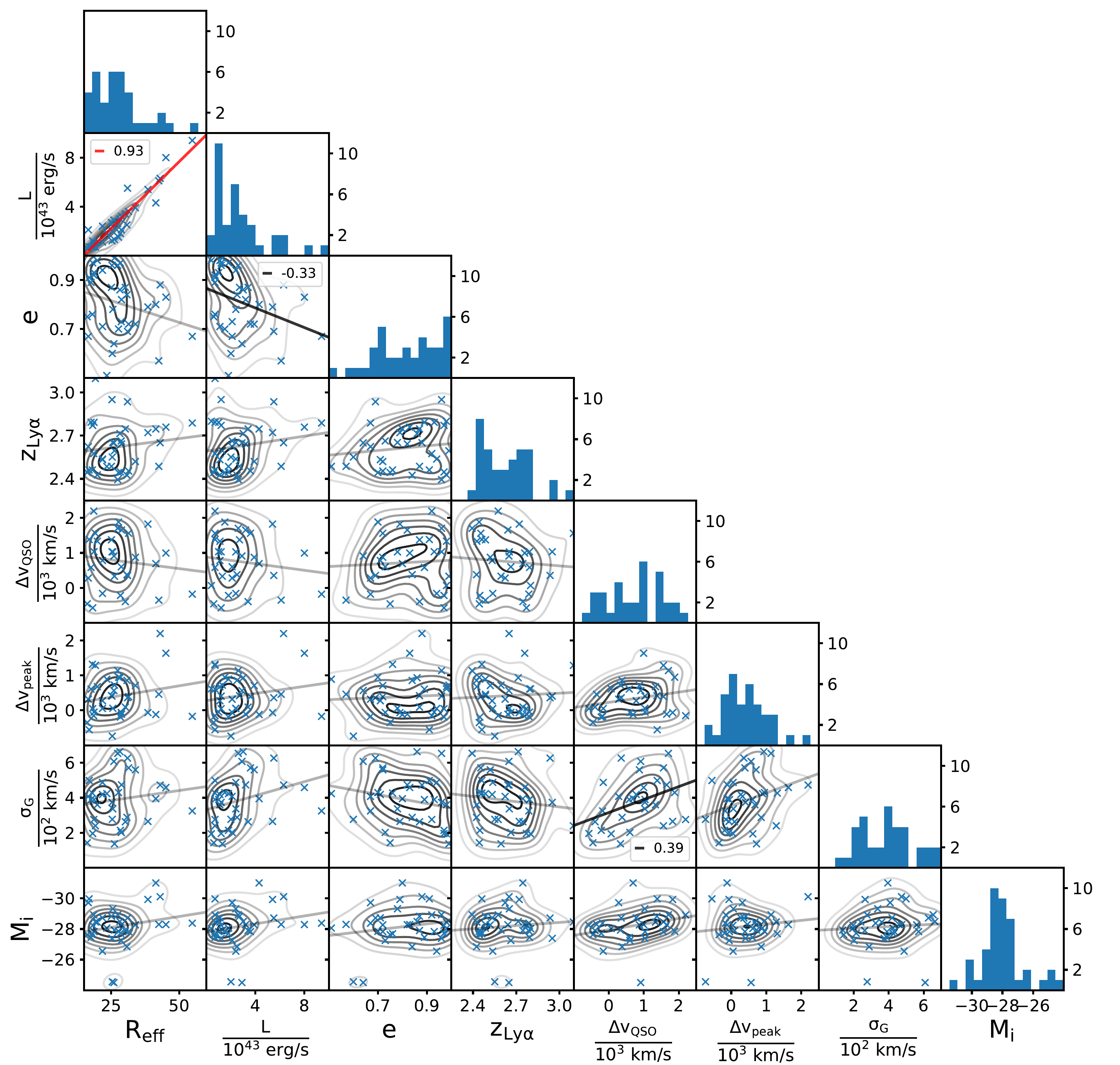}
\caption{ \small{Parameter space plots for the nebulae detected in the FLASHES Sample. $\mathrm{R_{eff}=\sqrt{Area/\pi}}$ is the effective size, $\mathrm{e}$ is the eccentricity ($\mathrm{0\leq e <1}$), $\mathrm{z_{Ly\alpha}}$ is the redshift of the nebular \LyA emission, $\mathrm{\Delta v_{QSO}}$ and $\mathrm{\Delta v_{peak}}$ are the velocity offsets with respect to the systemic redshift and peak of QSO \LyA emission, respectively, $\sigma_G$ is the standard deviation (in km s$^{-1}$) of the best-fit single-peaked Gaussian line profile, and $\mathrm{M_i}$ is the absolute i-band magnitude of the quasar. Contours in each plot show Gaussian Kernel Density Estimates of the 2D distribution. Black and red lines show linear regression models with $p<0.05$ and $p<0.01$, respectively. The r-values of these linear regressions are shown on the relevant tile. Faint grey lines indicate linear regression models with $p\geq0.05$ (i.e., no correlation clearly indicated.) } }
\label{fig:nebprops}
\end{figure*}

 {It is important to note that there are significant differences in extraction technique and sensitivity between our work, A19 and C19. Changes in morphology may equally be a result of changing sensitivity limits as of intrinsic differences in CGM properties. The different extraction techniques, in particular, make a one-to-one comparison of limiting sensitivity very difficult. We can, however, test whether the eccentricity itself depends on the limiting surface brightness used within our sample. In figure~\ref{fig:ecc_v_sb} we show the distribution of eccentricities calculated for different SNR isophotes ($SNR_{iso}=1.5,2.0,2.5$ and $3.0$) for data within $100~\mathrm{pkpc}$ of the QSO. Eccentricities are only calculated if there are at least $10$ spaxels within the isophotal threshold. The top panel shows the same data with the SNR isophotes converted to absolute surface brightness isophotes. Linear regression to the data in both panels does not indicate a significant correlation. This, combined with the fact that C19 report similar eccentricities having higher sensitivity than both the FLASHES Pilot survey and AB19, indicates that limiting sensitivity is at least not the primary driver of the increased eccentricity.} As the \LyA emission we are observing is likely powered by ionizing emission from the QSO, both the illumination and intrinsic distribution of gas play important roles in determining the morphology of the detected nebulae.  {These findings, combined with the finding from the previous section - that a lower covering factor seems to be driving the reduced average surface-brightness - paint a picture of a $z\sim2-3$ CGM that is increasingly patchy and asymmetric at lower redshifts.}  
 \\

\begin{deluxetable}{rcc}
\caption{Comparison of Eccentricity Distributions \label{tab:symtest}}
\tablehead{  \colhead{Test} & \colhead{K-S Statistic} & \colhead{p-value} }
\startdata
 FLASHES v. A19&  0.377 &    $1.9\times10^{-3}$\\
 FLASHES v. C19&  0.245 &    $4.4\times10^{-1}$\\
\enddata
\end{deluxetable}

\subsection{Relationships between Global Nebular Properties}
In Figure~\ref{fig:nebprops} we present a corner plot comparing some key measured properties of the detected nebulae. For each comparison, we test for any relationship between the parameters using a simple linear regression. If the result appears significant (i.e. has a p-value $<0.05$) - we plot the best-fit line and show the r-value of the linear regression, indicating the strength of the correlation ($-1\leq r \leq +1$). The strongest correlation found is no surprise - being between effective radius and luminosity. Visual inspection of this tile confirms a roughly quadratic relationship, as can be expected for these parameters.   {Eccentricity appears inversely related to luminosity, which can be explained if smaller detections tend to more eccentric (see Section~\ref{discussion:ecc}.)} We find a  {no significant} correlation between the absolute i-band magnitude of the QSO, $\mathrm{M_i}$, and the effective size.  {A weak correlation is found between the velocity offset from the systemic redshift ($\mathrm{\Delta v_{QSO}}$) and the global dispersion as measured from a Gaussian fit ($\mathrm{\sigma_G}$). It is not immediately obvious what might cause such a relationship, though it seems plausible that the dispersion and local absorption are both influenced by certain global properties of the surrounding CGM (such as the average temperature of Ly$\alpha$ emitting/absorbing gas). The correlation is not strong enough to motivate a thorough  study here, but presents an interesting element to test with the more sensitive deep survey data. Beyond these few instances, there appear to be no significant correlations between any of the other measured nebular properties.}\\

\subsection{Kinematics of the \texorpdfstring{\LyA}{Lyman-Alpha} Emission \label{analysis:kinematics}}

The flux-weighted centroid of the \LyA emission measured in our sample varies by many hundreds of $\mathrm{km~s^{-1}}$ from the systemic redshift of the QSO ($\mathrm{\sigma(\Delta v_{z})=994~km~s^{-1}}$) and from the peak of QSO \LyA emission ($\mathrm{\sigma(\Delta v_{peak})=606~km~s^{-1}}$). The spread with respect to the SDSS HeII $\lambda$1640 redshift is even more significant, with $\mathrm{\sigma(\Delta v_{HeII})=1130~km~s^{-1}}$. This spread, comparable to that reported in A19, highlights the challenge faced by narrow-band imaging searches for \LyA emission from the CGM around specific targets. All three velocity offset distributions, shown in Figure~\ref{fig:veldist}, present a clear bias towards the red.  { Some of this effect may be attributable to the re-absorption of blue-shifted emission (i.e. rest-frame $\lambda\leq1216~\mathrm{\AA}$) in the intervening IGM. However, it could also indicate that the majority of detections feature outflowing gas with a red-dominated line profile; e.g \cite{Gronke2015}.} \\

\begin{figure}[t]
\centering
\includegraphics[width=0.45\textwidth]{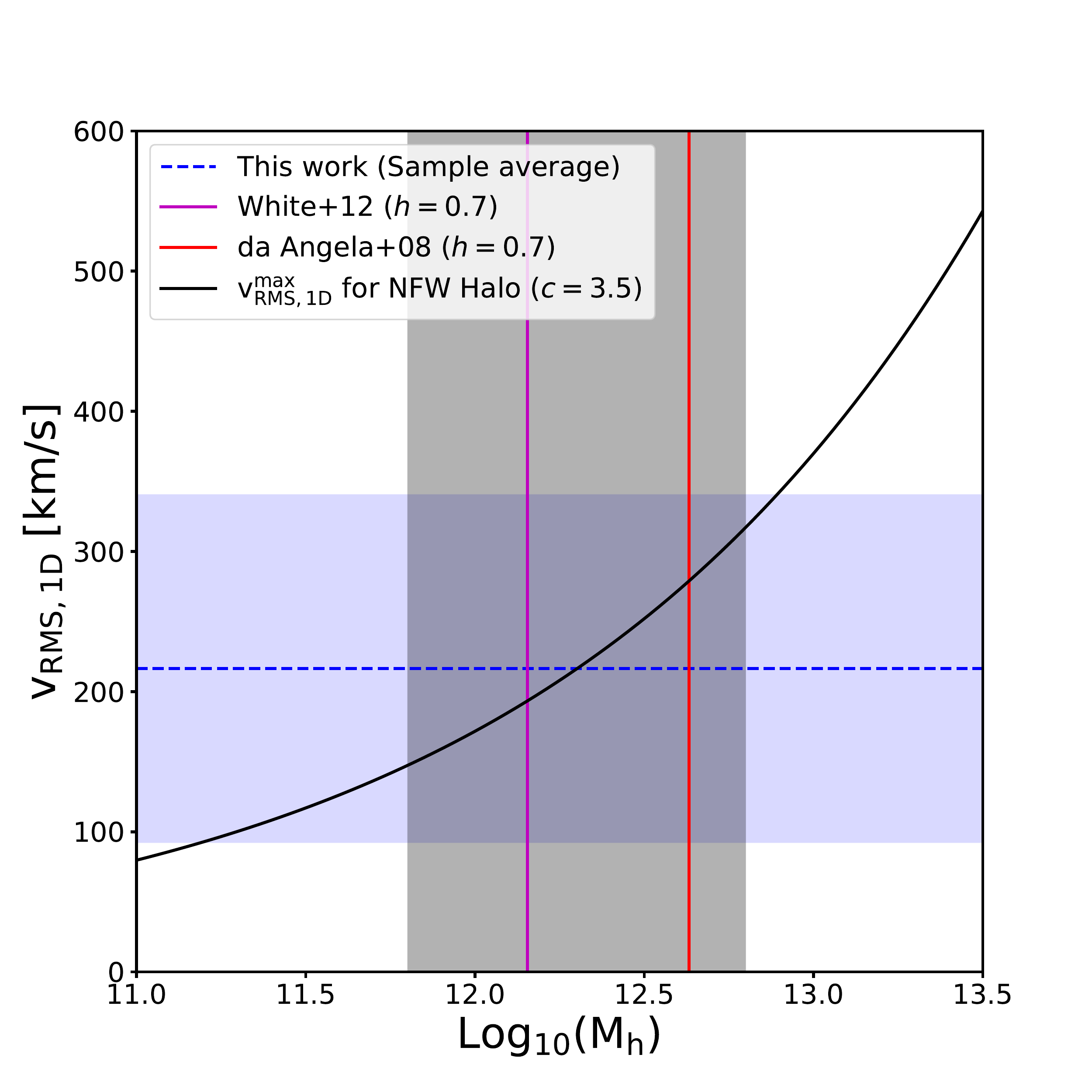}
\caption{ \small{A comparison of the RMS line-of-sight velocity detected in the FLASHES pilot survey with gravitational motions in an NFW halo. The dashed blue line and blue shaded region represent the average and $\mathrm{\pm1\sigma}$ spread in the line-of-sight RMS velocities of FLASHES pilot nebulae, respecitvely. The solid black curve shows the (maximum) RMS line-of-sight velocity of an NFW halo as a function of halo mass following \cite{Munari+2013} ($\mathrm{\sigma_{1D}=0.68v_{200}}$, where $\mathrm{v_{200}}$ is the circular velocity at the virial radius). The grey shaded region indicates the halo masses of high-luminosity QSOs (HLQSOs) at a redshift of $z=2.7$ in \cite{Steidel2012}. } }
\label{fig:velVnfw}
\end{figure}

The average dispersions of the nebulae, shown in the third column of Figure~\ref{fig:targplots}, appear to be in agreement with the finding of A19, in that nearly all targets have mean dispersions $\sigma_{avg}\lesssim400~ \mathrm{km~s^{-1}}$. As we note in Section~\ref{analysis}, the statistical second moments here provide upper limits in the presence of strong noise. However, this finding is supported by our line-fitting analysis of the integrated nebular spectra. As the top panel of Figure~\ref{fig:disphist} the global dispersions of the integrated spectra have a mean of $\mathrm{\sigma_v = 399~km~s^{-1}}$ even when modelled using a single Gaussian component. In the bottom panel of the same figure, we see that the vast majority of global dispersions above $\mathrm{400~km~s^{-1}]}$ disappear when multiple Gaussian components are allowed, indicating that these line-widths are the result of complex line shapes, attributable in part to both the superposition of spatially distinct kinematic components and intrinsically complex spectra (i.e. within a single spaxel).\\  

 {Approximately one third (15/37)} of the detected nebulae appear to be best fit by a single peak, while the plurality  {(17/37)} seem to be best described by a two-component fit, and the remaining few  {(5/37)} have more complex line shapes with three or more components. We note that these best-fit measurements, determined using the BIC, only represent the relative likelihood of the models considered, and the presence of considerable noise and occasional systematics such as bright sky-line residuals should be taken into account when interpreting these results. For example, for target 24, a bright sky line  ($\mathrm{Hg~\lambda4358.3}$) coincided almost exactly with the position of the red-shifted \LyA line. A small wavelength region around this line had to be masked before analyzing the data, so the line complexity here is likely artificial. \\

 {In Figure~\ref{fig:velVnfw}, we compare the measured RMS velocities to the line-of-sight RMS velocity ($\mathrm{v_{RMS,1D}}$) expected for a Nevarro-Frenk-White (NFW) halos \citep{NFW97} with concentration parameter $c=3.5$. We measure the RMS velocity of each nebula detected in the sample and find the average value to be $\mathrm{v_{RMS, avg}=208\pm128~km~s^{-1}}$, which corresponds to the values expected from a halo mass range of $\mathrm{Log_{10}(M_{h} [M_\odot])=12.2^{+0.7}_{-1.2}}$. \cite{Steidel2012} measured the halo masses for a sample of high-luminosity QSOs at a redshift of $\mathrm{z\simeq2.7}$, and found the range to be $\mathrm{Log_{10}(M_{h} [M_\odot])=12.3\pm0.5}$. An analysis of the clustering of $z\sim1.5$ QSOs in the 2dF QSO Redshift Survey by \cite{daAngela08} found that QSOs tend to inhabit $\mathrm{M_h\simeq3\times10^{12} h^{-1} M_\odot}$, regardless of luminosity or redshift, while \cite{white2012} studied the clustering of $2.2\leq z \leq 2.8$ QSOs in the Baryonic Oscillation Spectroscopic Survey and found their results to be consistent with QSO host halo masses of $\mathrm{M_h\simeq 10^{12} h^{-1} M_\odot}$. We thus find that the RMS velocity values among the FLASHES pilot detections are broadly consistent with those expected from gravitational motions in the host dark matter halos of QSOs at their redshift (median redshift $z\simeq2.63$.) It is important to note that there are many more effects contributing to the observed Ly$\alpha$ kinematics beyond gravitational motions in an ideal NFW halo; e.g. outflows, mergers, AGN feedback, and radiative transfer. This comparison was performed to test for any clear \emph{inconsistency} between the expected and measured kinematics. The fact that the results appear to be consistent with halo motions only tells us that we cannot directly rule out an interpretation of the moment maps as reflecting physical motions, not that this is the most appropriate interpretation. The FLASHES deep survey will provide us with an opportunity to perform more detailed modeling of kinematics including radiative transfer effects.}

\subsection{Coherence in the Ly$\alpha$ Moment Maps}

 { As we mentioned in Section~\ref{analysis:kinematics}, two targets - IDs 4 and 7 - appear to exhibit some coherent kinematic structure, with regions that are systematically red- or blue-shifted with respect to their flux weighted center. We test for the presence of systematic structure in two ways: first by measuring the specific projected angular momentum of each nebula and second by performing a simple comparison of 2D kinematic models with and without shear. }\\

\subsubsection{Specific Projected Angular Momentum}
 {We define the flux-weighted average of the projected specific angular momentum for each nebula as:}\\ 
\begin{equation}
    \langle~\vec{j}~\rangle_f = \frac{\sum_x \sum_y F(x,y)\vec{R}_{\perp}(x,y)\times \vec{v}_{z}(x,y)}{\sum_x \sum_y F(x,y)}
\end{equation}
 {where $\vec{R}_{\perp}(x,y)$ is the projected radius, in $\mathrm{pkpc}$, from the flux-weighted centroid of the nebula to the point $(x,y)$, $F(x,y)$ is the flux at that point, and $\vec{v}_z(x,y)$ is the line-of-sight velocity, in $\mathrm{km~s^{-1}}$. }\\

\begin{figure}[t]
\centering
\includegraphics[width=0.45\textwidth]{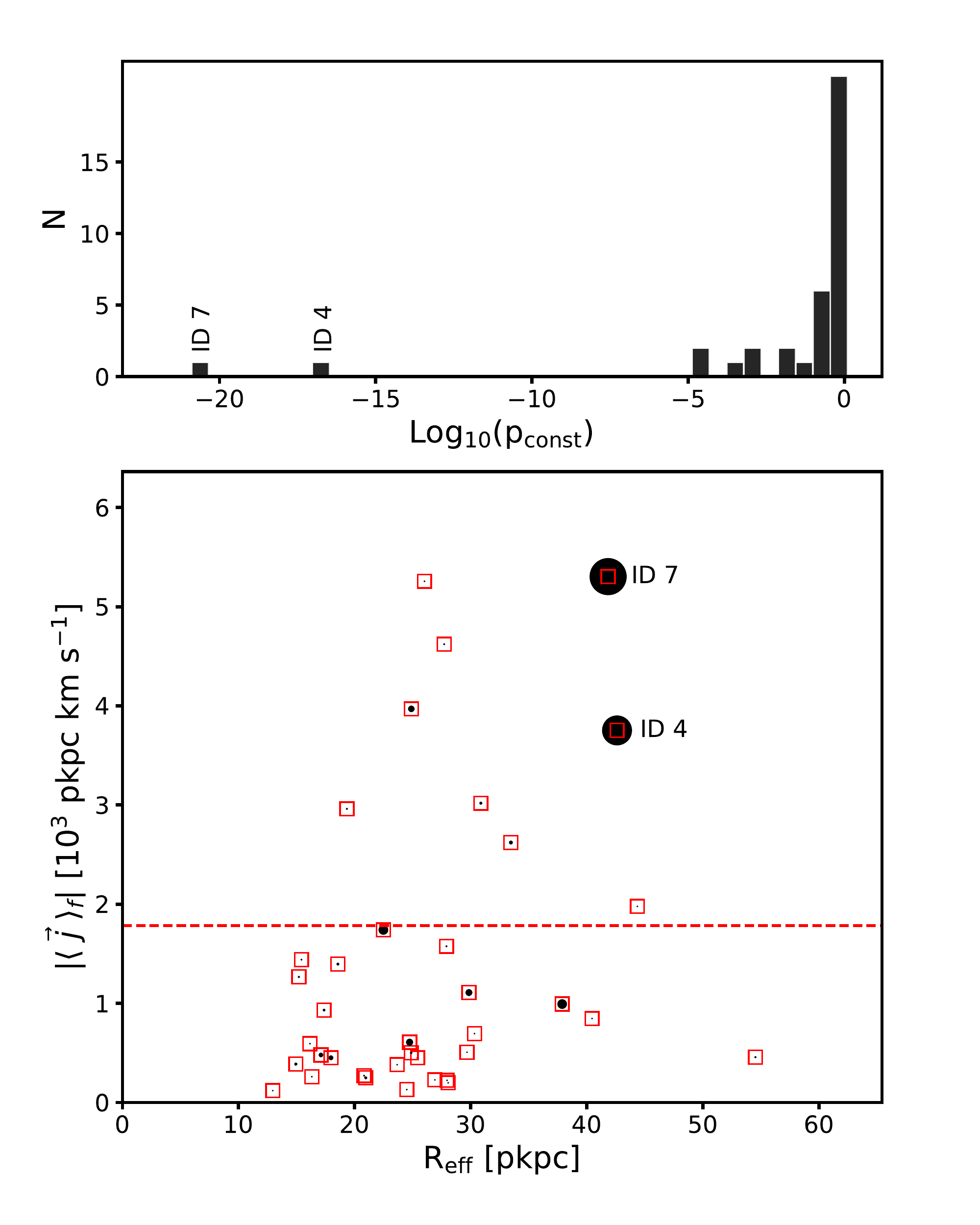}
\caption{ \small{ {Top: log likelihood that the flat model is more appropriate than a shared model, $\mathrm{Log_{10}(p_{const})}$. Bottom: average projected specific angular momentum, $\langle~\vec{j}~\rangle_f$, versus effective radius, $\mathrm{R_{eff}}$. The size of the circular markers is proportional to $\mathrm{-Log_{10}(p_{const})}$ (i.e. the relative likelihood of the shear model). Each target is also shown as a red square of fixed size, for clarity. The horizontal line represents the minimum resolvable angular momentum, discussed in the text.}} }
\label{fig:angmom}
\end{figure}

 {To determine whether a given measured value of $\langle~\vec{j}~\rangle_f$ is significant, we estimate a `minimum' value; i.e. that measured from two adjacent but spatially and kinematically distinct regions, A and B. The effective area of the seeing `disk' in an individual exposure is approximately $\theta_{slit}\times\theta_{seeing}$, where $\theta_{slit}\simeq2''.5$ is the angular width of a slit and $\theta_{seeing}~1-2''$ is the typical seeing (FWHM) at Palomar. As such, let us consider two adjacent $3''\times3''$ regions ($\mathrm{6\times6~px^2}$.) Our typical error on the average velocity in a region of this size is $\mathrm{\delta v_{reg} \sim 20~km~s^{-1}}$, taking covariance from binning into account (see  Section~\ref{analysis:covar}). Let us consider the average velocities of region A and B to be $\mathrm{-2\delta v_{reg}}$ and $\mathrm{+2\delta v_{reg}}$, respectively, such that they are kinematically separated. Finally, assuming a physical plate scale of $\mathrm{\delta R_\perp /\delta x=8~pkpc~px^{-1}}$, which is typical for our redshift range, we get $|\langle~\vec{j}~\rangle_{f}|_{min}=1783\mathrm{~pkpc~km~s^{-1}}$.}\\

\subsubsection{Flat vs. Sheared Model Comparison}
 {We perform a comparison of two basic models using the BIC (see Section~\ref{sec:spectralmodeling}); a flat model, $v(x,y)=v_0$, and a model with linear terms in $x$ and $y$, $v(x,y)=v_0+Ax+By$. This provides a qualitative test as to whether the moment map is flat or has any spatial dependence, to first order. As before, we use the BIC values to estimate the relatively likelihood of each model.  $p_{flat}$ represents the likelihood that the flat model is more appropriate, while $p_{shear}=1-p_{flat}$ indicates the likelihood that the shear model is more appropriate. For the majority of fields, the result is clearly in favor of the flat model ($p_{flat}>0.05$ - 27/37) or only weakly indicative of the sheared model ($p_{flat}>0.01$ - 30/37). A small number of targets indicate some significant likelihood that the shear model better represents the data ($0.01>p_{flat}>10^{-5}$, 5/37). However, for targets 4 and 7, there is a vanishing probability that the flat model is better ($p_{flat}\sim10^{-15}$ and $p_{flat}\sim10^{-21}$).}\\

 {Figure~\ref{fig:angmom} shows the detected nebulae in the parameter space of $\mathrm{R_{eff}}$ versus $|\langle~\vec{j}~\rangle_{f}|$. The size of the markers is indicative of the likelihood of a shear model being correct ($\mathrm{size=-Log_{10}}(p_{flat})$).  From this combined perspective, it is clear that targets 4 and 7 represent two targets which are (i) among the largest detections, (ii) have significant projected specific angular momentum, and (iii) have strong indications from the BIC values that the velocity map is sheared rather than flat. We thus conclude that there is strong evidence of coherent kinematics in these two fields, though we leave the physical interpretation and modeling of this effect as a topic for the deep component of the FLASHES Survey. }

\section{Conclusions} \label{conclusions}
We have conducted the first large IFS survey targeting the $z=2.3-3.1$ CGM in emission. We observed 48 quasar fields over a four-year period using PCWI on the Hale 5m telescope at Palomar Observatory. We find that:

\begin{enumerate}[I]
    \item  {Of the 48 quasars observed (to an average limiting surface brightness of $\sim6\times10^{-18}\mathrm{~erg~s^{-1}~cm^{-2}~arcsec^{-2}}$ in a $1''$ aperture), 37 exhibit extended \LyA emission on a wide range of scales, varying in flux-weighted radius over $\mathrm{R_{rms}=12-59kpc}$ and in maximum (radial) extent over $\mathrm{R_{max}=19-120~pkpc}$. The average flux-weighted projected radius of the nebulae is $\mathrm{R_{rms}^{avg}}=22~\mathrm{pkpc}$ and the spread in the distribution of these sizes is $\mathrm{\sigma(R_{rms})}=16~\mathrm{pkpc}$. The reported sizes are smaller than those in A19 or B16 by about $\mathrm{\Delta R_{max}\sim30~pkpc}$, and comparable to those in C19. }
    \item The circularly averaged radial profiles peak at $\mathrm{SB_{max}^{obs}}=6\times10^{-18}~\mathrm{erg~s^{-1}~cm^{-2}~arcsec^{-2}}$ in observed surface brightness and $\mathrm{SB_{max}^{adj}}=1\times10^{-15}~\mathrm{erg~s^{-1}~cm^{-2}~arcsec^{-2}}$ when adjusted for cosmological surface brightness dimming.
    \item The integrated nebular luminosities range from  {$L_{min}=0.4\times10^{43}~\mathrm{erg~s^{-1}}$ to $L_{max}=9.4\times10^{43}~\mathrm{erg~s^{-1}}$}
    \item  {The nebulae are asymmetric on average, with measured eccentricities ranging from $e=0.51$ to $e\sim1.0$, and a sample-wide mean eccentricity of $e_{avg}=0.82$. We find that the nebulae have a slightly higher eccentricity on average than those found by A19 around $z\gtrsim3$ quasars but the same mean value as those reported around $z\simeq2.1-2.3$ QSOS by C19.}
    \item  { The $\mathrm{S/N\geq2\sigma}$  covering factor profiles peak at $f_{c}\simeq30\%$ at small radii for the sample-wide average when non-detections are included and $\sim40\%$ when they are excluded.}.
    \item The flux-weighted average velocity of the nebulae varies by thousands of $\mathrm{km~s^{-1}}$ with respect to the systemic QSO redshift  {($\mathrm{\sigma(\Delta v_{QSO})}=994~ \mathrm{km~s^{-1}}$) and has a red-shifted bias ($\mathrm{\Delta v_{QSO,med}=+871~ \mathrm{km~s^{-1}}}$). The flux-weighted average velocity of the nebulae also varies significantly with respect to the \LyA peak of the QSO spectrum, albeit by a smaller but considerable amount ($\mathrm{\sigma(\Delta v_{peak})=606~km~s^{-1}}$) and has a lesser but still red-shifted bias ($\mathrm{\Delta v_{peak,med}=+390~ \mathrm{km~s^{-1}}}$)}. 
    \item Most of the integrated nebular emission line profiles are either single-peaked  {(15/37) or double-peaked (17/37)} with a few nebulae exhibiting more complex line shapes.
    \item Global dispersions for the nebulae range from  {$143-708$ km/s, with a mean of $399$ $\mathrm{km~s^{-1}}$ and standard deviation of $155$ km/s. The average RMS line-of-sight velocity is is found to be $\mathrm{v_{RMS, avg}=208\pm128~km~s^{-1}}$, consistent with that expected from QSO host halos with a mass range of  $\mathrm{Log_{10}(M_{h} [M_\odot])=12.2^{+0.7}_{-1.2}}$}.

\end{enumerate}
\section{Acknowledgements}

This work was supported by the National Science Foundation (NSF Award Number 1716907). The authors would like to thank Chuck Steidel (California Institute of Technology) and Ryan Trainor (Franklin \& Marshall College) for their insightful discussions and help with target selection, as well as Sebastiano Cantalupo (ETH Z{\"u}rich)  {and Heather Knutson (Caltech)}. We would also like to thank the staff at Palomar Observatory, for their continuous support over nearly seventy nights of observations since 2014. \textbf{Finally, we would like to thank our reviewer for their time, effort and thoughtful feedback.}

\thispagestyle{empty}

\thispagestyle{empty}

\bibliography{flashesI}{}
\bibliographystyle{aasjournal}

\appendix 

\section{PSF Asymmetry}
\begin{figure}[h]
\centering
\includegraphics[width=0.45\textwidth]{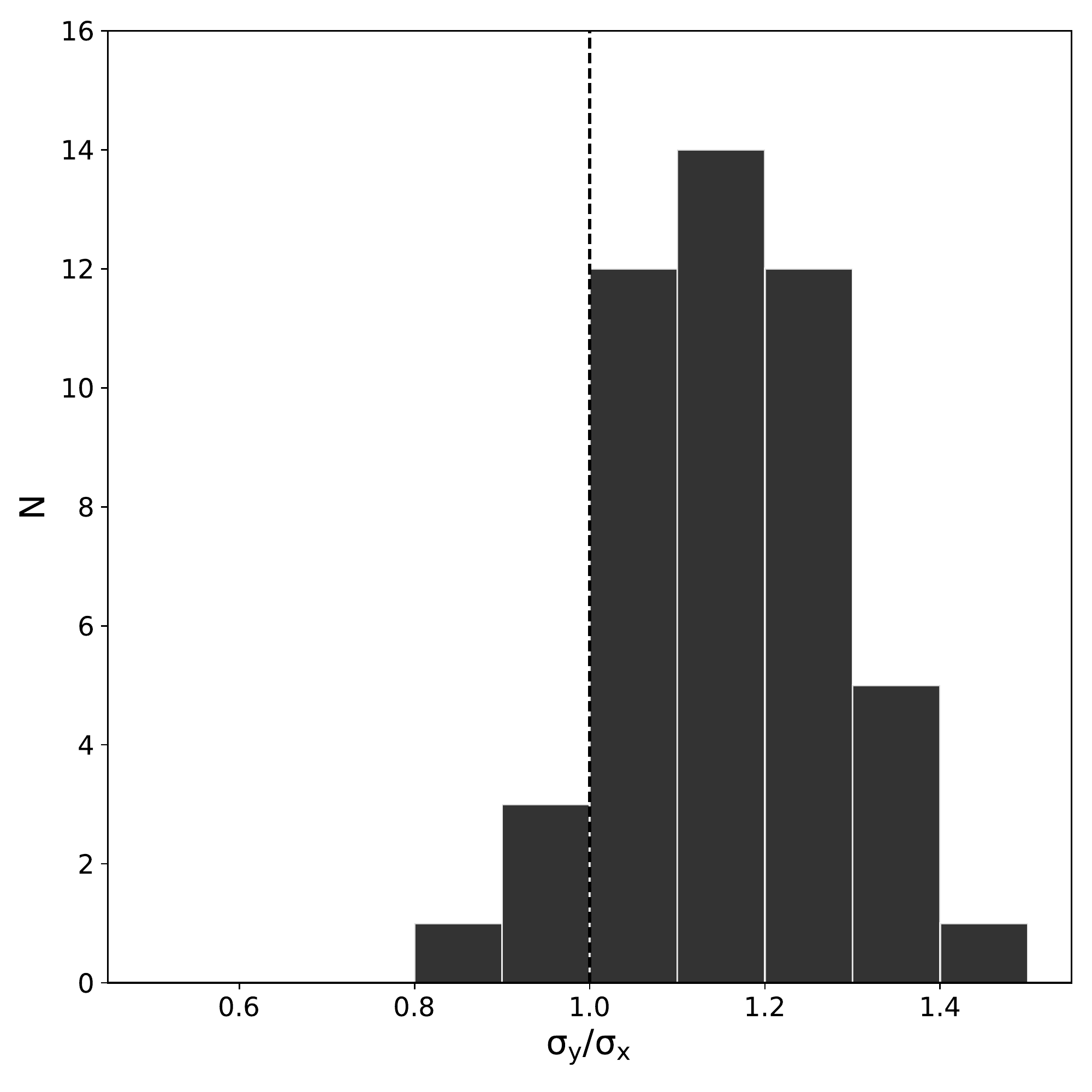}
\caption{Measured asymmetry (i.e. the y:x aspect ratio, determined by a 1D Gaussian fit to the collapsed PSF along each axis) of the PSF in the final pNB images, shown in Figure~\ref{fig:targplots}.}
\label{fig:psfasyms}
\end{figure}
\clearpage
\section{Off-band PSF Subtraction}
\begin{figure}[h]
\centering
\includegraphics[width=0.85\textwidth]{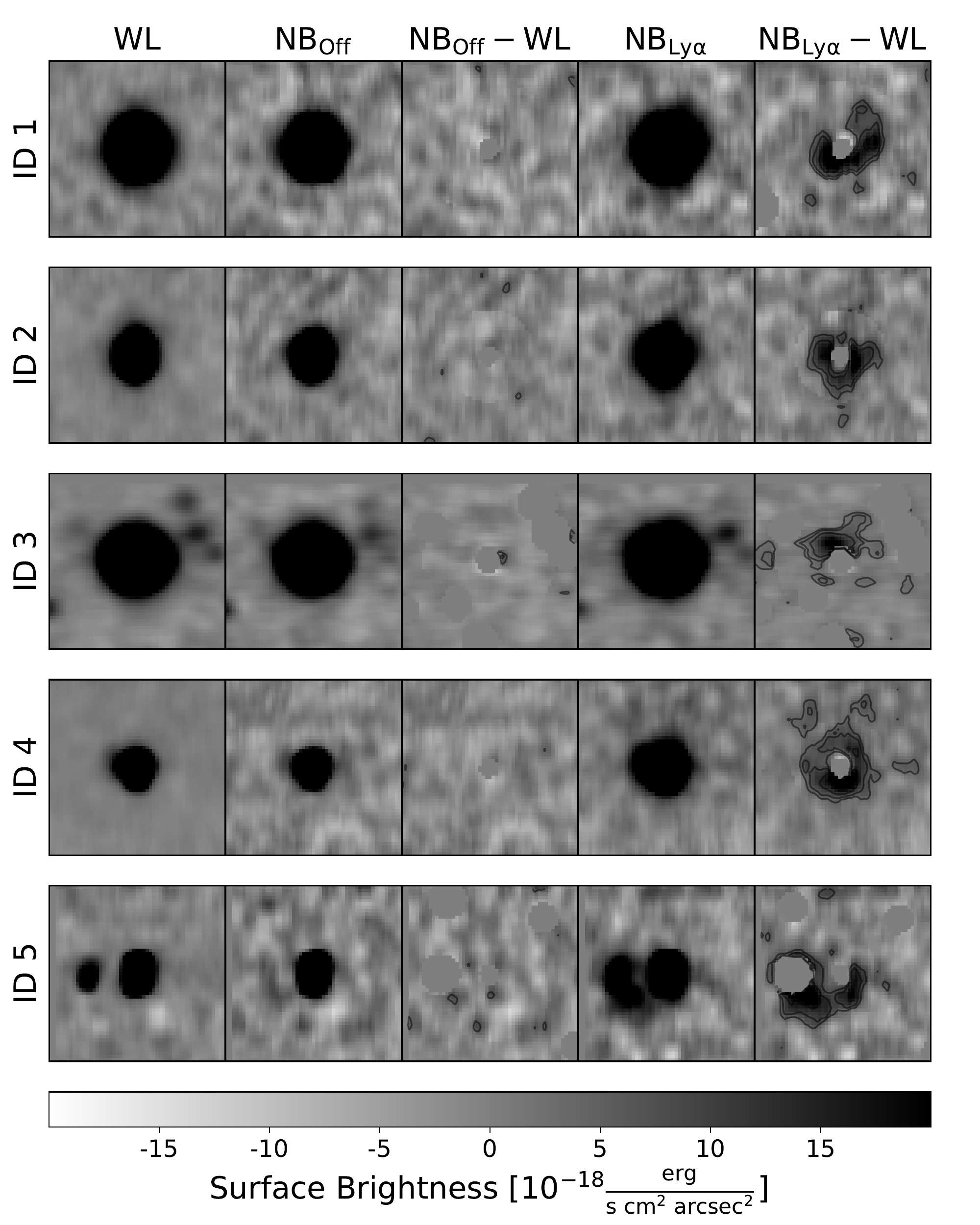}
\caption{ PSF subtraction performed for off-center (i.e. continuum) wavelengths for the first five targets. The columns, from left to right, show: the scaled white-light image, the off-center pNB image, the subtracted off-center pNB image, the Ly$\alpha$ pNB image (for comparison) and the subtracted Ly$\alpha$ pNB image. }
\label{fig:pnbappendix}
\end{figure}

\section{Extended Photometry Data for FLASHES Targets } \label{app:seds}

\startlongtable
\begin{deluxetable*}{lccccccccccccccc}
\tablecaption{Multi-band Photometric Data (AB Magnitudes) for the FLASHES Pilot Sample \label{tab:sed_data}}

\tablewidth{0pt}
\tablehead{ \colhead{Target} & \multicolumn{2}{c|}{GALEX\tablenotemark{a}} & \multicolumn{5}{c|}{SDSS DR12\tablenotemark{b}} & \multicolumn{3}{c|}{2MASS\tablenotemark{c}} & \multicolumn{4}{c|}{WISE\tablenotemark{d}} & FIRST\tablenotemark{e} \\ \tableline
\colhead{SDSS/HS} & \colhead{FUV}	& \colhead{NUV}	&\colhead{u}	&\colhead{g}	&\colhead{r}	&\colhead{i}	&\colhead{z}	&\colhead{H}	&\colhead{J}	&\colhead{K}	&\colhead{W1}	&\colhead{W2}	&\colhead{W3}	&\colhead{W4}	&\colhead{1.4GHz}
}

\startdata
1700+6416	&18.99	&18.77	&16.74	&16.05	&15.94	&15.84	&15.77	&16.64	&17.53	&16.80	&15.70	&15.49	&13.88	&13.09	&...\\
0006+1614	&...	&22.86	&19.18	&18.33	&18.13	&18.10	&17.84	&15.22	&16.15	&15.78	&18.00	&17.42	&16.00	&15.38	&...\\
0012+3344	&...	&20.81	&18.97	&18.32	&18.27	&18.26	&17.97	&17.37	&18.18	&17.46	&17.74	&17.40	&15.61	&14.53	&...\\
0013+1630	&...	&...	&18.93	&18.33	&18.26	&18.17	&17.93	&16.93	&18.22	&17.55	&17.26	&17.14	&15.74	&14.83	&...\\
0015+2927	&...	&...	&19.31	&18.15	&17.99	&18.01	&17.90	&17.23	&18.29	&17.46	&17.59	&17.38	&16.01	&15.46	&...\\
0041+1925	&...	&...	&20.95	&19.86	&19.70	&19.50	&19.32	&17.22	&18.20	&17.70	&19.34	&19.18	&17.35	&15.46	&...\\
0057+0346	&...	&20.65	&18.84	&18.18	&18.13	&18.06	&17.84	&...	&...	&...	&17.62	&17.34	&16.19	&15.30	&...\\
0103+1316	&...	&...	&17.32	&16.57	&16.37	&16.27	&16.16	&17.00	&18.37	&17.21	&16.37	&16.00	&14.02	&13.18	&...\\
0107+1104	&...	&...	&21.51	&20.96	&20.66	&20.68	&20.39	&15.85	&16.84	&16.38	&19.98	&20.22	&16.88	&15.62	&...\\
0108+1635	&...	&...	&18.1	&17.19	&17.00	&16.87	&16.67	&...	&...	&...	&16.56	&16.30	&14.75	&13.91	&...\\
0118+1950	&...	&...	&19.11	&18.14	&17.99	&18.01	&17.89	&16.04	&17.18	&16.73	&17.84	&17.64	&16.12	&15.21	&...\\
0126+1559	&...	&...	&19.77	&19.00	&18.82	&18.81	&18.60	&16.77	&18.10	&17.50	&18.62	&18.37	&16.78	&15.66	&...\\
0132+3326	&...	&...	&19.73	&19.10	&19.18	&19.10	&18.77	&...	&...	&...	&18.00	&18.02	&17.32	&15.27	&...\\
0137+2405	&...	&...	&24.93	&22.23	&21.80	&21.67	&22.06	&16.55	&17.75	&17.54	&20.30	&20.00	&16.98	&15.06	&...\\
0144+0838	&...	&...	&18.92	&18.38	&18.26	&18.27	&18.09	&...	&...	&...	&18.25	&17.46	&15.66	&14.92	&...\\
0205+1902	&...	&...	&18.31	&17.45	&17.27	&17.07	&16.90	&16.64	&17.53	&16.80	&16.75	&16.43	&14.58	&13.84	&...\\
0211+3117	&...	&...	&19.71	&19.00	&18.86	&18.86	&18.79	&17.30	&18.31	&17.75	&18.89	&18.54	&16.64	&15.14	&...\\
0214+1912	&...	&...	&18.77	&17.97	&17.91	&17.74	&17.39	&16.64	&17.53	&16.80	&16.79	&16.41	&14.97	&13.98	&...\\
0300+0222	&...	&22.04	&18.63	&18.04	&17.95	&17.89	&17.61	&16.64	&17.53	&16.80	&17.50	&17.13	&15.43	&14.45	&...\\
0303+3838	&...	&...	&20.52	&19.24	&18.96	&18.87	&18.70	&16.64	&17.53	&16.80	&18.44	&17.98	&16.17	&15.26	&...\\
0321+4132	&...	&...	&18.16	&17.22	&16.75	&16.59	&16.31	&16.64	&17.53	&16.80	&16.08	&15.71	&14.31	&13.59	&...\\
0639+3819	&...	&...	&21.43	&20.36	&20.34	&20.09	&19.69	&16.64	&17.53	&16.80	&19.18	&19.33	&16.92	&15.32	&...\\
0730+4340	&...	&22.43	&20.52	&19.19	&19.06	&19.00	&18.87	&16.20	&17.17	&16.79	&18.67	&18.41	&16.89	&15.70	&...\\
0735+3744	&...	&...	&20.32	&18.68	&18.56	&18.34	&18.13	&...	&...	&...	&17.86	&17.59	&15.78	&15.40	&...\\
0822+1626	&19.79	&19.45	&18.36	&17.88	&17.88	&17.90	&17.67	&16.48	&17.69	&16.39	&17.60	&17.24	&15.63	&14.93	&...\\
0834+1238	&...	&...	&18.95	&18.17	&18.02	&17.94	&17.82	&17.09	&18.00	&17.34	&17.78	&17.40	&15.49	&14.28	&...\\
0837+1459	&...	&...	&18.4	&17.74	&17.74	&17.72	&17.44	&...	&...	&...	&17.24	&16.86	&15.17	&14.41	&...\\
0851+3148	&...	&...	&22.58	&21.32	&21.64	&21.60	&21.47	&15.46	&16.68	&15.82	&20.66	&18.96	&14.97	&13.41	&...\\
0958+4703	&20.99	&21.63	&18.5	&17.73	&17.73	&17.65	&17.35	&...	&...	&...	&17.31	&17.19	&15.88	&14.84	&...\\
1002+2008	&...	&...	&20.01	&19.09	&18.94	&18.85	&18.64	&...	&...	&...	&18.50	&18.16	&15.29	&13.31	&...\\
1011+2941	&...	&...	&16.76	&16.17	&16.09	&16.02	&15.90	&...	&...	&...	&15.87	&15.64	&14.21	&13.45	&...\\
1112+1521	&...	&...	&19.58	&18.10	&17.96	&17.82	&17.58	&...	&...	&...	&17.23	&17.09	&16.43	&15.12	&...\\
1218+2414	&...	&...	&17.46	&16.91	&16.97	&16.94	&16.72	&16.34	&17.68	&16.51	&16.70	&16.40	&14.59	&13.85	&...\\
1258+2123	&...	&...	&22.27	&21.15	&21.33	&21.50	&20.88	&...	&...	&...	&20.54	&19.46	&15.81	&14.54	&...\\
1428+2336	&...	&...	&20.11	&18.82	&18.58	&18.44	&18.39	&16.78	&17.82	&16.87	&18.26	&17.86	&16.06	&15.22	&...\\
1532+3059	&...	&...	&17.9	&17.25	&17.17	&17.14	&16.98	&...	&...	&...	&16.86	&16.55	&15.20	&14.61	&...\\
1552+1757	&...	&...	&23.78	&21.55	&21.31	&21.31	&20.76	&15.28	&16.26	&15.73	&19.06	&19.03	&17.58	&15.19	&...\\
1625+4858	&...	&22.18	&19.52	&18.09	&17.94	&17.63	&17.41	&15.79	&16.57	&16.15	&17.25	&17.11	&15.94	&15.32	&12.86\\
1625+4858	&...	&22.18	&19.52	&18.09	&17.94	&17.63	&17.41	&15.79	&16.57	&16.15	&17.25	&17.11	&15.94	&15.32	&12.86\\
2151+0921	&...	&...	&18.96	&18.42	&18.38	&18.36	&18.10	&16.77	&17.67	&17.47	&18.21	&18.01	&16.77	&15.38	&...\\
2234+2637	&...	&...	&23.59	&22.03	&21.50	&21.00	&20.41	&16.04	&17.17	&16.18	&20.35	&20.21	&16.87	&15.59	&...\\
2241+1225	&...	&...	&18.73	&18.05	&17.93	&17.84	&17.70	&...	&...	&...	&17.60	&17.19	&15.53	&15.05	&...\\
2259+2326	&...	&...	&19.02	&18.26	&18.11	&17.99	&17.65	&...	&...	&...	&17.33	&16.96	&15.40	&14.49	&...\\
2328+0443	&...	&...	&22.67	&20.78	&21.14	&21.55	&20.76	&16.26	&17.33	&16.83	&19.99	&19.17	&16.00	&14.66	&...\\
2338+1504	&21.3	&21.66	&18.19	&17.68	&17.63	&17.50	&17.22	&...	&...	&...	&16.99	&16.69	&15.49	&14.98	&12.27\\
2339+1901	&...	&22.3	&18.12	&17.20	&17.12	&17.00	&16.59	&16.64	&17.53	&16.80	&16.04	&15.89	&14.96	&14.24	&...\\
2340+2418	&...	&...	&21.13	&20.69	&20.56	&20.53	&20.09	&16.90	&18.02	&16.91	&19.71	&20.07	&17.05	&14.98	&...\\
2350+3135	&...	&...	&22.94	&21.02	&20.67	&20.82	&20.65	&...	&...	&...	&19.92	&20.56	&17.34	&15.33	&...\\
\enddata
\tablenotetext{a}{GALEX DR5 \citep{Bianchi+2011}}
\tablenotetext{b}{SDSS DR12 \citep{Alam+2015}}
\tablenotetext{c}{2MASS Catalog \citep{Skrutskie+2006}}
\tablenotetext{d}{AllWISE Catalog \citep{Cutri+2013}}
\tablenotetext{e}{FIRST Survey}
\end{deluxetable*}

\section{Closing Window Calculation}\label{app:iterative_moments}

\begin{figure*}[h]
\centering
\includegraphics[width=0.75\textwidth]{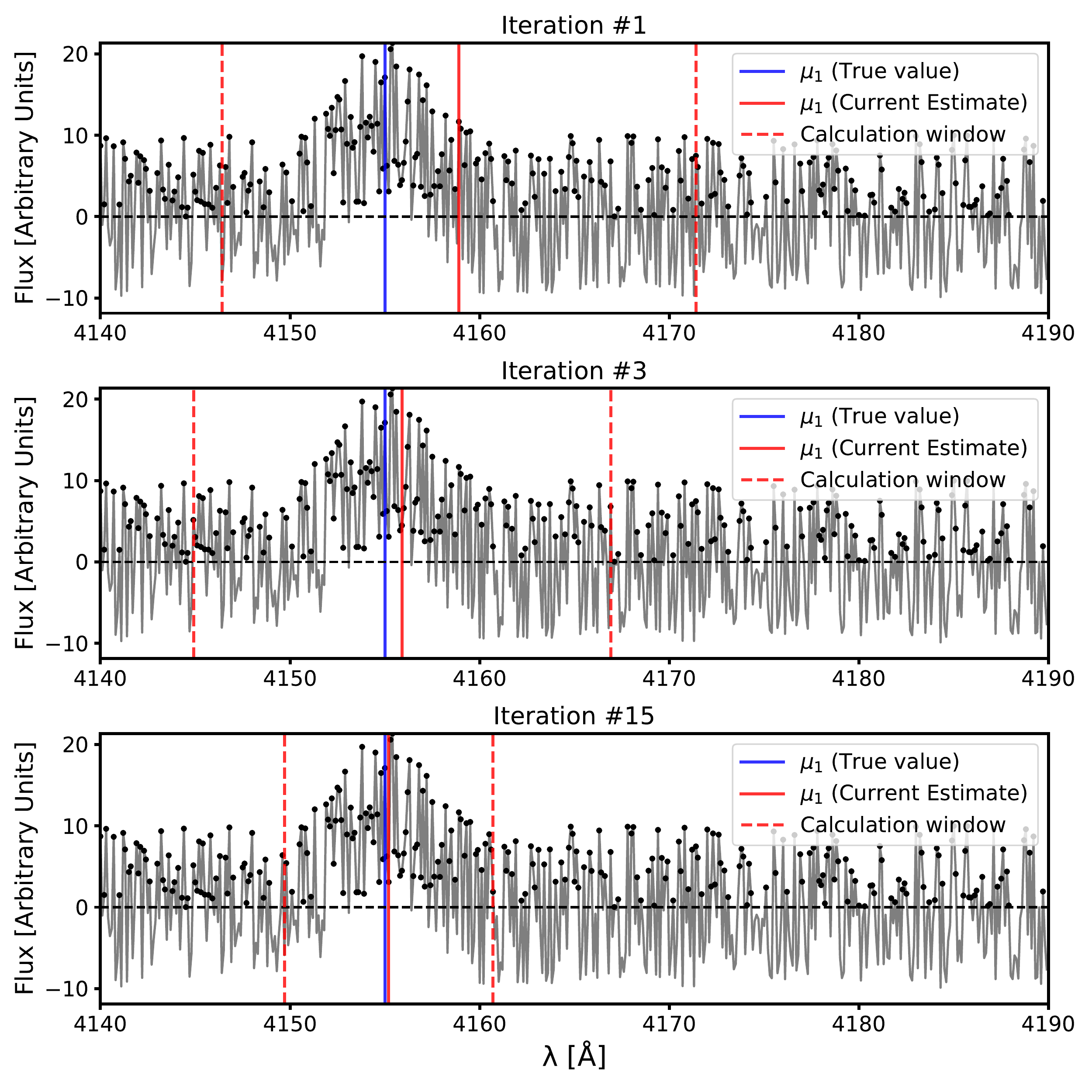}
\caption{ The iterative method, called the 'closing-window' method, used to calculate the first moments of noisy spectra. Each panel shows the noisy spectrum in grey, with positive values shown as black points. The vertical blue line indicates the true first moment of the simulated signal. From top to bottom, the panels shows the estimate of the first moment for the 1st, 3rd and final iterations, respectively. The vertical red line shows the current estimate of $\mathrm{\mu_1}$ at each step, while the dashed vertical lines show the size of the window used to perform the calculation. The window size starts at $25$\AA~in order to explore the full range of the pNB bandwidth. Upon each iteration, the window center is updated to the most recently calculated value of $\mathrm{\mu_1}$, and the window size is decreased by $\mathrm{\Delta \lambda=1}$\AA~ until a minimum size of $10$\AA~is achieved. The shrinking window size helps to mitigate the influence of noise on the calculation, while the iterative process allows the calculation to converge on the true value. }
\label{fig:clwindow}
\end{figure*}

\end{document}